\documentclass[10pt]{article}
\usepackage{amssymb}
\usepackage{amsmath}
\usepackage{amsthm}
\usepackage{latexsym}
\usepackage[dvips]{epsfig}
\usepackage{mathrsfs}
\usepackage{eufrak}

\theoremstyle{plain}

\newtheorem{lemma}{Lemma}
\newtheorem{theorem}{Theorem}

\font\SYM=msbm10
\newcommand{\Real}{\mbox{\SYM R}}
\newcommand{\Complex}{\mbox{\SYM C}}
\newcommand{\Natural}{\mbox{\SYM N}}
\newcommand{\Integer}{\mbox{\SYM Z}}
\newcommand{\Sphere}{\mbox{\SYM S}}

\font\tenscr=rsfs10 scaled1100
\font\sevenscr=rsfs7 
\font\fivescr=rsfs5 
\skewchar\tenscr='177
\skewchar\sevenscr='177
\skewchar\fivescr='177
\newfam\scrfam
\textfont\scrfam=\tenscr
\scriptfont\scrfam=\sevenscr
\scriptscriptfont\scrfam=\fivescr

\newcommand{\TT}[3]{T_{#1 \phantom{#2} #3}^{\phantom{#1} #2}}
\newcommand{\updn}[3]{#1^{#2}_{\phantom{#2}#3}}
\newcommand{\dnup}[3]{#1_{#2}^{\phantom{#2}#3}}

\newcommand{\updnup}[4]{#1^{#2\phantom{#3}#4}_{#3}}

\begin{document}


\title{\textbf{Regularity conditions at spatial infinity revisited}}

\author{Juan Antonio Valiente Kroon \thanks{E-mail address:
 {\tt j.a.valiente-kroon@qmul.ac.uk}} \\
School of Mathematical Sciences,\\ Queen Mary, University of London,\\
Mile End Road,\\
London E1 4NS,
\\United Kingdom.}

\maketitle

\begin{abstract}
  The regular finite initial value problem at infinity is used to
  obtain regularity conditions on the freely specifiable parts of
  initial data for the vacuum Einstein equations with non-vanishing
  second fundamental form. These conditions ensure that the solutions
  of the propagation equations implied by the conformal Einstein
  equations at the cylinder at spatial infinity extend smoothly (and
  in fact analytically) through the critical sets where spatial
  infinity touches null infinity. In order to ease the analysis the
  conformal metric is assumed to be analytic, although the results
  presented here could be generalised to a setting where the conformal
  metric is only smooth. The analysis given here is a generalisation
  of the analysis on the regular finite initial value problem first
  carried out by Friedrich, for initial data sets with
  non-vanishing second fundamental form.
\end{abstract}


\section{Introduction}
The regular finite initial value problem near spatial infinity
introduced in \cite{Fri98a} provides an exceptional tool for analysing
the properties of the gravitational field in the regions of spacetime
``close'' to both spatial and null infinity. This initial value
problem makes use of the so-called extended conformal Einstein field
equations and general properties of conformal structures, and it is
such that both the equations and the data are regular at the conformal
boundary. The formalism of the regular finite initial value problem
has been used to analyse the behaviour of the conformal field
equations on the \emph{critical sets} where null infinity ``touches''
spatial infinity. The analysis in \cite{Fri98a} has shown that the
solutions to conformal field equations develop a certain type of
logarithmic singularities at the critical sets. These singularities
form an intrinsic part of the conformal structure. It could be the
case that these logarithmic singularities do not affect the regularity
of the rest of the spacetime ---and in particular the smoothness of
null infinity--- but the hyperbolic nature of the propagation
equations suggests otherwise. In \cite{Val04a} it has been shown that
there is another class of logarithmic singularities forming at the
critical sets. Subsequent generalisations of these last calculations
carried out in \cite{Val04d,Val04e,Val05a} seem to point out that
stationary solutions play a prominent role in the discussion of the
structure of spatial infinity.

One of the crucial features of the formalism of the regular finite
initial value problem at spatial infinity is that it allows to relate
the behaviour of solutions to the conformal field equations at the
critical sets with properties of the initial data. In particular,
under the assumption of a time symmetric initial data set with an
analytic conformal metric, it was possible to find conditions on the
free initial data near spatial infinity which ensure that for
solutions developing from from these data, singularities cannot occur.

Given the need to systematise and gain further insight into the
results for initial data with a non-vanishing second fundamental form
given in \cite{Val04e,Val05a}, in this article a generalisation of the
discussion of \cite{Fri98a} is carried out: it is studied how to construct
conditions on free data of a class of non-time symmetric initial data
so that their development does not have the type of logarithmic
singularities discussed in \cite{Fri98a}. The understanding of the
logarithmic singularities in \cite{Val04a,Val04d,Val04e,Val05a} is an
outstanding open problem.

In order to state the main result of this article, it is recalled that
if one uses the so-called \emph{conformal Ansatz} to construct
asymptotically Euclidean solutions to the vacuum Einstein constraint
equations under the assumption that the initial data is maximal, then
the freely specifiable data on, say, a compact mainfold $\mathcal{S}$
is given in terms of a pair of symmetric tensors
$(h_{ij},\mathring{\Phi}_{ij})$, where $h_{ij}$ is the \emph{conformal
  metric} of the initial hypersurface $\mathcal{S}$ and
$\mathring{\Phi}_{ij}$ is $h$-tracefree ---throughout this article
$i,j,k,\ldots$ will denote spatial tensorial indices taking the values $0,1,2$.
Following the discussion in \cite{DaiFri01}, the tensor
$\mathring{\Phi}_{ij}$ is determined in a neighbourhood
$\mathcal{B}_a\subset \mathcal{S}$ of \emph{the point at infinity}
$i\in\mathcal{S}$ by providing a real number $A$ (\emph{the
  expansion}), constant 3-vectors $P^i$ (\emph{the linear momentum}),
$J^i$ (\emph{the angular momentum}), $Q^i$ (\emph{the conformal
  momentum}), and a complex function $\lambda$ which is $C^\infty$ in
$\mathcal{B}_a\setminus \{i\}$. It can be seen that the real part of
$\lambda$ gives a contribution, $\mathring{\Phi}_{ij}[\lambda^{(R)}]$,
with properties similar to the contributions by $A$, $P^i$ and $Q^i$,
hence it can be thought of as containing the \emph{higher mass
  multipoles} of $\mathring{\Phi}_{ij}$. On the other hand, the
imaginary part of $\lambda$ gives a contribution
$\mathring{\Phi}_{ij}[\lambda^{(I)}]$ with properties analogous to
$J^i$. Accordingly, it can be interpreted as giving the \emph{higher
  angular momentum multipoles} of $\mathring{\Phi}_{ij}$.

\bigskip In terms of $h$-normal coordinates $x^i$ based on $i$, some
technical assumptions will be made on the freely specifiable data. The
precise details of these assumptions will be discussed in the main
text. The conformal metric $h_{ij}$ which in normal coordinates
centred on $i$ takes the form $h_{ij}=-\delta_{ij} + \hat{h}_{ij}$ is
taken to be in $C^\omega(\mathcal{B}_a)$ (the set of analytic
functions on $\mathcal{B}_a$), where in addition the technical
condition $\delta^{ij}\hat{h}_{ij}=0$ will be assumed. Furthermore,
$P^i=0$ and $r\lambda^{(R)}, r\lambda^{(I)} \in
E^\infty(\mathcal{B}_a)$, where a function $f$ is said to be in
$E^\infty(\mathcal{B}_a)$ if it can be written as $f=f^{(1)} + r
f^{(2)}$ with $f^{(1)}, f^{(2)} \in E^\infty(\mathcal{B}_a)$, and
$r^2= (x^1)^2+(x^2)^2 + (x^3)^2$.  This last assumption is a necessary
condition for the solution of the Licnerowicz equation to be expanded
in terms of powers of $r$ only. In addition, it will be assumed that
both $\lambda$ and $h_{ij}$ are related in such a way that the
solutions to the momentum constraint are of the form $r^{-3}\Phi_{ij}$
with $\Phi_{ij}\in E^\infty(\mathcal{B}_a)$.

\bigskip
In order to express the \emph{regularity conditions} product of
the present analysis, a couple of further tensors will be required:
\begin{eqnarray*}
&& b_{ij} \equiv \updn{\epsilon}{kl}{j}\left( D_k r_{il}-\frac{1}{4} h_{il} D_k {}^{(3)}r \right), \\
&& c_{ij}[\lambda^{(R)}] \equiv \dnup{\epsilon}{(j}{kl} D_{|k} \chi_{l|i)}[\lambda^{(R)}],
\end{eqnarray*}
with $\chi_{ij}[\lambda^{(R)}]$ the part of the second fundamental
form constructed out of $\mathring{\Phi}_{ij}[\lambda^{(R)}]$
---obtained from $\mathring{\Phi}_{ij}[\lambda^{(R)}]$ by multiplying
by a conformal factor---, and $r_{ij}$, ${}^{(3)}r$ ,$D_k$, respectively the
Ricci tensor, the Ricci scalar and the Levi-Civita connection of
$h_{ij}$, and $\epsilon_{ijk}$ denotes the volume form on
$\mathcal{S}$. The tensor $b_{ij}$ is the Hodge dual of the
\emph{Coton(-York-Bach) tensor} of $h_{ij}$.

\begin{theorem}\label{thm:main0}
  For the class of data under consideration, the solution to the
  regular finite initial value problem at spatial infinity is smooth
  through the critical sets only if
\begin{eqnarray*}
&& \mathscr{C}( D_{i_p} \cdots D_{i_1} b_{jk})(i)=0 \quad p=0,1,2,\ldots \\
&& \mathscr{C}( D_{i_q} \cdots D_{i_1} c_{jk}[\lambda^{(R)}]\updn{\epsilon}{i_1 j}{l})(i)=0 \quad q=0,1,2,\ldots
\end{eqnarray*}
is satisfied by the initial data. If the above conditions are
violated at some order $p$ or $q$, then the solution will develop
logarithmic singularities at the critical sets.
\end{theorem}

In the above expression $\mathscr{C}$ denotes the operation of taking
the symmetric-tracefree part of the relevant tensor. An alternative
statement of the theorem written in the more natural language
of space spinors will be given in the main text. It is a non-obvious
result ---to be proved in the main text--- that the condition on the
tensor $c_{ij}$ is actually a condition on the freely specifiable
tensor $\mathring{\Phi}_{ij}$.

\subsubsection*{Overview of the article}
The article is structured as follows. In section
\ref{section:preeliminaries}, in order to fix notation and conventions
some relevant preliminaries concerning the construction of spacetimes
from an initial value problem are reviewed. These include the
construction of asymptotically Euclidean solutions to the vacuum
Einstein constraint equations from free data on a compact manifold.
There is also a brief review of the solutions to the momentum
constraint on flat space and also in the more general case where the
conformal metric is not flat. Of particular relevance is subsection
\ref{section:freedata} where the assumptions made on the class of
initial data being considered are spelled out.

Section \ref{section:manifold_Ca} contains a brief discussion of the
construction of the bundle manifold $\mathcal{C}_a$ which provides a
convenient alternative representation of the asymptotic region
of the initial manifold: one in which the point at infinity $i$ is
blown up to a set which is topologically a 2-sphere. In particular
subsections \ref{subsection:normal_exp_i} and
\ref{subsection:normal_exp_I0} discuss aspects of normal
expansions at infinity. The ideas of these subsections are essential
for the sequel. The techniques in section \ref{section:manifold_Ca} are slight
adaptations of the original constructions given in the
seminal reference \cite{Fri98a} ---there is a more recent detailed 
discussion in \cite{Fri04}. The purpose of this section is to
introduce necessary notation and to serve as a quick reference for the
reader.

Section \ref{section:spinor_mc} discusses the solutions of the
momentum constraint of the bundle manifold $\mathcal{C}_a$. This
section builds from an analysis given in references \cite{Val04e} and
\cite{Val05a}. However, the particular level of detail required for
our analysis, in particular with regards to the non-flat case, is 
not to be available elsewhere in the literature.

Section \ref{section:weyltensor} builds on the results of the previous
section to provide a detailed analysis of the normal expansions of the
Weyl spinor near infinity. The main results of this section are presented in
theorems \ref{thm:symmetries} and \ref{thm:vanishing}. These theorems
can be regarded as the generalisation of theorem 4.1 in
\cite{Fri98a}, for initial data sets with a
non-vanishing second fundamental form.

Section \ref{section:Fgauge} gives a brief overview of the ideas
behind the so-called \emph{regular finite initial value problem at
  spatial infinity}. Again, the discussion is kept to a minimum and
has the purpose of introducing notation and ideas which will be of
relevance in the sequel. In particular, subsection
\ref{subsection:regularityconds} contains the crucial result of
\cite{Fri98a} stating that the solution to the conformal propagation
equations will generically develop logarithmic singularities at the
sets where null infinity touches spatial infinity. How to eliminate
such singularities by means of conditions under the class of initial
data under consideration is the concern of the rest of the article.

Section \ref{section:main} provides the main result of the article,
theorem \ref{thm:main}, which recasts the technical regularity
conditions in terms of conditions on the Coton tensor of the conformal
metric and on what is essentially the curl of the second fundamental
form. This theorem brings together the discussions of section
\ref{section:weyltensor} and \ref{section:Fgauge}.

Section \ref{section:extensions} discusses the possibility of
generalisations of the main result to other classes of data. A
particularly desirable generalisation would be to of a class of data
including stationary solutions.

Finally section \ref{section:conclusions} discusses the connections of the
conditions obtained in the main result to similar ---but crucially not
identical--- conditions which have been obtained from an analysis of
purely radiative spacetimes ---see \cite{Fri88} and also the
concluding remarks in \cite{Fri98a}.

The article contains two appendices. The first one contains a number
of spinorial identities which are used through out the main text. The
second appendix contains a discussion of the asymptotic expansions of
the solutions of the vacuum Einstein constraint equations for the
class of data under consideration. The appendix
\ref{appendix:momentum} briefly reviews some relevant results obtained
in \cite{DaiFri01} and extends some of the analysis therein as the
data required to render our analysis non-trivial turns out to be
slightly more general than the one considered in the aforementioned
reference.

\bigskip
One of the main challenges of the analysis presented in
this article is to bring together results and ideas obtained in
different frameworks and cast them in a common language which allows,
in turn, to obtain new results and hopefully also valuable new
insights into the structure of the gravitational field near spatial
infinity.

\section{Preliminaries}
\label{section:preeliminaries}

This article is concerned with properties of asymptotically flat
spacetimes $(\tilde{\mathcal{M}},\tilde{g})$ solving the Einstein
vacuum field equations
\begin{equation}
\tilde{R}_{\mu\nu}=0. \label{Einstein:eqns}
\end{equation}
The metric $\tilde{g}_{\mu\nu}$ will be assumed to have signature
$(+,-,-,-)$ and $\mu,\nu,\ldots$ are spacetime indices taking the
values $0,\ldots,3$ . The spacetime $(\tilde{\mathcal{M}},\tilde{g})$
will be thought of as the development of some initial data prescribed
on an asymptotically Euclidean Cauchy hypersurface
$\tilde{\mathcal{S}}$. The data on $\tilde{\mathcal{S}}$ are given in
terms of a 3-metric $\tilde{h}_{ij}$ of signature $(-,-,-)$ and a
symmetric tensor field $\tilde{\chi}_{ij}$ representing the second
fundamental form induced by $\tilde{g}_{\mu\nu}$ on
$\tilde{\mathcal{S}}$. As mentioned in the introductions
$i,j,k,\ldots$ will be spatial tensorial indices taking the values
$1,2,3$. The Einstein vacuum field equations imply the constraint
equations
\begin{eqnarray}
&& {}^{(3)}\tilde{r} - \updn{\tilde{\chi}}{i}{i} +\tilde{\chi}^{ij} \tilde{\chi}_{ij}=0, \\
&& \tilde{D}^i \tilde{\chi}_{ij}  -\tilde{D}_j \updn{\tilde{\chi}}{i}{i}=0,
\end{eqnarray}
where $\tilde{D}$ denotes the Levi-Civita connection and
$^{(3)}\tilde{r}$ the Ricci scalar of the metric $\tilde{h}_{ij}$. The
initial data $(\tilde{\mathcal{S}},\tilde{h}_{ij},\tilde{\chi}_{ij})$
will be assumed to be asymptotically Euclidean ---for simplicity the
situation with only one asymptotically flat end will be considered. In
the asymptotically flat end it will be assumed that coordinates
$\{y^i\}$ can be introduced such that
\[
\tilde{h}_{ij} =-\left(1+\frac{2m}{|y|}  \right) \delta_{ij} + \mathcal{O}\left( \frac{1}{|y|^2}\right), \quad \tilde{\chi}_{ij}=\mathcal{O}\left( \frac{1}{|y|^2}\right) \quad \mbox{ as } |y|\rightarrow \infty,
\]
with $|y|^2 =(y^1)^2 +(y^2)^2 +(y^3)^2$ and $m$ a constant ---the ADM
mass of $\tilde{\mathcal{S}}$. In addition to these asymptotic
flatness requirements, it will be assumed that there is a
3-dimensional, orientable, smooth \emph{compact} manifold
$(\mathcal{S},h)$, a point $i \in \mathcal{S}$, a diffeomorphism
$\Phi:\mathcal{S}\setminus \{i\} \longrightarrow \tilde{\mathcal{S}}$
and a function $\Omega \in C^2(\mathcal{S}) \cap
C^\infty(\mathcal{S})$ with the properties
\begin{subequations}
\begin{eqnarray}
&& \Omega(i)=0, \quad D_{j}\Omega(i)=0, \quad D_j D_k \Omega(i)=-2h_{jk}(i), \label{asymptotic:1} \\
&&  \Omega >0 \mbox{ on } \mathcal{S}\setminus \{ i\}, \label{asymptotic:2} \\
&& h_{ij} = \Omega^2 \Phi_* \tilde{h}_{ij}. \label{asymptotic:3}
\end{eqnarray}
\end{subequations}
The last condition shall be, sloppily, written as $h_{ij} = \Omega^2
\tilde{h}_{ij}$ ---that is, $\mathcal{S}\setminus \{i\}$ will be
identified with $\tilde{\mathcal{S}}$. Under these assumptions
$(\tilde{\mathcal{S}},\tilde{h})$ will be said to be
\emph{asymptotically Euclidean and regular}. Suitable punctured
neighbourhoods of the point $i$ will be mapped into the asymptotic end
of $\tilde{\mathcal{S}}$. It should be clear from the context whether
$i$ denotes a point or a tensorial index.

\subsection{The constraint equations}
In order to discuss the asymptotic properties of the solutions of the
vacuum Einstein equations (\ref{Einstein:eqns}), the latter will be
rewritten in terms of a suitable conformal factor $\Omega$ and a
conformal metric $g_{\mu\nu}$ such that
\[
g_{\mu\nu}=\Omega^2 \tilde{g}_{\mu\nu}.
\]
The first and second fundamental forms determined by the metrics $g_{\mu\nu}$ and $\tilde{g}_{\mu\nu}$ on $\tilde{\mathcal{S}}$ are related to each other via
\[
h_{ij} =\Omega^2 \tilde{h}_{ij}, \quad \chi_{ij}=\Omega(\tilde{\chi}_{ij}+\Sigma \tilde{h}_{ij}),
\]
where $\Sigma$ denotes the derivative of $\Omega$ in the direction of the future directed $g$-unit normal of $\mathcal{S}$. If $\chi=h^{ij}\chi_{ij}$, $\tilde{\chi}=\tilde{h}^{ij}\tilde{\chi}_{ij}$, one has that
\[
\Omega\chi =\tilde{\chi} +3\Sigma.
\]
In terms of the fields $\Omega$, $h_{ij}$ and
$\chi_{ij}$, the constraint equations take the form
\begin{eqnarray}
&& 2\Omega D_i D^i \Omega -3 D_i\Omega D^i \Omega +\frac{1}{2} \Omega^2 {}^{(3)}r -3 \Sigma^2 -\frac{1}{2} \Omega^2 \left( \chi^2 -\chi_{ij}\chi^{ij}\right) + 2 \Omega \Sigma \chi=0, \label{conformal:Hamiltonian}\\
&& \Omega^3 D^{i}(\Omega^{-2}\chi_{ij})-\Omega \left( D_j \chi -2 \Omega^{-1}D_j \Sigma \right)=0, \label{conformal:momentum}
\end{eqnarray}
where $D$ denotes the Levi-Civita connection and ${}^{(3)}r$ the Ricci scalar
of the metric $h$. In the sequel it shall be assumed that
\begin{equation}
\Sigma=0, \quad \chi=0 \quad \mbox{ on } \tilde{\mathcal{S}}. \label{assumption:maximal}
\end{equation}
That is, the hypersurface $\tilde{\mathcal{S}}$ will be assumed to be
maximal with respect to both $\tilde{h}_{ij}$ and
$h_{ij}$.

\subsubsection{The conformal constraint equations}
\label{section:conformalconstraints}

We now briefly review the conformal constraints implied by the
\emph{metric conformal field equations} on a hypersurface
$\tilde{\mathcal{S}}$ ---see e.g. \cite{Fri83} . The latter are written
in terms of the fields $g_{\mu\nu}$, $\Omega$ and
\begin{eqnarray*}
&& S\equiv \frac{1}{4} \nabla_{\mu} \nabla^\mu \Omega +\frac{1}{24} R\Omega, \\
&& L_{\mu\nu} \equiv \frac{1}{2} R_{\mu\nu} -\frac{1}{12} Rg_{\mu\nu}, \\
&& \updn{W}{\mu}{\nu\lambda\rho} \equiv \Omega^{-1} \updn{C}{\mu}{\nu\lambda\rho},
\end{eqnarray*}
as
\begin{eqnarray*}
&& \updn{R}{\mu}{\nu\lambda\rho} = 2\left( \updn{g}{\mu}{[\lambda} L_{\rho]\nu} -g_{\nu[\lambda} \dnup{L}{\rho]}{\mu}  \right) +\updn{C}{\mu}{\nu\lambda\rho}, \\
&& 2\Omega S-\nabla_\mu \Omega \nabla^\mu \Omega =0, \\
&& \nabla_\mu \nabla_\nu \Omega = -\Omega L_{\mu\nu} + Sg_{\mu\nu}, \\
&& \nabla_\mu S = -L_{\mu\nu} \nabla^\nu \Omega, \\
&& \nabla_\lambda L_{\rho\nu}-\nabla_\rho L_{\lambda\nu}=\nabla_\mu\Omega \updn{W}{\mu}{\nu\lambda\rho}, \\
&& \nabla_\mu \updn{W}{\mu}{\nu\lambda\rho}=0,
\end{eqnarray*}
where $\nabla_\mu$ denotes the Levi-Civita covariant derivative of
$h_{\mu\nu}$, and $R_{\mu\nu\lambda\rho}$, $R_{\mu\nu}$, $R$ are the
associated Riemann tensor, Ricci tensor and Ricci scalar of
$g_{\mu\nu}$, while $C_{\mu\nu\lambda\rho}$ denotes its Weyl tensor.
The tensor $W_{\mu\nu\lambda\rho}$ will be called the \emph{rescaled
  Weyl tensor}. It will be of crucial relevance in the present
article. The constraints implied by these equations on
$\tilde{\mathcal{S}}$ will be discussed using a $g$-orthonormal frame
field $\{e_a\}$, $a=0,1,2,3$ such that $e_0$ corresponds to the
$g$-normal to $\tilde{\mathcal{S}}$. Furthermore, let $\nabla_a\equiv
e^\mu_a \nabla_\mu$. In what follows, the (frame) indices
$a,b,c,\ldots$ will be assumed to take the values $1$, $2$, $3$. Set
$\Sigma = \nabla_0 \Omega$ and
$W^*_{\mu\nu\lambda\rho}=\frac{1}{2}W_{\mu\nu\alpha\beta}
\updn{\epsilon}{\alpha\beta}{\lambda\rho}$. One further defines
\begin{eqnarray*}
&& L_a\equiv L_{a0}, \\
&& w_{abcd} \equiv W_{abcd}, \quad w_{ab}\equiv W_{a0b0}, \quad w^*_{ab} \equiv W^*_{a0b0}, \quad w_{abc}\equiv W_{a0bc},
\end{eqnarray*}
satisfying
\begin{eqnarray*}
&&w_{abcd} =-2\left( h_{a[c} w_{d]b} + h_{b[d} w_{c]a} \right), \quad w^*_{ad} \updn{\epsilon}{d}{bc}=w_{abc}, \quad w^*_{ad}=-\frac{1}{2}w_{abc}\dnup{\epsilon}{d}{bc}, \\
&& w_{ab}=w_{ba}, \quad \updn{w}{a}{a}=0, \quad w^*_{ab}=w^*_{ba}, \quad w_{a}^{*a}=0, \\
&& w_{abc}=-w_{acb}, \quad \updn{w}{a}{ac}=0, \quad w_{[abc]}=0,
\end{eqnarray*}
where the indices are moved with $h_{ab}=-\delta_{ab}$, and
$\epsilon_{abc}$ is the totally antisymmetric 3-dimensional tensor with
$\epsilon_{123}=1$. The tensors $w_{ab}$ and $w^*_{ab}$ represent,
respectively, the $n$-electric and $n$-magnetic parts of
$\updn{W}{i}{jkl}$ on $\tilde{\mathcal{S}}$. Using the Gauss' and
Codazzi equations on $\tilde{\mathcal{S}}$ one obtains the following
interior equations (\emph{the conformal constraint equations}):
\begin{subequations}
\begin{eqnarray}
&& 2\Omega S-\Sigma^2 -D_a\Omega D^a \Omega =0, \label{cc1}\\
&& D_a D_b \Omega = -\Sigma \chi_{ab} -\Omega L_{ab} + S h_{ab}, \\
&& D_a \Sigma =\dnup{\chi}{a}{c}D_c \Omega -\Omega L_a, \\
&& D_a S = -D^b\Omega L_{ba} -\Sigma L_a, \\
&& D_a L_{bc} -D_b L_{ac} = D^e \Omega w_{ecab} -\Sigma w_{cab} -\chi_{ac} L_b +\chi_{bc} L_a, \\
&& D_a L_b -D_bL_a = D^e \Omega w_{eab} + \dnup{\chi}{a}{c} L_{bc} -\dnup{\chi}{b}{c}L_{ac}, \\
&& D^cw_{cab} =\updn{\chi}{c}{a} w_{bc} -\updn{\chi}{c}{b}w_{ac}, \\
&& D^a w_{ab}= \chi^{ac} w_{abc}. \label{cc2}
\end{eqnarray}
\end{subequations}
These equations can be read as conformal constraints for the fields
$\Omega$, $\Sigma$, $S$, $h_{ab}$, $\chi_{ab}$, $L_a$, $L_{ab}$,
$w_{ab}$, $w^*_{ab}$. Alternatively, ---and this is the point of view
which shall be adopted here--- if one has \emph{physical data},
$(\tilde{h}_{ij},\tilde{\chi}_{ij})$ solving the vacuum constraint
equations and a conformal factor $\Omega$, and moreover, a choice of
$\Sigma$ and $R$ has  been made, then one can use the equations to
calculate the initial data for the conformal field equations.

\subsection{Solving the constraint equations}
\label{section:constraint_eqns}
Under the assumptions (\ref{assumption:maximal}) the conformal constraint equations take the form
\begin{eqnarray}
&& \left( \Delta -\frac{1}{8} {}^{(3)}r  \right)\vartheta = \frac{1}{8} \chi_{ij}\chi^{ij} \vartheta, \quad \mbox{ with } \vartheta=\Omega^{-1/2}, \label{Hamiltonian:1}\\
&& D^i\left(\Omega^{-2}\chi_{ij}\right)=0,
\end{eqnarray}
with $\Delta =D_k D^k$. In what follows it shall be assumed
that the above equations are solved using an adaptation of the
so-called \emph{conformal method}. Namely:
\begin{enumerate}
\item choose a smooth, negative definite metric $h_{ij}$ on a 3-dimensional, orientable, smooth compact manifold $\mathcal{S}$ and pick a point $i\in \mathcal{S}$. Set $\tilde{\mathcal{S}}=\mathcal{S}\setminus\{i\}$.
\item Find a smooth, symmetric tensor field $\psi_{ij}$ on $\tilde{\mathcal{S}}$ which is trace-free with respect to $h_{ij}$ and satisfies
\begin{equation}
D^i \psi_{ij}=0. \label{divergence:psi}
\end{equation}
The tensor $\psi_{ij}$ can be obtained by means of a York-splitting. Given a smooth, symmetric, trace-free tensor $\mathring{\Phi}_{ij}$ on $\tilde{\mathcal{S}}$, set
\begin{eqnarray*}
&& \psi_{ij} =D_i v_j + D_j v_i -\frac{2}{3} h_{ij} D_{k} v^k + \mathring{\Phi}_{ij}, \\
&& \phantom{\psi_{ij}}= (\mathcal{L} v)_{ij} + \mathring{\Phi}_{ij}
\end{eqnarray*}
where $v_i$ is some 1-form on $\tilde{\mathcal{S}}$ and
$(\mathcal{L} v)_{ij}$ is the conformal Killing operator of $h_{ij}$. Given the
above Ansatz, then equation (\ref{divergence:psi}) implies the
following elliptic equation for $v_i$:
\begin{equation}
\label{elliptic:eqn:v}
\Delta v_j +\frac{1}{3} D_j D_k v^k +r_{jk} v^k =-D^k\mathring{\Phi}_{kj},
\end{equation}
which under suitable conditions can be solved.

\item Setting $\chi_{ij}=\vartheta^{-4}\psi_{ij}$ in equation (\ref{Hamiltonian:1}) one obtains the \emph{Licnerowicz equation}
\begin{equation}
\label{Licnerowicz:eqn}
\left( \Delta -\frac{1}{8} {}^{(3)}r \right)\vartheta =\frac{1}{8} \psi_{ij} \psi^{ij} \vartheta^{-7},
\end{equation}
that is, an elliptic equation for $\vartheta$.
\end{enumerate}

The fields $h_{ij}$, $\Omega=\vartheta^{-2}$ and
$\chi_{ij}=\Omega^2 \psi_{ij}$ so constructed render
a solution to the conformal constraints (\ref{conformal:Hamiltonian})
and (\ref{conformal:momentum}). It is important to recall that if $\phi$ is a positive scalar field on $\mathcal{S}$ then the transitions
\begin{equation}
\label{data:conformal_freedom}
h_{ij}\rightarrow \phi^4 h_{ij}, \quad \psi_{ij}\rightarrow \phi^{-2} \psi_{ij}, \quad \Omega \rightarrow \phi^2 \Omega, \quad \chi_{ij}\rightarrow \phi^2 \chi_{ij},
\end{equation}
provide another solution to the conformal constraint with the same
physical data $\tilde{h}_{ij}$, $\tilde{\chi}_{ij}$. This freedom in
the conformal gauge will be used in the sequel to obtain
simplifications in the analysis of asymptotic expansions.

Given a 3-metric $h_{ij}$ consider a suitable $a>0$ such that
$\mathcal{B}_a\equiv\mathcal{B}_a(i)$ is a strictly convex normal
neighbourhood of $i$ and let $x^i$ be normal coordinates with origin
at $i$ based on an $h$-orthonormal frame $e_a$. The asymptotic
condition (\ref{asymptotic:1}) implies
\[
|x|\vartheta\rightarrow 1 \quad \mbox{ as } x\rightarrow 0.
\]
Accordingly, one has that
\[
\chi_{ij} =\mathcal{O}(1), \quad \psi_{ij}=\left( \frac{1}{|x|^4} \right) \mbox{ as }x\rightarrow 0,
\]
where the following transformation rules have been taken into account:
\[
\tilde{\chi}_{ij} =\Omega^{-1} \chi_{ij} =\Omega \psi_{ij}.
\]

Under suitable conditions, the conformal factor $\vartheta$
admits, in a neighbourhood $\mathcal{B}_a$ with $a>0$, the decomposition
\begin{equation}
\label{parametrisation_theta}
\vartheta = \frac{U}{|x|} +W,
\end{equation}
where
\[
\left( \Delta -\frac{1}{8} {}^{(3)}r \right)\left( \frac{U}{|x|} \right) =4\pi \delta(i),
\]
with $\delta(i)$ the \emph{Dirac's delta distribution} with support on $i$ and
\[
\left( \Delta -\frac{1}{8} {}^{(3)}r \right) W= \frac{1}{8} \psi_{ij} \psi^{ij}\left( \frac{U}{|x|}+W \right)^{-7}, \quad W(i)=\frac{m}{2}.
\]
The function $U$ describes the \emph{local geometry} in
$\mathcal{B}_a$ and can be calculated by means of Hadamard's
parametrix construction. If $h_{ij}=-\delta_{ij}$ ---the conformally
flat case--- then $U=1$. The function $W$ contains global information
about $(\mathcal{S},h_{ij},\chi_{ij})$.

\subsection{The momentum constraint in flat space}
\label{section:tensor_mc}

The analysis of the solutions of the momentum constraint in flat space
given in \cite{DaiFri01} is now briefly recalled. In section
\ref{section:spinor_mc} it will be recast in a form ---in terms of
space spinors--- which is convenient for the application discussed in
this article.

In this section assume that $\mathcal{S}$ is flat in at least
a neighbourhood $\mathcal{B}_a$ of $i$. Let $x^k$ be a Cartesian
coordinate system with origin at $i$. In these coordinates one has
that the metric is given by $h_{ij}=-\delta_{ij}$. We shall often write
$r=|x|$. Let $n^i\equiv x^i/|x|$ and complement it with complex
vectors $m^i$, $\overline{m}^i$ to form a basis of the tangent bundle of
$\Real^3$, by requiring that
\[
m_i m^i = \overline{m}_i \overline{m}^i = n_i m^i = n_i \overline{m}^i =0, \quad m_i \overline{m}^i =-1.
\]
Recall that in this construction one has freedom of performing
rotations $m_i\mapsto e^{\mbox{i}\theta} m_i $ with $\theta$
independent of $r$. The metric $h_{ij}$ can be written in the form
\[
h_{ij}= n_i n_j +\overline{m}_i m_j +m_i \overline{m}_j,
\]
while an arbitrary trace-free tensor $\mathring{\psi}_{ij}$ can be written as
\begin{equation}
\mathring{\psi}_{ij}= \frac{1}{r^3}\left( \xi (3 n_i n_j -\delta_{ij}) + \sqrt{2}\eta n_{(i} \overline{m}_{j)} + \sqrt{2}\overline{\eta} n_{(i} m_{j)} + \overline{\mu} m_i m_j + \mu \overline{m}_i \overline{m}_j \right), \label{mathring:ansatz}
\end{equation}
with
\[
\xi = \frac{1}{2} r^3 \mathring{\psi}_{ij} n^i n^j, \quad \eta= \sqrt{2} r^3 \mathring{\psi}_{ij} n^i m^j, \quad \mu = r^3 \mathring{\psi}_{ij} m^i m^j.
\]
In the previous expression $\xi$ is real, while $\eta$ and $\mu$ are
complex functions of spin weight 1 and 2 respectively. Let $\tilde{\lambda}$
be an arbitrary complex $C^\infty$ function on
$\mathcal{B}_a$. Let $\lambda\equiv \eth^2
\tilde{\lambda}$, where $\eth$ denotes the \emph{eth}-operator ---see
e.g. \cite{PenRin86,Ste91}. Let $\lambda^{(R)}\equiv
\mbox{Re}(\lambda)$ and $\lambda^{(I)}\equiv
\mbox{i}\,\mbox{Im}(\lambda)$. Let
\begin{subequations}
\begin{eqnarray}
&& \psi_P^{ij} \equiv \frac{3}{2 r^4} \left(-P^i n^j - P^j n^i -( \delta^{ij} -5 n^i n^j) P^k n_k \right),\\
&& \psi_J^{ij} \equiv \frac{3}{r^3} \left( n^i \epsilon^{jkl} J_k n_l + n^j \epsilon^{ikl} J_k n_l  \right), \\
&& \psi_A^{ij}\equiv \frac{A}{r^3} \left( 3n^in^j -\delta^{ij}  \right), \\
&& \psi_Q^{ij} \equiv \frac{3}{2r^2} \left( Q^i n^j + Q^j n^i -(\delta^{ij} -n^i n^j) Q^k n_k  \right),
\end{eqnarray}
\end{subequations}
where $P^i$ (\emph{the linear momentum}), $J^i$ (\emph{the angular
  momentum}), $Q^i$ (\emph{conformal momentum}) and $A$ are real constant
vectors, respectively, scalars. Furthermore, let
\begin{subequations}
\begin{eqnarray}
&& \xi = \overline{\eth}^2 \lambda^{(R)}, \label{xi}\\
&& \eta = -2 r \partial_r \overline{\eth} \lambda^{(R)} + \overline{\eth} \lambda^{(I)} , \label{eta}\\
&& \mu = 2 r \partial_r (r \partial_r \lambda^{(R)}) -2 \lambda^{(R)} +\eth \overline{\eth} \lambda^{(R)} - r \partial_r \lambda^{(I)}, \label{mu}
\end{eqnarray}
\end{subequations}
and denote by $\mathring{\psi}_{ij}[\tilde{\lambda}]$ the tensor
obtained by substituting (\ref{xi})-(\ref{mu}) into
(\ref{mathring:ansatz}). Then
\begin{equation}
\mathring{\psi}_{ij}= \mathring{\psi}_{ij}[P]+\mathring{\psi}_{ij}[J]+\mathring{\psi}_{ij}[A]+ \mathring{\psi}_{ij}[Q]+\mathring{\psi}_{ij}[\tilde{\lambda}],
\label{solution:flat:momentum}
\end{equation}
satisfies the \emph{flat space momentum constraint}
\begin{equation}
\partial^i \mathring{\psi}_{ij}=0.
\label{flat:momentum}
\end{equation}
Conversely, any smooth solution to equation (\ref{flat:momentum}) is
of the form (\ref{solution:flat:momentum}) ---cfr. theorem 14 in
\cite{DaiFri01}.

In the present article the complex function $\tilde{\lambda}$ which is smooth
on $\mathcal{B}_a\setminus\{i\}$ will be taken to be of the form
\begin{equation}
\label{Ansatz:lambda}
\tilde{\lambda} = \tilde{\lambda}^{(1)} + \frac{1}{r} \tilde{\lambda}^{(2)},
\end{equation}
with $\lambda^{(1)}, \; \lambda^{(2)}\in C^\infty(\mathcal{B}_a)$. In analogy to equation (\ref{Ansatz:lambda}) let also
\[
\lambda = \lambda^{(1)} + \frac{1}{r} \lambda^{(2)}.
\]
Note that the factor $1/r$ makes the function $\tilde{\lambda}$
non-smooth at $i$. The functions $\lambda^{(1)}$ and $\lambda^{(2)}$
are calculated using the $\eth$ operator. Accordingly, they are
non-smooth.

\subsection{Assumptions on the freely specifiable data}
\label{section:freedata}
In order to obtain a non-trivial outcome from the analysis to be
discussed in the sequel, one has to choose a suitable class of initial
data. If the initial data is to be constructed following the procedure
discussed in section \ref{section:constraint_eqns}, then the
freely specifiable data is given in terms of a negative-definite
3-metric $h_{ij}$ and an $h$-trace free symmetric tensor
$\mathring{\Phi}_{ij}$.

\subsection{The conformal metric}
For ease of the presentation, the results presented here will be
restricted to the case of conformal metrics $h_{ij}$ which are
\emph{analytic} in a neighbourhood, $\mathcal{B}_a$, of spatial
infinity.  This assumption is non-essential and with long recursive
arguments it should be possible to extend all the results presented here
to the smooth ---i.e. $C^\infty$--- setting. The analyticity of
$h_{ij}$ is explicitly used in the proof of part (ii) of
theorem \ref{thm:symmetries} and of part (i) of theorem \ref{thm:vanishing},
where the expansions of a certain tensor associated to the massless
part of the Weyl tensor are calculated.

It should be mentioned that in some senses the assumption of
analyticity of the conformal metric on $\mathcal{B}_a$ is as
restrictive as that of smoothness. The reason behind is that generic
stationary spacetimes do not have a smooth conformal metric. Instead,
as shown in \cite{Dai01b}, the metric is of the form
\[
h_{ij} = h^{(1)}_{ij} + r^3h^{(2)}_{ij},
\]
with $h^{(1)}_{ij}$ and $h^{(2)}_{ij}$ analytic tensors. Due to the
presence of the term $r^3$, the metric in this case will be
$C^{2,\alpha}(\mathcal{B}_a)$, $0<\alpha<1$. The regularity of this
conformal metric is conjectured to be optimal for strictly stationary
data ---i.e. stationary data which is not static.

\bigskip Throughout, it will be assumed that in \emph{normal
  coordinates} centred at $i$ one has that
\[
h_{ij}= -\delta_{ij} + \hat{h}_{ij}, \quad h^{ij}= -\delta^{ij} + \hat{h}^{ij},
\]
with $\hat{h}_{ij}\in C^\omega(\mathcal{B}_a)$, $\hat{h}^{ij}\in
C^\omega(\mathcal{B}_a)$. It is recalled that the normal coordinates satisfy
\[
x^i \hat{h}_{ij}=0, \quad x_i \hat{h}^{ij}=0.
\]
The latter relations imply that
\[
x_l x^ i\updn{\Gamma}{l}{ij}=0.
\]
The analysis of the solutions to the momentum constraint performed in
the appendix \ref{appendix:momentum} requires the technical assumption
\begin{equation}
\label{technical_condition}
\delta^{ij}\hat{h}_{ij}= \delta_{ij} \hat{h}^{ij}=0.
\end{equation}

The conformal gauge freedom (\ref{data:conformal_freedom}) can be used
to obtain a representative in the conformal class of $h_{ij}$ which
is in the so-called \emph{cn (conformal normal)-gauge} ---see
\cite{Fri98a}. Working with a metric in the cn-gauge renders a number
of useful simplifications in the present discussion. In order to
define the cn-gauge, consider on $\mathcal{B}_a$ solutions
$(x^i(t),b_i(t))$ of the \emph{conformal geodesic equations}
\begin{subequations}
\begin{eqnarray}
&& \dot{x}^k D_k \dot{x}^i = -2 (b_k \dot{x}^k) \dot{x}^i + (h_{kl}\dot{x}^k \dot{x}^l) h^{mi} b_m, \label{cfg1}\\
&& \dot{x}^k D_k b_i = (b_k \dot{x}^k) b_i -\frac{1}{2} (h^{kl} b_k b_l) h_{mi}\dot{x}^{m} + l_{ik} \dot{x}^k, \label{cfg2}
\end{eqnarray}
\end{subequations}
with initial conditions
\[
x(0)=i, \quad h_{ij} \dot{x}^i\dot{x}^j =-1, b_i(0)=0,
\]
where $x^i(t)$ denotes a curve in $\mathcal{B}_a$ through $i$ and
$b_i(t)$ is a 1-form along that curve. If neighbourhood
$\mathcal{B}_a$ of $i$ is taken to be small enough, then there is a
unique conformal rescaling of the form given by (\ref{data:conformal_freedom}) such that the
\emph{conformal metric} is analytic, keeps the metric and
connection unchanged at $i$, and such that if the solution to the
conformal geodesic equations (\ref{cfg1}) and (\ref{cfg2}) is written
in terms of the rescaled metric one has
\begin{equation}
b_i \dot{x}^i=0, \quad \mbox{ on } \mathcal{B}_a. \label{cngauge}
\end{equation}
A metric in the conformal gauge of $h_{ij}$ satisfying (\ref{cngauge})
will be said to be in the \emph{cn-gauge}. From here on, it will
always be assumed that the metric $h_{ij}$ is in the cn-gauge.

An important consequence of working in the cn-gauge is that both the
Ricci scalar and Ricci tensor of $h_{ij}$ vanish at $i$. Accordingly,
\[
\hat{h}_{ij}=O(r^3), \quad \partial_k h_{ij}=O(r^2).
\]
In addition, it will be proved ---see lemma
\ref{metric:cngauge}--- that if $h_{ij}$ is in the cn-gauge then
$\hat{h}_{ij}=r^2\check{h}_{ij}$ so that
\[
h_{ij}= -\delta_{ij} + r^2\check{h}_{ij}, \quad h^{ij}= -\delta^{ij} + r^2\check{h}^{ij},
\]
where $\check{h}_{ij}=O(r)$ is analytic. Note that because of the
condition (\ref{technical_condition}) one has that $\delta^{ij}
\check{h}_{ij}=0$.

\subsubsection{Free data for the second fundamental form}
The $h$-trace free symmetric tensor
$\mathring{\Phi}_{ij}$, containing the free data for the second fundamental form will be constructed out of the tensor
$\mathring{\psi}_{ij}$, solution of the flat space momentum constraint
given in equation (\ref{solution:flat:momentum}). Thus, let
\[
\mathring{\Phi}_{ij} \equiv \mathring{\psi}_{ij} -\frac{1}{3}h_{ij} h^{kl} \mathring{\psi}_{kl}.
\]
Thus, $\mathring{\Phi}_{ij}$ is specified by prescribing the constant
vectors $P^i$, $J^i$, $Q^i$, the constant scalar $A$, and the function
$\lambda$. In order to obtain a vector $v^i$ and a scalar $\vartheta$
solving the equation (\ref{Licnerowicz:eqn}) admitting an expansion in
powers of $r$ near $i$ one has to set $P^i=0$ and consider a $\lambda$
of the form given in equation (\ref{Ansatz:lambda}). A more extended
discussion of these issues is given in \cite{DaiFri01}. However, it
turns out that if one admits a contribution to the second fundamental
form of the type $\mathring{\psi}_{ij}$ with $\lambda$ given by
equation (\ref{Ansatz:lambda}), the solutions to the elliptic equation
(\ref{elliptic:eqn:v}) will not have an asymptotic expansion in
$\mathcal{B}_a$ consisting purely of powers of $r$. In order for this
to be the case, $\mbox{Im}(\lambda^{(1)})$ and
$\mbox{Re}(\lambda^{(2)})$ have to be related to the conformal metric
$h_{ij}$ in a particular way. This issue is discussed in more detail
in the appendix. Throughout the main body of the article, it will be
assumed that such conditions are satisfied, and accordingly, the
solutions $v^i$ of equation (\ref{elliptic:eqn:v}) have expansions in
$\mathcal{B}_a$ purely in powers of $r$ ---see below for the precise
details. Examples of classes of data where these are valid are axially
symmetric data and conformally flat data ---in which case $v^i=0$.

\subsubsection{Some consequences of the assumptions on the freely specifiable data}
\label{consequences:freedata}
In order to make precise the idea that a given function or tensor
field over $\mathcal{S}$ admits an expansion around $i$ in terms of
powers of $r$, introduce the following function spaces:
\begin{subequations}
\begin{eqnarray*}
&& E^\infty(\mathcal{B}_a) =\left\{f=f_1 + rf_2 \;|\; f_1, \; f_2\in C^\infty(\mathcal{B}_a)    \right\}, \\
&& \mathcal{Q}_\infty (\mathcal{B}_a)= \left\{v^i \in C^\infty(\mathcal{B}_a, \Real^3) \;|\; x_i v^i = r^2 v, \;\; v\in C^\infty(\mathcal{B}_a)   \right\}.
\end{eqnarray*}
\end{subequations}

Theorem 15 in \cite{DaiFri01} states that if the function $\lambda$
determining the higher multipoles of $\mathring{\psi}_{ij}$ is of
the form given by equation (\ref{Ansatz:lambda}) ---i.e. $\lambda r
\in E^\infty(\mathcal{B}_a)$ and $P^i=0$, then $r^8 \mathring{\psi}_{ij}
\mathring{\psi}^{ij} \in E^\infty(\mathcal{B}_a)$, from where it
follows from their main result ---theorem 1--- that in the conformally
flat case ($h_{ij}=-\delta_{ij}$) one has that
\[
\vartheta = 1/r + W, \quad W\in E^\infty(\mathcal{B}_a).
\]
In the non-conformally flat case, if $h_{ij}$ and $\lambda$ are such
that the solutions, $v^i$, of equation (\ref{elliptic:eqn:v}) are of
the form
\[
v^i = r^s v_1^i + v_2^i,
\]
for some integer $s$ with $v_1^i\in \mathcal{Q}_\infty
(\mathcal{B}_a)$, $ v^i_2 \in C^\infty(\mathcal{B}_a)$, and such that
$r^8\psi_{ij} \psi^{ij} \in
E^\infty(\mathcal{B}_a)$, with
\[
\psi_{ij}= \mathring{\Phi}_{ij}[A,J,Q,\lambda^{(1)},\lambda^{(2)}/r] + (\mathcal{L}_h v)_{ij}[A,J,Q] + (\mathcal{L}_h v)_{ij}[\lambda^{(1)}]+ (\mathcal{L}_h v)_{ij}[\lambda^{(2)}],
\]
then theorem 1 in \cite{DaiFri01} renders
\[
\vartheta = \frac{U}{r} +W,
\]
with $U\in C^\omega(\mathcal{B}_a)$ and $W\in
E^\infty(\mathcal{B}_a)$. Furthermore, because of the use of the
cn-gauge, $U=1+O(r^4)$. The discussion in appendix
\ref{appendix:momentum} renders a detailed description of the
structure of the vectors $v^i[A,J,Q,\lambda^{(1)},\lambda^{(2)}/r]$.

\section{The Manifold $\mathcal{C}_a$} \label{section:manifold_Ca} In
\cite{Fri98a} a representation of the region of spacetime close to
null infinity and spatial infinity has been introduced. The standard
representation of this region of spacetime depicts $i^0$ as a point.
In contrast, the representation introduced in \cite{Fri98a} depicts
spatial infinity as a cylinder ---\emph{the cylinder at spatial
  infinity}. The technical and practical grounds for introducing this
description have been discussed at length in that seminal reference.
The original construction in \cite{Fri98a} was carried out for the
class of time symmetric metrics with analytic conformal metric
$h_{ij}$. However, as discussed in \cite{Val04e,Val05a}, the
construction can be adapted to settings without a vanishing second
fundamental form. The purpose of this section is, primarily, to introduce
notation that will be used in the sequel and to provide enough
background material to follow the discussion in the sequel. In any
case, the reader is referred to \cite{Fri98a,Fri04} for a thorough
discussion of the details.

\subsection{The construction of the manifold}
Starting on the initial hypersurface $\mathcal{S}$, Friedrich's
construction makes use of a blow up of the point $i\in \mathcal{S}$ to
the 2-sphere $\Sphere^2$. This blow up requires the introduction of a
particular bundle of spin-frames over $\mathcal{B}_a$.  In what
follows a space spinor formalism analogous to a tensorial 3+1
decomposition will be used to this end. Consider the (unphysical,
conformally rescaled) spacetime $(\mathcal{M},g_{\mu\nu})$ obtained as
the development of the initial data set
$(\mathcal{S},h_{ij},\chi_{ij})$. Let $SL(\mathcal{S})$ be the set of
spin dyads $\delta=\{\delta_A\}_{A=0,1}=\{o_A,\iota_A\}$ on
$\mathcal{S}$ which are normalised with respect to the alternating
spinor $\epsilon_{AB}$ in such a way that $\epsilon_{01}=1$.

The set $SL(\mathcal{S})$ has a natural bundle structure where
$\mathcal{S}$ is the base space,  and its structure group is given by
\[
SL(2,\Complex)=\{t^A_{\phantom{A}B}\in GL(2,\Complex)\;|\; \epsilon_{AC}t^A_{\phantom{A}B} t^C_{\phantom{C}D}=\epsilon_{BD}\},
\]
acting on $SL(\mathcal{S})$ by $\delta\mapsto \delta\cdot t=\{\delta_A t^A_{\phantom{A}B}\}_{B=0,1}$.
Now, let $\tau=\sqrt{2}e_0$, where $e_0$ is the future $g$-unit normal of $\mathcal{S}$ and
\[
\tau_{AA'}=g(\tau,\delta_A\overline{\delta}_{A'})=\epsilon_{A}^{\phantom{A}0}\epsilon_{A'}^{\phantom{A'}0'}+\epsilon_{A}^{\phantom{A}1}\epsilon_{A'}^{\phantom{A'}1'}
\]
is its spinorial counterpart --- that is, $\tau=\tau^a e_a =
\sigma^a_{AA'}\tau^{AA'}e_a$ where $\sigma^a_{AA'}$ denote the
\emph{Infeld-van der Waerden symbols} and $\{e_a\}$, $a=0,\ldots,3$ is
an orthonormal frame. The spinor $\tau_{AA'}$ enables the introduction
of space-spinors ---sometimes also called $SU(2)$ spinors, see
\cite{Ash91,Fra98a,Som80}. It defines a sub-bundle $SU(\mathcal{S})$
of $SL(\mathcal{S})$ with structure group
\[
SU(2,\Complex)=\{ t^A_{\phantom{A}B}\in SL(2,\Complex)\; |\; \tau_{AA'}t^A_{\phantom{A}B}\overline{t}^{A'}_{\phantom{A'}B'}=\tau_{BB'}\},
\]
and projection $\pi$ onto $\mathcal{S}$. The spinor $\tau^{AA'}$ allows to introduce \emph{spatial van der Waerden symbols} via
\[
\sigma_{a}^{AB}=\sigma^{(A}_{a\phantom{(A}A'}\tau^{B)A'}, \quad \sigma^a_{AB}=\tau_{(B}^{\phantom{(B}A'}\sigma^{a}_{\phantom{a}A)A'}, \quad i=1,2,3.
\]
The latter satisfy
\[
h_{ab}=\sigma_{aAB}\sigma_b^{AB}, \quad -\delta_{ab}\sigma^a_{AB}\sigma^b_{CD}=-\epsilon_{A(C}\epsilon_{D)B}\equiv h_{ABCD},
\]
with $h_{ab}=h(e_a,e_b)=-\delta_{ab}$. The bundle $SU(\mathcal{S})$ can be
endowed with a $\mathfrak{su}(2,\Complex)$-valued \emph{connection form}
$\check{\omega}^A_{\phantom{A}B}$ compatible with the metric
$h_{ij}$ and 1-form $\sigma^{AB}$, the \emph{solder form} of
$SU(S)$. The solder form satisfies by
construction
\begin{equation}
\label{metric}
h\equiv h_{ij} dx^i \otimes dx^j = h_{ABCD} \sigma^{AB}\otimes \sigma^{CD},
\end{equation}
where $\sigma^{AB}=\sigma^{AB}_{i} dx^i$ ---note that the $\sigma^{AB}_i$ are not the spatial Infeld-van der Waerden symbols, $\sigma^{AB}_a$.

Now, given a spinorial dyad $\delta\in SU(\mathcal{S})$ one can define
an associated vector frame via $e_a=e_a(\delta)=\sigma^{AB}_a
\delta_A\tau_B^{\phantom{B}B'}\overline{\delta}_{B'}$, $a=1,2,3$.  We
shall restrict our attention to dyads related to frames
$\{e_j\}_{j=0,\cdots,3}$ on $\mathcal{B}_a$ such that $e_3$ is tangent
to the $h$-geodesics starting at $i$. Let $\check{H}$ denote the
horizontal vector field on $SU(\mathcal{S})$ projecting to the
radial vector $e_3$.

\medskip
The fibre $\pi^{-1}(i)\subset SU(\mathcal{S})$ (the fibre ``over'' $i$) can be
parametrised by choosing a fixed dyad $\delta^*$ and then letting the
group $SU(2,\Complex)$ act on it. Let $(-a,a)\ni \rho \mapsto
\delta(\rho,\updn{t}{A}{B})\in SU(\mathcal{S})$ be the integral curve to the vector
$\check{H}$ satisfying $\delta(0,\updn{t}{A}{B})=\delta(\updn{t}{A}{B})\in \pi^{-1}(i)$. With
this notation one defines the set
\[
\mathcal{C}_a =\big\{ \delta(\rho,\updn{t}{A}{B})\in SU(\mathcal{B}_{a}) \;\big|\; |\rho|<a, \; \updn{t}{A}{B}\in SU(2,\Complex)\big\},
\]
which is a smooth submanifold of $SU(\mathcal{S})$ diffeomorphic to
$(-a,a)\times SU(2,\Complex)$. The vector field $\check{H}$ is such
that its integral curves through the fibre $\pi^{-1}(i)$ project onto
the geodesics through $i$. From here it follows that the projection
map $\pi$ of the bundle $SU(\mathcal{S})$ maps $\mathcal{C}_a$ into
$\mathcal{B}_a$.

Let, in the sequel, $\mathcal{I}^0\equiv \pi^{-1}(i)=\{\rho=0\}$ denote
the fibre over $i$. It can be seen that $\mathcal{I}^0\approx
SU(2,\Complex)$. On the other hand, for $p\in \mathcal{B}_a\setminus
\{i\}$ it turns out that $\pi^{-1}(p)$ consists of an orbit of $U(1)$
for which $\rho=|x(p)|$, and another for which $\rho=-|x(p)|$, where
$x^i(p)$ denote normal coordinates of the point $p$. In order to
understand better the structure of the manifold $\mathcal{C}_a$ it is
useful to quotient out the effect of $U(1)$. It turns out that
$\mathcal{I}^0/U(1)\approx \Sphere^2$. Hence, one has an extension of
the physical manifold $\tilde{S}$ by blowing up the point $i$ to
$\Sphere^2$.

The manifold $\mathcal{C}_a$ inherits a number of structures from
$SU(\mathcal{S})$. In particular, the solder and connection forms can
be pulled back to smooth 1-forms on $\mathcal{C}_a$. These shall be
again denoted by $\sigma^{AB}$ and
$\check{\omega}^{A}_{\phantom{A}B}$. They satisfy the \emph{structure
  equations} relating them to the so-called \emph{curvature form} determined by the \emph{curvature spinor}
\[
r_{ABCDEF}=\bigg( \frac{1}{2}s_{ABCE}-\frac{1}{12}r h_{ABCE}\bigg) \epsilon_{DF} +\bigg( \frac{1}{2} s_{ABDF}-\frac{1}{12}r h_{ABDF} \bigg)\epsilon_{CE},
\]
where $s_{ABCD}=s_{(ABCD)}$ is the spinorial counterpart of the
tracefree part of the Ricci tensor of $h_{ij}$ and ${}^{(3)}r$ its Ricci
scalar. These satisfy the \emph{3-dimensional Bianchi identity}
\[
D^{AB}s_{ABCD}=\frac{1}{6}D_{CD}{}^{(3)}r.
\]

\subsection{Calculus on $\mathcal{C}_a$}
In the sequel $\updn{t}{A}{B}$ and $\rho$ will be used as coordinates
on $\mathcal{C}_a$. Consequently, one has that
$\check{H}=\partial_\rho$. Vector fields relative to the
$SU(2,\Complex)$-dependent part of the coordinates are obtained by
looking at the basis of the (3-dimensional) Lie algebra
$\mathfrak{su}(2,\Complex)$ given by
\[
u_1=\frac{1}{2}\left(
\begin{array}{cc}
0 & \mbox{i} \\
\mbox{i} & 0
\end{array}\right), \quad
u_2=\frac{1}{2}\left(
\begin{array}{cc}
0 & -1 \\
1 & 0
\end{array}\right), \quad
u_3=\frac{1}{2}\left(
\begin{array}{cc}
\mbox{i} & 0 \\
0 & -\mbox{i}
\end{array}\right).
\]
In particular, the vector $u_3$ is the generator of $U(1)$. Denote by
$Z_i$, $i=1,2,3$ the Killing vectors generated on $SU(\mathcal{S})$ by
$u_i$ and the action of $SU(2,\Complex)$. The vectors $Z_i$ are
tangent to $\mathcal{I}^0$. On $\mathcal{I}^0$ one sets
\begin{equation}
\label{diffops:X}
X_+=-(Z_2+\mbox{i}Z_1), \quad X_-=-(Z_2-\mbox{i}Z_1), \quad X=-2\mbox{i}Z_3,
\end{equation}
and extend these vector fields to the rest of $\mathcal{C}_a$ by demanding them
to commute with $\check{H}=\partial_\rho$. For latter use it is noted that
\[
[X,X_+]=2X_+, \quad [X,X_-]=-2X_-, \quad [X_+,X_-]=-X.
\]
The vector fields are complex conjugates of each other in the sense
that for a given real-valued function $W$,
$\overline{X_-W}=X_+W$. More importantly, it can be seen that for
$p\in \mathcal{B}_a\setminus\{i\}$ the projections of the fields $\check{H}$,
$X_\pm$ span the tangent space at $p$.

A frame $c_{AB}=c_{(AB)}$ dual to the solder forms
$\sigma^{CD}$ is defined so that it does not pick components along the fibres
---i.e. along the direction of $X$. These requirements imply
\begin{equation}
\label{solder_form}
\langle \sigma^{AB}, c_{CD} \rangle= h^{AB}_{\phantom{AB}CD}, \quad c_{CD}=c^1_{CD}\partial_\rho + c^+_{CD}X_+ + c^-_{CD}X_-,
\end{equation}
where $\langle \cdot , \cdot \rangle$ denotes the action of a 1-from on a
vector. Let $\alpha^\pm$ and $\alpha$ be 1-forms on $\mathcal{C}_a$
annihilating the vector fields $\partial_\tau$, $\partial_\rho$ and
having with $X_\pm$ the non-vanishing pairings
\[
\langle \alpha^+,X_+ \rangle = \langle \alpha^-, X_-\rangle = \langle \alpha, X \rangle =1.
\]
Furthermore, from the properties of the solder form $\sigma^{AB}$ one finds that
\begin{equation}
\label{frame_fields}
c^1_{AB}=x_{AB}, \quad c^+_{AB}=\frac{1}{\rho}z_{AB}+\check{c}^+_{AB}, \quad c^-_{AB}=\frac{1}{\rho}y_{AB}+\check{c}^-_{AB},
\end{equation}
with constant spinors $x_{AB}$, $y_{AB}$ and $z_{AB}$ given by
\[
x_{AB}\equiv \sqrt{2} o_{(A} \iota_{B)}, \quad y_{AB} \equiv -\frac{1}{\sqrt{2}}\iota_A \iota_B, \quad z_{AB}=\frac{1}{\sqrt{2}} o_A o_B,
\]
and analytic spinor fields satisfying
\[
\check{c}^\alpha_{AB}= O(\rho), \quad \check{c}^\alpha_{01}=0, \quad \alpha=1,\pm.
\]
Accordingly, one can write
\[
\check{c}^\pm_{AB} = \check{c}^\pm_y y_{AB} + \check{c}^\pm_z z_{AB}.
\]

Furthermore, using the structure equations it can be shown that in
fact $\check{c}^1_{AB}=0$. By virtue of the relations
(\ref{solder_form}) and (\ref{frame_fields}), the solder forms
$\sigma^{AB}$ descend to forms $n_i \mbox{d}x^i$, $m_i \mbox{d}x^i$,
$\overline{m}^i \mbox{d}x^i$ spanning the tangent space of the points of
$\mathcal{B}_a$ with non-vanishing pairings
\[
n^in_i=-1, \quad m_i \overline{m}^i =-1,
\]
and such that $n^i=x^i/r$. Note that $n^i$, $m^i$ and $\overline{m}^i$ are
not smooth functions with respect to the coordinates $x^i$.

The connection coefficients are defined by contracting the connection
form $\check{\omega}^A_{\phantom{A}B}$ with the frame $c_{AB}$. In
general, one writes
\begin{eqnarray*}
&&\gamma_{CD\phantom{A}B}^{\phantom{CD}A} \equiv \langle \check{\omega}^A_{\phantom{A}B},c_{CD} \rangle = \frac{1}{\rho} \gamma_{CD\phantom{A}B}^{*\phantom{D}A}+ \check{\gamma}_{CD\phantom{A}B}^{\phantom{CD}A}, \\
&& \gamma^*_{ABCD}=\frac{1}{2}(\epsilon_{AC}x_{BD}+\epsilon_{BD}x_{AC}).
\end{eqnarray*}
The spinor $\gamma^*_{ABCD}$ denotes the singular part of the
connection coefficients. The regular part of the connection can be
related to the frame coefficients $c_{AB}$ via commutator equations.
The smooth part $\check{\gamma}_{ABCD}$ of the connection coefficients
can be seen to satisfy
\[
\check{\gamma}_{11CD}=0, \quad \check{\gamma}_{ABCD}=O(\rho).
\]
Furthermore, from an analysis of the structure equations one obtains
that $\check{\gamma}_{1100}=-\check{\gamma}_{0011}$. Let $f$ be a
smooth function on $\mathcal{C}_a$
\[
D_{AB}f=c_{AB}(f).
\]
Similarly, let $\mu_{AB}$ represent both a smooth spinor field on
$\mathcal{C}_a$. Then the covariant derivative of $\mu_{AB}$ is given
by
\[
D_{AB}\mu_{CD}= c_{AB}(\mu_{CD})- \gamma_{AB\phantom{E}C}^{\phantom{AB}E}\mu_{ED}-\gamma_{AB\phantom{E}D}^{\phantom{AB}E}\mu_{CE}.
\]
Analogous formulae hold for higher valence spinors.

\subsection{Normal expansions at $i$}
\label{subsection:normal_exp_i}
In \cite{Fri98a} a certain type of expansions of analytic fields near
$i$ has been discussed. Although the conformal metric will be assumed
to be analytic on $\mathcal{B}_a$, most of the other fields ---in
particular objects derived from the second fundamental form--- will
at most be smooth. Thus, the ideas of \cite{Fri98a} have to be adapted
to the smooth setting. This can be readily done.

If $f\in C^\infty(\mathcal{B}_a)$ then one has
that
\[
f\sim \sum_{k\geq 0} f_{i_1\cdots i_k} x^{i_1}\cdots x^{i_k},
\]
which is to be interpreted as
\[
f = \sum_{k=0}^m f_{i_1\cdots i_k} x^{i_1}\cdots x^{i_k} + f_R, \quad
\]
with $f_R\in C^\infty(\mathcal{B}_a)$, $f_R=o(r^m)$, for all $m\geq
0$. The term $\sum_{k=0}^m f_{i_1\cdots i_k} x^{i_1}\cdots x^{i_k}$,
with $f_{i_1\cdots i_k}$ constant vectors, is the Taylor polynomial of
degree $m$ of $f$.

Now, suppose that $\updn{T}{i_1\cdots i_r}{j_1\cdots j_s}$ is a smooth
tensorial field of rank$(r,s)$ on $\mathcal{B}_a$ with components
$\updn{T}{*a_1\cdots a_r}{b_1\cdots b_s}$ with respect to the frame
$e^*_a$ on which the normal coordinates $x^i$ are based. Let $V=
x^j\partial_j$ be the radial vector which is tangent to geodesics
through $i$ and satisfying $V_i V^i=-1$. Let also $n^i$ be defined by
$V^i=|x|n^i$. By construction one has that $V^k D_k e^*_a=0$, thus
following the procedure described in section 3.3 of \cite{Fri98a} one
obtains
\begin{equation}
\label{tensor:expansion1}
\updn{T}{*a_1\cdots a_r}{b_1\cdots b_s}(q) = \sum_{p=0}^m \frac{1}{p!}|x|^p n^{l_p}(q) \cdots n^{l_1}(q) D_{l_p} \cdots D_{l_1} \updn{T}{*a_1\cdots a_r}{b_1\cdots b_s}(i) + \left( \updn{T}{*a_1\cdots a_r}{b_1\cdots b_s} \right)_{\mathcal{R}},
\end{equation}
with $q\in \mathcal{B}_a$, and $( \updn{T}{*a_1\cdots a_r}{b_1\cdots
  b_s})_{\mathcal{R}} \in C^\infty(\mathcal{B}_a)$ and $(
\updn{T}{*a_1\cdots a_r}{b_1\cdots b_s})_{\mathcal{R}}=o(r^m)$. The
first term in the right-hand side of expression
(\ref{tensor:expansion1}) will the called the \emph{analytic part} of
$\updn{T}{*a_1\cdots a_r}{b_1\cdots b_s}$.

An analogous expansion can be obtained for smooth analytic spinor
fields.  Suppose $\xi_{A_1B_1\cdots A_lB_l}$ denotes the components of
a smooth even rank spinorial field with respect to the spin frame
$\delta^*_A$ associated to $e^*_a$. In analogy to expression 
(\ref{tensor:expansion1}) one can introduce the expansion
\begin{equation}
\xi^*_{A_1B_1\cdots A_lB_l} (q)= \sum_{p=0}^m \frac{1}{p!}|x|^p n^{C_pD_p}\cdots n^{C_1D_1} D_{C_pD_p}\cdots D_{C_1D_1} \xi^*_{A_1B_1\cdots A_lB_l} + \left( \xi^*_{A_1B_1\cdots A_lB_l}   \right)_\mathcal{R}, \label{spinor:expansion1}
\end{equation}
with $n^{AB}=v^{AB}(q)$, $q\in \mathcal{B}_a$ and the derivatives of
the spinor field are evaluated at the point $i$, and $(
\xi^*_{A_1B_1\cdots A_lB_l})_\mathcal{R}\in C^\infty(\mathcal{B}_a)$,
$( \xi^*_{A_1B_1\cdots A_lB_l})_\mathcal{R}=o(r^m)$. Again, the first
term of the right-hand side of equation (\ref{spinor:expansion1}) will
be referred to as the \emph{analytic part}.

An analysis of the decomposition in terms of irreducible spinors of
the summands in (\ref{spinor:expansion1}) will be important in the
sequel. The derivatives in the expression (\ref{spinor:expansion1})
can be replaced by symmetrised derivatives (in the pair of indices
${}_{C_kD_k}$) as they are contracted with the same spinor $n^{AB}$. In
these symmetrised derivatives, the contraction of indices ${}_{C_j}$, ${}_{C_k}$
with $j\neq k$ renders an expression antisymmetric in the indices
${}_{D_j}$ and ${}_{D_k}$.  Accordingly, the decomposition of every sumand in
(\ref{spinor:expansion1}) with $p\geq 2$ into symmetric, irreducible
parts ---with respect to the indices ${}_{C_pD_p\cdots C_1D_1}$--- renders
an expansion
\begin{eqnarray}
&& n^{C_pD_p}\cdots n^{C_1D_1} D_{C_pD_p}\cdots D_{C_1D_1} \xi^*_{A_1B_1\cdots A_lB_l} \nonumber \\
&&\hspace{1cm}= n^{C_pD_p}\cdots n^{C_1D_1}\bigg( \xi^*_{p,0;C_pD_p\cdots C_1D_1A_1B_1\cdots A_lB_l} +\xi^*_{p,1;C_pD_p\cdots C_3D_3A_1B_1\cdots A_lB_l}h_{C_1D_1C_2D_2}+ \cdots\bigg), \nonumber
\end{eqnarray}
where the dots indicate terms at least quadratic in $h_{ABCD}$. In particular one has that
\begin{eqnarray*}
&& \xi^*_{p,0;C_pD_p\cdots C_1D_1A_1B_1\cdots A_lB_l} =D_{(C_pD_p}\cdots D_{C_1D_1)}\xi^*_{A_1B_1\cdots A_lB_l}, \\
&& \xi^*_{p,0;C_pD_p\cdots C_1D_1A_1B_1\cdots A_lB_l} =\frac{1}{2p-1} \sum_{1\leq k< h \leq p} D_{(C_pD_p} \cdots D_{|EF|} \cdots D^{EF} \cdots  D_{C_3D_3)} \xi^*_{A_1B_1\cdots A_lB_l}, \nonumber
\end{eqnarray*}
with $D_{EF}$ and $D^{EF}$ assumed to be, respectively, in the $k$th and $h$th position. More generally, one has the symmetry
\[
\xi^*_{p,i;C_pD_p\cdots C_{2i+1}D_{2i+1}A_1B_1\cdots A_lB_l}=\xi^*_{p,i;(C_pD_p\cdots C_{2i+1}D_{2i+1})(A_1B_1\cdots A_lB_l)},
\]
with $i=0,\ldots p/2$ if $p$ is even and $i=0,\ldots, (p-1)/2$ if $p$ is odd. The complete decomposition of the coefficients in (\ref{spinor:expansion1}) renders
\begin{eqnarray}
&& \updnup{\xi}{*}{p,i;C_pD_p\cdots C_{2i+1}D_{2i+1}}{A_1B_1\cdots A_lB_l} \nonumber \\
&&\hspace{1cm} =  \updnup{\xi}{0}{p,i;C_pD_p\cdots C_{2i+1}D_{2i+1}}{A_1B_1\cdots A_lB_l} + \updnup{\xi}{1}{p,i;(C_pD_p\cdots C_{2i+1}}{(A_1B_1\cdots A_l}\dnup{\epsilon}{D_{2i+1})}{B_l)} \nonumber \\
&&\hspace{1.5cm} + \updnup{\xi}{2}{p,i;(C_pD_p\cdots D_{2i+3}}{(A_1B_1\cdots B_{l-1}}\dnup{\epsilon}{C_{2i+1}}{A_l}\dnup{\epsilon}{D_{2i+1})}{B_l)}+\cdots,
\end{eqnarray}
with
\begin{eqnarray}
&& \xi^0_{p,i;C_pD_p\cdots C_{2i+1}D_{2i+1}A_1B_1\cdots A_lB_l}=\xi^*_{p,i;(C_pD_p\cdots C_{2i+1}D_{2i+1}A_1B_1\cdots A_lB_l)}, \\
&& \xi^1_{p,i;C_pD_p\cdots C_{2i+1}D_{2i+1}A_1B_1\cdots A_lB_l}= c^1_{p,i,l} \updnup{\xi}{*}{p,i;(C_pD_p\cdots C_{2i+1}|E|A_1B_1\cdots A_l)}{E},
\end{eqnarray}
for some real coefficients $c^1_{p,i,l}$.

\subsection{An orthonormal basis for functions on $SU(2,\Complex)$}
The lift of the expansion (\ref{spinor:expansion1}) from
$\mathcal{B}_a$ to $\mathcal{C}_a$ introduces in a natural way a class
of functions associated with unitary representations of
$SU(2,\Complex)$. Namely, given $\updn{t}{A}{B}\in SU(2,\Complex)$,
define
\begin{eqnarray*}
&& \TT{m}{j}{k}(\updn{t}{A}{B}) = \binom{m}{j}^{1/2} \binom{m}{k}^{1/2} \updn{t}{(B_1}{(A_1}\cdots \updn{t}{B_m)_j}{A_m)_k}, \\
&& \TT{0}{0}{0}(\updn{t}{A}{B})=1,
\end{eqnarray*}
with $j,k=0,\ldots,m$ and $m=1,2,3,\ldots$. The expression $(A_1\cdots
A_m)_k$ means that the indices are symmetrised and then $k$ of them are
set equal to $1$, while the remaining ones are set to $0$. Details about the
properties of these functions can be found in \cite{Fri86a,Fri98a}.
The functions $\sqrt{m+1}\TT{m}{j}{k}$ form a complete orthonormal set
in the Hilbert space $L^2(\mu,SU(2,\Complex))$, where $\mu$ denotes
the normalised Haar measure on $SU(2,\Complex)$. In particular, any
analytic complex-valued function $f$ on $SU(2,\Complex)$ admits an
expansion
\[
f(\updn{t}{A}{B})=\sum_{m=0}^\infty \sum_{j=0}^m \sum_{k=0}^m f_{m,k,j} \TT{m}{k}{j}(\updn{t}{A}{B}),
\]
with complex coefficients $f_{m,k,j}$. Under complex conjugation the
functions transform as
\[
\overline{\TT{m}{j}{k}} =(-1)^{j+k} \TT{m}{m-j}{m-k}.
\]
The action of the differential operators (\ref{diffops:X}) on the
functions $\TT{m}{k}{j}$ is given by
\begin{eqnarray*}
&& X \TT{m}{k}{j}= (m-2j) \TT{m}{k}{j}, \\
&& X_+ \TT{m}{k}{j} = \sqrt{j(m-j+1)} \TT{m}{k}{j-1}, \quad X_-\TT{m}{k}{j}=-\sqrt{(j+1)(m-j)}\TT{m}{k}{j+1}.
\end{eqnarray*}
A function $f$ is said to have spin weight $s$ if
\[
Xf =2s f.
\]
Such a function has a \emph{simplified} expansion of the form
\[
f=\sum_{m\geq|2s|}^\infty \sum_{k=0}^m f_{m,k} \TT{m}{k}{m/2-s}.
\]

\subsection{Normal expansions at $\mathcal{I}^0$}
\label{subsection:normal_exp_I0}
In the sequel it will be necessary to be able to relate fields in
$\mathcal{C}_a$ with fields in $\mathcal{B}_a$. Crucially, one will
require to be able to \emph{lift} smooth fields defined on
$\mathcal{B}_a$ to $\mathcal{C}_a$. As in section
(\ref{subsection:normal_exp_i}) consider normal coordinates $x^i$ on
$\mathcal{B}_a$ centred on $i$ based on the orthonormal frame $c^*_a =
\sigma^{AB}_a c^*_{AB} =\sigma_a^{AB}\delta^*_A \delta^*_B$. In terms
of $\rho$ and $\updn{t}{A}{B}$ on $\mathcal{C}_a$ and the normal
coordinates $x^i$, the projection $\pi^\prime$ has the local
expression
\[
\pi^\prime: (\rho,\updn{t}{A}{B}) \rightarrow x^i(\rho,\updn{t}{A}{B}) = \sqrt{2} \rho \sigma^i_{CD} \updn{t}{C}{0} \updn{t}{D}{1}.
\]
This expression can be used to pull-back fields (lift) to
$\mathcal{C}_a$. In particular the pull-back of $|x|$ is $\rho$.

The procedure to lift expansions of the type given by
(\ref{spinor:expansion1}) has been discussed in \cite{Fri98a}.
Starting from the analytic part of (\ref{spinor:expansion1}) one has
to perform the following operations:
\begin{itemize}
\item[(i)] the transition $\xi^*_{A_1\cdots A_l} \rightarrow
  \xi^*_{B_1\cdots B_l} \updn{t}{B_1}{A_1} \cdots \updn{t}{B_l}{A_l}$;
\item[(ii)] the replacement $|x|\rightarrow \rho$ and
  $n^{AB}\rightarrow \sqrt{2} \updn{t}{(A}{0} \updn{t}{B)}{1}$;
\item[(iii)] decompose all the spinor-valued coefficients into sums of
  products of symmetric coefficients with $\epsilon_{AB}$'s.
  Contractions of $\epsilon_{AB}$'s with a pair of $\updn{t}{C}{D}$
  yields factors of $0$ or $1$, while the remaining expressions assume
  the form of expansions in terms of the functions $\TT{m}{k}{l}$.
\end{itemize}
Applying the previous procedure to the analytic part of the expansion
(\ref{spinor:expansion1}) of a spinorial field $\xi_{A_1B_1\cdots A_l
  B_l}$ on $\mathcal{B}_a$ one obtains the expansion of the
spinor-valued function $\xi_{A_1B_1\cdots A_l B_l}$ on
$\mathcal{C}_a$. Denote by $\xi_j=\xi_{(A_1B_1\cdots A_l B_l)_j}$,
$0\leq j \leq l$ its essential components. The function $\xi_j$ has
spin weight $s=l-j$ and a unique expansion of the form
\begin{equation}
\xi_j = \sum^m_{p=0} \xi_{j,p}\rho^p + (\xi_j)_\mathcal{R}, \label{expansion:Ca1}
\end{equation}
for all $m$, with
\begin{equation}
\label{expansion:Ca2}
\xi_{j,p} = \sum_{q= \max\{|l-j|, l-p\}}^{p+l} \sum_{k=0}^{2q} \xi_{j,p;2q,k} \TT{2q}{k}{2q-l+j},
\end{equation}
and complex coefficients $\xi_{j,p;2q,k}$. In particular one has that
\[
\xi_{j,p;2q+2l,k}= (\sqrt{2})^p {\binom{2p+2l}{k}}^{1/2} {\binom{2p+2l}{p+j}}^{-1/2} D_{(C_pD_p} \cdots D_{C_1D_1} \xi^*_{A_1B_1 \cdots A_P B_p)}(i).
\]
Hence, one has the symmetry
\[
\xi_{0,p;2p+2l,k}= \xi_{2l,p;2p+2l,k}
\]
which will play an important role in the sequel. Another lengthy, but
straightforward calculation shows that
\begin{equation}
\label{expansion:Ca3}
\xi_{j,p;2p+2,k} = K_{p,j,k} D_{(A_pB_p} \cdots D_{A_1|E|} \dnup{\xi}{ABC)_j}{E}(i),
\end{equation}
with $K_{p,j,k}$ a constant depending on $p$, $j$, $k$. If $l$ is even, then $\xi_{A_1\cdots A_l}$ is associated to a real spatial tensor if and only if the following reality conditions hold:
\[
\xi_j =(-1)j \overline{\xi}_{2l-j}, \quad \xi_{j,p;2q,k} =(-1)^{r+q+k} \overline{\xi}_{2l-j,p;2q,2q-k}.
\]

\bigskip The discussion in the following sections will require to
consider smooth spinorial fields $\xi_{A_1 B_1 \cdots A_r B_r}$ with
essential components $\xi_j=\xi_{(A_1B_1\cdots A_rB_r)_j}$, $0\leq j
\leq 2r$ of spin weight $s=r-j$ with expansions which are more general
than those given in equations (\ref{expansion:Ca1}) and
(\ref{expansion:Ca2}). Accordingly, they do not descend to smooth
spinors on $\mathcal{B}_a$. In this case, instead of
(\ref{expansion:Ca2}), one considers the more general expression
\[
\xi_{j,p} = \sum_{q=|r-j|}^{q(p)} \sum_{k=0}^2q \xi_{j,p;2q,k} \TT{2q}{k}{q-r+j},
\]
where in principle $0\leq |r-j| \leq q(p) \leq \infty$. In this case
one will speak of an \emph{expansion of type} $q(p)$. If $f$, $g$ have
expansion types $q(p)$, $q'(p)$, respectively, then the product $fq$
will have expansion type $\max_{0\leq j \leq p} \{ q'(j) + q(p-j)\}$,
while the sum $f+g$ will have expansion type $\max\{q(p),q'(p)\}$. In
particular, $\rho f$ will have expansion type $q(p)-1$.

\subsection{Consequences of the cn-gauge for normal expansions}
\label{section:cngauge}

In \cite{Fri98a} a number of consequences for quantities derived from
an analytic conformal metric $h_{ij}$ in the cn-gauge has been deduced 
---see lemma 4.7 in the aforementioned reference.

\begin{lemma}
\label{lemma:expansion_type1}
In the cn-gauge one has that
\begin{eqnarray*}
&& \mbox{\em type}(r)= p, \quad  r=O(\rho^2),\\
&& \mbox{\em type}(s_{ABCD}) =p+1, \quad  s_{ABCD}=O(\rho), \\
&& \mbox{\em type}(\check{\gamma}_{ABCD}) =p, \quad \check{\gamma}_{ABCD}=O(\rho^2),\\
&& \mbox{\em type}(\check{c}^\pm_{AB})=p, \quad \check{c}^\pm_{AB}=O(\rho^2),\\
&& \mbox{\em type}(U-1) =p-1, \quad  U=1+O(\rho^4).
\end{eqnarray*}
\end{lemma}

In addition, from the discussion in section
\ref{consequences:freedata} one has that under the
assumptions being made it follows that the function $W$ in
equation(\ref{parametrisation_theta}) satisfies $W\in
E^\infty(\mathcal{B}_a)$. The procedure of subsection
\ref{subsection:normal_exp_I0} gives

\begin{lemma}
\label{lemma:expansion_type2}
\[
 \mbox{\em type}(W)= p,  \quad  W=\frac{m}{2} + O(\rho)
\]
\end{lemma}

\textbf{Remark.} It is worth noticing that by means of an adequate choice of
the \emph{centre of mass} it is possible to set $W=m/2 +O(\rho^2)$
---see e.g. \cite{Val04e}, although this fact will not be used
here.

An important consequence of lemma \ref{lemma:expansion_type1}, of
crucial relevance in the analysis of solutions of the momentum
constraint and in particular of equation (\ref{elliptic:eqn:v}) is the
following lemma already mentioned in section \ref{section:freedata}.

\begin{lemma} \label{metric:cngauge}
In normal coordinates based around $i$ an analytic metric in the cn-gauge
is of the form
\[
h_{ij} = -\delta_{ij} + r^2 \check{h}_{ij},
\]
with $\check{h}_{ij}=O(r)$.
\end{lemma}

\textbf{Proof.} As seen in section \ref{section:manifold_Ca}, the lift
of the conformal metric $h_{ij}$ to $\mathcal{C}_a$ can be expressed
in terms of the soldering forms $\sigma^{AB}$ as $h = h_{ABCD}
\sigma^{AB} \otimes \sigma^{CD}$. Now, using the explicit
decomposition
\[
\sigma^{AB}= -x^{AB} \mbox{d}\rho + (-2\rho y^{AB} + \check{\sigma}^y_+ y^{AB} + \check{\sigma}^z_+ z^{AB}) \alpha^+ + (-2\rho z^{AB} + \check{\sigma}^z_- z^{AB} + \check{\sigma}^y_- y^{AB}) \alpha^-.
\]
one finds that
\begin{eqnarray*}
&& \check{\sigma}_+^y = \frac{2\rho^2}{\xi} \left( \rho \check{c}^-_z \check{c}^+_y - \rho \check{c}^-_y \check{c}^+_z - \check{c}^+_z \right), \quad \check{\sigma}^z_+ = \frac{2\rho}{\xi}\check{c}^-_z, \\
&&  \check{\sigma}^y_- = \frac{2\rho}{\xi}\check{c}^+_y, \quad \check{\sigma}_-^z = \frac{2\rho^2}{\xi} \left( \rho \check{c}^-_z \check{c}^+_y - \rho \check{c}^-_y \check{c}^+_z - \check{c}^-_y \right),
\end{eqnarray*}
with
\[
\xi= \rho^2\check{c}^-_z \check{c}^+_y -\rho^2 \check{c}^-_y \check{c}^+_z -\rho \check{c}^+_x -\rho \check{c}^-_y -1.
\]
In the cn-gauge $\check{c}^\pm_{AB}$ has expansion type $p$ and hence
$\sigma_\pm^{AB}$ has expansion type $p-2$. Furthermore, because in
the cn-gauge $c_{AB}^\pm=O(\rho^2)$ one concludes that
$\sigma_\pm^{AB}=O(\rho^4)$. Accordingly one can write
\[
\check{\sigma}^{AB}_\pm = \rho^3 \tilde{\sigma}^{AB}_\pm, \quad \tilde{\sigma}^{AB}_\pm=O(\rho).
\]
Now, it is noticed that
\begin{eqnarray*}
&& \mathring{h} = h_{ABCD} \left(-x^{AB} \mbox{d}\rho - 2\rho y^{AB} \alpha^+ -2 \rho z^{AB} \alpha^-  \right) \otimes \left( -x^{CD} \mbox{d}\rho - 2\rho y^{CD} \alpha^+ -2 \rho z^{CD} \alpha^-\right), \\
&& \phantom{\mathring{h}}= -\mbox{d}\rho \otimes \mbox{d}\rho - 2 \rho^2 (\alpha^+\otimes \alpha^- + \alpha^- \otimes \alpha^+),
\end{eqnarray*}
on $\mathcal{C}_a$ descends to the flat metric (in Cartesian coordinates)
\[
\mathring{h} = -\delta_{ij} \mbox{d}x^i \otimes \mbox{d}x^j,
\]
for the pull-back of the standard metric on the 2-sphere is given by
$2(\alpha^+\otimes \alpha^- + \alpha^- \otimes \alpha^+)$. Thus the
1-forms $-x^{AB}\mbox{d}\rho$, $- 2\rho y^{AB} \alpha^+$ and $-2 \rho
z^{AB} \alpha^-$ on $\mathcal{C}_a$ descend, respectively, to 1-forms
on $\mathcal{B}_a$, $n_i \mbox{d}x^i$, $\mathring{m}_i \mbox{d}x^i$
and $\overline{\mathring{m}}_i \mbox{d}x^i$ with $n^i=x^i/r$ and such that
\[
\mathring{h} = \left( n_i n_j + \mathring{m}_i \overline{\mathring{m}}_j +   \mathring{m}_j \overline{\mathring{m}}_i  \right) \mbox{d}x^i \otimes \mbox{d}x^j.
\]
Similarly, the 1-forms $\check{\sigma}^{AB}_+ \alpha^+$ and
$\check{\sigma}^{AB}_- \alpha^-$ on $\mathcal{C}_a$ descend to the
1-forms $r^2 \tilde{m}_j \mbox{d}x^j$ and $r^2 \overline{\tilde{m}}_j
\mbox{d}x^j$ on $\mathcal{B}_a$, with $\tilde{m}_j$ and
$\overline{\tilde{m}}_j$ analytic. Thus, one has that
\begin{eqnarray*}
&& h = \left( n_i n_j + \mathring{m}_i \overline{\mathring{m}}_j +   \mathring{m}_j \overline{\mathring{m}}_i  \right) \mbox{d}x^i \otimes \mbox{d}x^j + r^2 \left(\mathring{m}_i \overline{\tilde{m}}_j  +\overline{\mathring{m}}_i \tilde{m}_j  + \mathring{m}_j \overline{\tilde{m}}_i  +\overline{\mathring{m}}_j \tilde{m}_i\right)\mbox{d}x^i \otimes \mbox{d}x^j \\
&& \hspace{2cm} + r^4 \left( \tilde{m}_i \overline{\tilde{m}}_j + \tilde{m}_j \overline{\tilde{m}}_i   \right)\mbox{d}x^i \otimes \mbox{d}x^j.
\end{eqnarray*}
Now, $\mathring{m}_i= O(1)$ is not a smooth function of the normal
coordinates, however, by assumption $h_{ij}$ is smooth, and
consequently the combination
\[
\mathring{m}_i \overline{\tilde{m}}_j  +\overline{\mathring{m}}_i \tilde{m}_j  + \mathring{m}_j \overline{\tilde{m}}_i  +\overline{\mathring{m}}_j \tilde{m}_i,
\]
must be analytic on $\mathcal{B}_a$. Hence, the metric is of the required form.
\hfill $\Box$

\section{The solutions to the momentum constraint on $\mathcal{C}_a$}
\label{section:spinor_mc}
In this section the lifts of the solutions to the momentum constraint
introduced in section \ref{section:tensor_mc} will be analysed. The procedure described here has been previously implemented in \cite{Val04e,Val07a}.

\subsection{The conformally flat case}
In section \ref{section:tensor_mc} the solutions of the flat space
momentum constraint have been discussed by introducing a certain frame
$\{ n^i, \,m^i, \,\overline{m}^i\}$. This frame is related to an
orthonormal frame by means of the relations
\[
e^i_1=\frac{1}{\sqrt{2}}(m^i +\overline{m}^i), \quad e^i_2 =\frac{\mbox{i}}{\sqrt{2}}(m^i-\overline{m}^i), \quad e^i_3 = n^i.
\]
Inverting these relations one can readily rewrite $\{ n^i, \,m^i,
\,\overline{m}^i\}$ in terms of the orthonormal frame
$\{e^i_{a}\}_{a=1,2,3}$. This leads to a direct transcription into
spinorial objects. Let $\{n_a, \, m_a, \,\overline{m}_a \}$ denote the
components of the frame $\{ n^i, \,m^i, \,\overline{m}^i\}$ with
respect to $\{e^i_a\}$, $a=1,2,3$. Using the spatial Infeld
symbols one obtains to the following transcription rules:
\[
n_a \rightarrow x_{AB}, \quad m_a \rightarrow \sqrt{2} z_{AB}, \quad \overline{m}_a \rightarrow \sqrt{2} y_{AB}.
\]
Thus, formula (\ref{mathring:ansatz}) lifts to
 \begin{eqnarray}
&&\rho^3\mathring{\psi}_{ABCD}=\xi(3x_{AB}x_{CD}+h_{ABCD})+\sqrt{2}\overline{\eta}_1(x_{AB}y_{CD}+x_{CD}y_{AB})+\sqrt{2}\eta_1(x_{AB}z_{CD}+x_{CD}z_{AB}) \nonumber \\
&&\hspace{3cm}+2\overline{\mu}_2(y_{AB}y_{CD})+2\mu_2 (z_{AB}z_{CD}). \label{spinor:ansatz1}
\end{eqnarray}
Alternatively, one could rewrite the previous formula in terms of the
totally symmetric spinors $\epsilon^i_{ABCD}$, $i=0,1,\ldots,4$ as:
\begin{equation}
\label{spinor:ansatz2}
\rho^3 \mathring{\psi}_{ABCD}= \mathring{\psi}_0 \epsilon^0_{ABCD} + \mathring{\psi}_1 \epsilon^1_{ABCD}   + \mathring{\psi}_2 \epsilon^2_{ABCD}  + \mathring{\psi}_3 \epsilon^3_{ABCD}  + \mathring{\psi}_4 \epsilon^4_{ABCD}
\end{equation}
with
\begin{eqnarray*}
&& \mathring{\psi}_0 \equiv \mu, \quad \mathring{\psi}_1 \equiv 2\sqrt{2} \eta_1, \quad  \mathring{\psi}_2 \equiv 6\xi, \\
&& \mathring{\psi}_3 \equiv -2\sqrt{2} \overline{\eta}, \quad \mathring{\psi}_4 \equiv \overline{\mu}.
\end{eqnarray*}
In order to show the equivalence of the expressions
(\ref{spinor:ansatz1}) and (\ref{spinor:ansatz2}) one makes use of the
spinorial identities given in appendix \ref{appendix:spinors}.
Following the notation of section \ref{section:tensor_mc} one writes
\[
\mathring{\psi}_{ABCD}=\mathring{\psi}^A_{ABCD}+\mathring{\psi}^P_{ABCD}+\mathring{\psi}^Q_{ABCD}+\mathring{\psi}^J_{ABCD}.
\]
The non-vanishing components of $\mathring{\psi}^A_{ABCD}$ are given by
\[ \label{psi A}
 \mathring{\psi}_2^A=-\frac{A}{\rho^3}T_{0\phantom{0}0}^{\phantom{0}0}.
\]
Those of $\psi^P_{ABCD}$ are
\begin{eqnarray*}
&& \mathring{\psi}_1^P=\frac{3}{\rho^4}(P_2+\mbox{i}P_1)T_{2\phantom{0}0}^{\phantom{2}0}-\frac{3\sqrt{2}}{\rho^4}P_3T_{2\phantom{1}0}^{\phantom{2}1}-\frac{3}{\rho^4}(P_2-\mbox{i}P_1)T_{2\phantom{2}0}^{\phantom{2}2}, \\
&& \mathring{\psi}_2^P=\frac{9\sqrt{2}}{2\rho^4}(P_2+\mbox{i}P_1)T_{2\phantom{0}1}^{\phantom{2}0}-\frac{9}{\rho^4}P_3T_{2\phantom{1}1}^{\phantom{2}1}-\frac{9\sqrt{2}}{2\rho^4}(P_2-\mbox{i}P_1)T_{2\phantom{2}1}^{\phantom{2}2}, \\
&& \mathring{\psi}_3^P=-\frac{3}{\rho^4}(P_2-\mbox{i}P_1)T_{2\phantom{2}2}^{\phantom{2}2}-\frac{3\sqrt{2}}{\rho^4}P_3T_{2\phantom{1}2}^{\phantom{2}1}+\frac{3}{\rho^4}(P_2+\mbox{i}P_1)T_{2\phantom{2}2}^{\phantom{2}0}.
\end{eqnarray*}
While in the case of $\mathring{\psi}_{ABCD}^Q$ one has,
\begin{subequations}
\begin{eqnarray*}
&&\mathring{\psi}_1^Q=\frac{3}{\rho^2}(Q_2+\mbox{i}Q_1)T_{2\phantom{0}0}^{\phantom{2}0}-\frac{3\sqrt{2}}{\rho^2}Q_3T_{2\phantom{1}0}^{\phantom{2}1}-\frac{3}{\rho^2}(Q_2-\mbox{i}Q_1)T_{2\phantom{2}0}^{\phantom{2}2}, \label{psi Q1}\\
&&\mathring{\psi}_2^Q=\frac{9\sqrt{2}}{2\rho^2}(Q_2+\mbox{i}Q_1)T_{2\phantom{0}1}^{\phantom{2}0}-\frac{9}{\rho^2}Q_3T_{2\phantom{1}1}^{\phantom{2}1}-\frac{9\sqrt{2}}{2\rho^2}(Q_2-\mbox{i}Q_1)T_{2\phantom{2}1}^{\phantom{2}2} \label{psi Q2},\\
&&\mathring{\psi}_3^Q=-\frac{3}{\rho^2}(Q_2-\mbox{i}Q_1)T_{2\phantom{2}2}^{\phantom{2}2}-\frac{3\sqrt{2}}{\rho^2}Q_3T_{2\phantom{1}2}^{\phantom{2}1}+\frac{3}{\rho^2}(Q_2+\mbox{i}Q_1)T_{2\phantom{2}2}^{\phantom{2}0}. \label{psi Q3}
\end{eqnarray*}
\end{subequations}
And finally for $\mathring{\psi}_{ABCD}^J$,
\begin{subequations}
\begin{eqnarray*}
&&\mathring{\psi}_1^J=\frac{6}{\rho^3}(-J_1+\mbox{i}J_2)\TT{2}{0}{0}+\frac{6\sqrt{2}}{\rho^3}\mbox{i}J_3\TT{2}{1}{0}-\frac{6}{\rho^3}(J_1+\mbox{i}J_2)\TT{2}{2}{0}, \label{psi J1}\\
&&\mathring{\psi}_2^J=0, \label{psi J2}\\
&&\mathring{\psi}_3^J=\frac{6}{\rho^3}(J_1+\mbox{i}J_2)\TT{2}{2}{2}-\frac{6\sqrt{2}}{\rho^3}\mbox{i}J_3\TT{2}{1}{2}-\frac{6}{\rho^3}(-J_1+\mbox{i}J_2)\TT{2}{0}{2}. \label{psi J3}
\end{eqnarray*}
\end{subequations}
In the above expressions $A$, $P_1$, $P_2$, $P_3$, $J_1$, $J_2$,
$J_3$, $Q_1$, $Q_2$, $Q_3\in \Real$. With regards to the part of
$\mathring{\psi}_{ABCD}$ derived from the complex function one has
that
\begin{subequations}
\begin{eqnarray*}
&& \xi= X_{-}^2 \lambda^{(R)},  \label{soln_lambda_1}\\
&& \eta_1=-2\rho\partial_\rho X_{-}\lambda^{(R)} +X_{-}\lambda^{(I)}, \label{soln_lambda_2} \\
&& \mu_2=2\rho\partial_\rho(\rho\partial_\rho \lambda^{(R)})+X_+X_-\lambda^{(R)}-2\lambda^{(R)}-\rho\partial_\rho\lambda^{(I)}, \label{soln_lambda_3}
\end{eqnarray*}
\end{subequations}
with
\[
\lambda=X_+^2\tilde{\lambda}=X_+^2\mbox{Re}(\tilde{\lambda})+X_+^2\mbox{Im}(\tilde{\lambda})=\lambda^{(R)}+\lambda^{(I)}.
\]

By direct inspection of the previous expressions of the components of
$\mathring{\psi}_{ABCD}$ one has the following lemma.

\begin{lemma}
\label{lemma:mathringpsi_type}
The lifts to $\mathcal{C}_a$ of the solutions to the flat momentum
constraint satisfy
\begin{eqnarray*}
&& \mathring{\psi}_{ABCD}[A] = \rho^{-3} \Xi_{ABCD}[A], \quad \mbox{\em type}(\Xi_{ABCD}[A])=p, \\
&& \mathring{\psi}_{ABCD}[J] = \rho^{-3} \Xi_{ABCD}[J], \quad \mbox{\em type}(\Xi_{ABCD}[J])=p+1, \\
&& \mathring{\psi}_{ABCD}[Q] = \rho^{-3} \Xi_{ABCD}[Q], \quad \mbox{\em type}(\Xi_{ABCD}[Q])=p, \\
&& \mathring{\psi}_{ABCD}[\lambda^{(1)}] = \rho^{-3} \Xi_{ABCD}[\lambda^{(1)}], \quad \mbox{\em type}(\Xi_{ABCD}[\lambda^{(1)}])=p, \\
&& \mathring{\psi}_{ABCD}[\lambda^{(2)}/r] = \rho^{-3}
\Xi_{ABCD}[\lambda^{(2)}/r], \quad \mbox{\em
type}(\Xi_{ABCD}[\lambda^{(2)}/r])=p+1.
\end{eqnarray*}

\end{lemma}

\bigskip
Note that because $\mathring{\psi}_{ABCD}$ is totally symmetric, then
its projection to $\mathcal{B}_a$ is associated to an $h$-trace free
symmetric tensor. In other words, borrowing the terminology of section
\ref{section:freedata}, $\mathring{\Phi}_{ab}=\sigma_a^{AB} \sigma_b^{CD}
  \mathring{\psi}_{ABCD}$.

\subsection{The non-conformally flat case}
A detailed discussion of the solutions, $v^i$, of equation
(\ref{elliptic:eqn:v}) has been given in appendix
\ref{appendix:momentum}. From this analysis one can readily deduce the
expansion types of the lift to $\mathcal{C}_a$ of the various parts of
$v^i$.

Consider the vector $v^i[\mbox{Re}(\lambda^{(2)})/r]$ produced by the
seed $\mathring{\psi}_{ij}[\mbox{Re}(\lambda^{(2)})/r]$. In
appendix \ref{appendix:momentum} it is shown that under the
assumptions of section \ref{section:freedata} one has
\[
v^i[\mbox{Re}(\lambda^{(2)})/r]= v_1^i[\mbox{Re}(\lambda^{(2)})/r]
+ v^i_2[\mbox{Re}(\lambda^{(2)})/r],
\]
with $v_1^i[\mbox{Re}(\lambda^{(2)})/r]\in
\mathcal{Q}_\infty(\mathcal{B}_a)$,
$v_1^i[\mbox{Re}(\lambda^{(2)})/r]=O(r^{4})$, and
$v_2^i[\mbox{Re}(\lambda^{(2)})/r]\in C^\infty(\mathcal{B}_a)$. In
what follows, the affixed $[\mbox{Re}(\lambda^{(2)})/r]$ will be
suppressed for ease of reading. Noting that because of $x_i v^i =
r^2 u$, $u\in C^\infty(\mathcal{B}_a)$, one can write
\[
v_1^i = ru n^i + w m^i + \overline{w} \overline{m}^i,
\]
with $w$ a $C^\infty$ complex function. Thus, one has that
\[
v^i = r n^i + (w m^i + \overline{w} \overline{m}^i).
\]
Accordingly, the lift of $v^i$ to $\mathcal{C}_a$ is readily found to be
\[
v_{AB} = \rho u x_{AB} +  (w z_{AB} + \overline{w} y_{AB}) + v^2_{AB},
\]
with $v$, $u$, $v^2_{AB}$ of expansion type $p+1$. Write $u_{AB}=ux_{AB}$, $w_{AB}=w z_{AB}$ and
$\tilde{w}_{AB}= \overline{w} y_{AB}$. A calculation renders
\begin{eqnarray*}
&&D_{(AB} v_{CD)} = u x_{(AB}x_{CD)} + \rho D_{(AB} u_{CD)} + D_{(AB} w_{CD)} + D_{(AB} \tilde{w}_{CD)} +D_{(AB} v^2_{CD)} \\
&&\phantom{D_{(AB} v_{CD)}} = \rho^{-3} \bigg(  \rho^3 u x_{(AB}x_{CD)} + \rho^4 D_{(AB} u_{CD)} + \rho^3 D_{(AB} w_{CD)} + \rho^3 D_{(AB} \tilde{w}_{CD)} + \rho^3D_{(AB} v^2_{CD)} \bigg).
\end{eqnarray*}
From the discussion in the previous paragraph it follows that
$\mbox{type}(\rho^3 u)=p-2$, $\mbox{type}(\rho^4 D_{(AB}
u_{CD)})=p-2$, $\mbox{type}(\rho^3 D_{(AB}w_{CD)})=p-1$ and
$\mbox{type}(\rho^3 D_{(AB}v^2_{CD)})=p-1$. Accordingly, one can write
\[
D_{(AB}v_{CD)}[\mbox{Re}(\lambda^{(2)})/r] = \rho^{-3}
\Phi_{ABCD}^v[\mbox{Re}(\lambda^{(2)})/r], \quad
\mbox{type}(\Phi_{ABCD}^v[\mbox{Re}(\lambda^{(2)})/r])= p-1.
\]

Using similar arguments one obtains the following

\begin{theorem}
\label{thm:v}
Under the assumptions of section \ref{section:freedata} on the initial
data, the lifts $v_{AB}$ of the vectors $v^i$ solving equation
(\ref{elliptic:eqn:v}) are of the form:
\begin{eqnarray*}
&& D_{(AB}v_{CD)}[A]= \rho^{-3} \Phi^v_{ABCD}[A], \quad \mbox{\em type}(\Phi^v_{ABCD}[A])=p-1, \\
&& D_{(AB}v_{CD)}[A]= \rho^{-3} \Phi^v_{ABCD}[J], \quad \mbox{\em type}(\Phi^v_{ABCD}[J])=p, \\
&& D_{(AB}v_{CD)}[Q]= \rho^{-3} \Phi^v_{ABCD}[Q], \quad \mbox{\em type}(\Phi^v_{ABCD}[Q])=p, \\
&& D_{(AB}v_{CD)}[\mbox{\em Re}(\lambda^{(1)})] = \rho^{-3} \Phi_{ABCD}^v[\mbox{\em Re}(\lambda^{(1)})], \quad \mbox{\em type}(\Phi_{ABCD}^v[\mbox{\em Re}(\lambda^{(1)})])= p-1, \\
&& D_{(AB}v_{CD)}[\mbox{\em Im}(\lambda^{(1)})] = \rho^{-3} \Phi_{ABCD}^v[\mbox{\em Im}(\lambda^{(1)})], \quad \mbox{\em type}(\Phi_{ABCD}^v[\mbox{\em Im}(\lambda^{(1)})])= p-1,\\
&& D_{(AB}v_{CD)}[\mbox{\em Re}(\lambda^{(2)})/r] = \rho^{-3} \Phi_{ABCD}^v[\mbox{\em Re}(\lambda^{(2)})/r], \quad \mbox{\em type}(\Phi_{ABCD}^v[\mbox{\em Re}(\lambda^{(2)})/r])= p-1, \\
&& D_{(AB}v_{CD)}[\mbox{\em Im}(\lambda^{(2)})/r] = \rho^{-3}
\Phi_{ABCD}^v[\mbox{\em Im}(\lambda^{(2)})/r], \quad \mbox{\em
type}(\Phi_{ABCD}^v[\mbox{\em Im}(\lambda^{(2)})/r])= p.
\end{eqnarray*}
\end{theorem}

\section{Structure of the Weyl tensor on $\mathcal{C}_a$}
\label{section:weyltensor}

The rescaled Weyl tensor plays a fundamental role in the
discussion of the asymptotic properties of the gravitational
field. Anticipating the discussion of section \ref{section:Fgauge}
an analysis of its structure near infinity at the level of initial
data using the manifold $\mathcal{C}_a$ is now given. The
discussion of this section is a generalisation of the analysis of
section 4 in reference \cite{Fri98a}.

The spinorial counterpart of the rescaled tensor
$W_{\mu\nu\lambda\rho}$ is given in terms of
\[
W_{AA'BB'CC'DD'}=  \phi_{ABCD}\epsilon_{A'B'} \epsilon_{C'D'} + \overline{\phi}_{A'B'C'D'}\epsilon_{AB}\epsilon_{CD},
\]
where $\phi_{ABCD}$ is the spinorial counterpart of a self-dual
tensor. At the level of initial data, $\phi_{ABCD}$, is fully
described in terms of the spinors $w_{ABCD}$ and $w^*_{ABCD}$ ---the
spinorial counterparts of the spatial tensors $w_{ab}$ and
$w^*_{ab}$ introduced in section
\ref{section:conformalconstraints}. One has that
\[
\phi_{ABCD}= w_{ABCD} + \mbox{i} w^*_{ABCD}.
\]
Using the conformal constraint equations (\ref{cc1})-(\ref{cc2}) of
section \ref{section:conformalconstraints} and transcribing into space
spinorial language one finds
\begin{subequations}
\begin{eqnarray}
&&\hspace{-1cm} w_{ABCD}=  \Omega^{-2} D_{(AB} D_{CD)} \Omega + \Omega^{-1} s_{ABCD} + \Omega^{-1} \left( \updn{\chi}{EF}{EF} \chi_{(ABCD)} - \updn{\chi}{EF}{(AB} \chi_{CD)EF}  \right), \label{electric}\\
&&\hspace{-1cm} w^*_{ABCD} = -\mbox{i} \Omega^{-1} \sqrt{2} \updn{D}{F}{(A} \chi_{BCD)F}, \label{magnetic}
\end{eqnarray}
\end{subequations}
where $\chi_{ABCD}$ is the spinorial counterpart of the second
fundamental form $\chi_{ij}$. For the class of initial data under
consideration (maximal) $\updn{\chi}{EF}{EF}=0$. If $\psi_{ABCD}$
is the spinorial counterpart of the rescaled second fundamental
form $\psi_{ij}$ then
\[
\chi_{ABCD}=\Omega^2 \psi_{ABCD}.
\]
As the magnetic part, $w^*_{ABCD}$, is linear
in $\psi_{ABCD}$, it will prove useful to consider the following splitting of
$w^*_{ABCD}$:
\[
w^*_{ABCD} = w^*_{ABCD}[A,J,Q]+w^*_{ABCD}[\lambda^{(R)}] + w^*_{ABCD}[\lambda^{(I)}].
\]
In this last formula $w^*_{ABCD}[\lambda^{(R)}]$ denotes the part of
$w^*_{ABCD}$ obtained from $\psi_{ij}[\lambda^{(R)}]$ while
$w^*_{ABCD}[\lambda^{(I)}]$ is the part calculated from
$\psi_{ij}[\lambda^{(I)}]$.

\subsection{The massive and massless parts of $\phi_{ABCD}$}
Using formula (\ref{parametrisation_theta}), the conformal factor
$\Omega$ can be written as
\[
\Omega = \frac{|x|^2}{(U+|x|W)^2}.
\]
Let now
\[
\Omega^\prime \equiv \frac{|x|^2}{U^2}.
\]
In terms of the latter quantity one can calculate the \emph{massless part} of $\phi_{ABCD}$. Namely,
\begin{equation}
\phi'_{ABCD} = \phi^{\flat \prime}_{ABCD} + \phi^{\sharp \prime}_{ABCD}, \label{phi:massless}
\end{equation}
with
\begin{eqnarray*}
&& \phi^{\flat \prime}_{ABCD}= \frac{1}{|x|^4}\bigg( U^2 D_{(AB} D_{CD)} |x|^2 -4 U D_{(AB} |x|^2 D_{CD)}U -2|x|^2 D_{(AB} D_{CD)}U \\
&& \hspace{3cm} + 6 |x|^2 D_{(AB} U D_{CD)} U + |x|^2 U^2 s_{ABCD}  \bigg), \\
&& \phi^{\sharp \prime}_{ABCD}= -\frac{|x|^6}{U^6} \updn{\psi}{EF}{(AB} \psi_{CD)EF} + \frac{2\sqrt{2}}{U^2}\updn{D}{F}{(A}|x|^2 \psi_{BCD)F} -4\frac{\sqrt{2}|x|^2}{U^3} \updn{D}{F}{(A}U \psi_{BCD)F} \\
 && \hspace{3cm} +\frac{\sqrt{2}|x|^2}{U^2}\updn{D}{F}{(A}\psi_{BCD)F}.
\end{eqnarray*}
From here it is direct to obtain expressions for $w^\prime_{ABCD}$ and $w^{*\prime}_{ABCD}$. One finds that
\begin{subequations}
\begin{eqnarray}
&& w^\prime_{ABCD}=\phi^{\flat \prime}_{ABCD} -\frac{|x|^6}{U^6} \updn{\psi}{EF}{(AB} \psi_{CD)EF}, \label{electric:massless}\\
&& w^{*\prime}_{ABCD}= \phi^{\sharp \prime}_{ABCD}+\frac{|x|^6}{U^6} \updn{\psi}{EF}{(AB} \psi_{CD)EF}. \label{magnetic:massless}
\end{eqnarray}
\end{subequations}

The \emph{massive part} is given by
\begin{equation}
\phi^\bullet_{ABCD} = \phi^{\flat \bullet}_{ABCD} + \phi^{\sharp
\bullet}_{ABCD}, \label{phi:massive}
\end{equation}
with
\begin{eqnarray*}
&& \phi^{\flat \bullet}_{ABCD}=\frac{1}{|x|^4}\bigg( -\frac{3}{2|x|} UW D_{(AB}|x|^2 D_{CD)}|x|^2 + U W |x| D_{(AB} D_{CD)} |x|^2  \\
&& \hspace{3cm} +2|x| \big(  W D_{(AB} |x|^2 D_{CD)}U -3 U D_{(AB} |x|^2 D_{CD)} W \big) \\
&& \hspace{3cm} +2|x|^3 \big( -U D_{(AB}D_{CD)}W -W D_{(AB}D_{CD)} U + 6 D_{(AB} U D_{CD)} W + UW s_{ABCD} \big) \\
&& \hspace{3cm} +|x|^4 \big( -2 W D_{(AB} D_{CD)} W + 6 D_{(AB} W D_{CD)} W + W^2 s_{ABCD} \big) \bigg) \\
&& \phi^{\sharp \bullet}_{ABCD}=\frac{|x|^6}{U^6 (U+|x|W)^6} \big( 6U^5 |x| W + 15 U^4 |x|^2 W^2 +20 U^3 |x|^3 W^3 +15 U^2 |x|^2 W^4 \\
 && \hspace{5cm}+6 U|x|^3 W^5 +|x|^3 W^6\big) \updn{\psi}{EF}{(AB} \psi_{CD)EF} \\
&& \hspace{3cm}-\frac{2\sqrt{2}}{U^2(U+|x|W)^2}\big(2U|x|W + |x|^2 W^2  \big) \updn{D}{F}{(A}|x|^2 \psi_{BCD)F} \\
&& \hspace{3cm}+\frac{4\sqrt{2}|x|^2}{U^3(U+|x|W)^3}\big( 3U |x|^2 W^2 + £ U^2 |x| W + |x|^3 W^3   \big) \updn{D}{F}{(A}U \psi_{BCD)F} \\
&& \hspace{3cm}+\frac{4\sqrt{2}|x|^2W}{(U+|x|W)^3} \updn{D}{F}{(A}|x| \psi_{BCD)F} +\frac{4\sqrt{2}|x|^3}{(U+|x|W)^3} \updn{D}{F}{(A}W \psi_{BCD)F}\\
&& \hspace{3cm} -\frac{\sqrt{2}|x|^2}{U^2(U+|x|W)^2} \big(2U|x|W + |x|^4 W^2\big) \updn{D}{F}{(A}\psi_{BCD)F},
\end{eqnarray*}
respectively, the time symmetric and non-time symmetric parts of the
massive part of the Weyl spinor.

It will turn out necessary to refine further the decomposition of the massless
magnetic part of $\phi_{ABCD}$ ---which will be denoted by
$w^{*\prime}_{ABCD}$. Due to the linearity of the momentum
constraint, it is possible to consider individually the
various parameters in the second fundamental form. These will be
denoted by $w^{*\prime}_{ABCD}[A]$, $w^{*\prime}_{ABCD}[J]$, etc.

From expression (\ref{phi:massive}) for the massive part of
the Weyl spinor one can see that unless the ADM mass of the data
vanishes one has $\phi^\bullet_{ABCD}=O(|x|^{-3})$. Thus, in order
to discuss the behaviour of $\phi^\bullet_{ABCD}$ near $i$ one
needs to introduce a suitable rescaling \footnote{It is worth
noticing that on the other
  hand, under suitable assumptions $\phi^\prime_{ABCD}$ is an analytic
  spinor on $\mathcal{B}_a$. These considerations will be retaken in
  section \ref{section:conclusions}.}. To this end let $\kappa =|x|
\kappa'$ with $\kappa'(i)=1$ smooth. Consider the lifts to
$\mathcal{C}_a$ of the spinorial fields
\[
\breve{\phi}_{ABCD} = \kappa^3 \phi_{ABCD}, \quad
\breve{\phi}'_{ABCD}=\kappa^3 \phi'_{ABCD}, \quad
\breve{\phi}^\bullet_{ABCD}= \kappa^3 \phi^\bullet_{ABCD},
\]
and so on. Let $\breve{\phi}_{ABCD}$ denote any of the aforementioned
spinorial fields. From general principles one would expect
$\breve{\phi}_{ABCD}$ to be of expansion type $p+2$. That is, its
essential components $\breve{\phi}_j$, $j=0,\ldots,4$, have normal
expansions near $\mathcal{I}^0$ of the form
\[
\breve{\phi}_j \sim \sum_{p=0}^\infty \sum_{q=|2-j|}^{2p+4} \sum_{k=0}^{2q} \frac{1}{p!} \breve{\phi}_{j,p;2q,k} \TT{2q}{k}{q+j-2} \rho^p,
\]
with $\breve{\phi}_{j,p;2q,k}\in \Complex$. The symbol $\sim$ is to be
understood in the sense described in section
\ref{subsection:normal_exp_i}. It turns out that the normal expansions
have a more restricted form. The following generalisation of parts (i)
and (ii) of theorem 4.1 in \cite{Fri98a} will be proved.

\begin{theorem} \label{thm:symmetries}
The analytic lifts
  $\breve{\phi}_{ABCD}$, $\breve{\phi}'_{ABCD}$ and
  $\breve{\phi}^\bullet_{ABCD}$ to $\mathcal{C}_a$ have expansion type
  $p$, whereas $\breve{w}^{*\prime}_{ABCD}[A,J,Q]$ is of expansion type
  $p-1$. In addition:
\begin{itemize}
\item[(i)] The expansion coefficients
$\breve{\phi}^\bullet_{j,p;2q,k}$ of
$\breve{\phi}^\bullet_j=\breve{\phi}^\bullet_{(ABCD)_j}$ satisfy
\[
\breve{\phi}^\bullet_{0,p;2p,k} = \breve{\phi}^\bullet_{4,p;2p,k},
\quad p=0,1,2,\ldots, \quad k=0,\ldots, 2p.
\]

\item[(ii)] The expansion coefficients $\breve{w}'_{j,p;2p,k}$ satisfy the antisymmetry condition
\[
\breve{w}'_{0,p;2p,k} = -\breve{w}'_{4,p;2p,k},  \quad p=0,1,2,\ldots, \quad k=0,\ldots, 2p.
\]

\item[(iii)] The expansion coefficients $\breve{w}^{*\prime}_{j,p;2p,k}[A,Q,\mbox{\em Re}(\lambda)]$ satisfy the antisymmetry condition
\[
\breve{w}^{*\prime}_{0,p;2p,k}[A,Q,\mbox{\em Re}(\lambda)] = -\breve{w}^{*\prime}_{4,p;2p,k}[A,Q,\mbox{\em Re}(\lambda)],  \quad p=0,1,2,\ldots, \quad k=0,\ldots, 2p.
\]

\item[(iv)] The expansion coefficients $\breve{w}^{*\prime}_{j,p;2p,k}[J,\mbox{\em Im}(\lambda)]$ satisfy the symmetry condition
\[
\breve{w}^{*\prime}_{0,p;2p,k}[J,\mbox{\em Im}(\lambda)] = \breve{w}^{*\prime}_{4,p;2p,k}[J,\mbox{\em Im}(\lambda)],  \quad p=0,1,2,\ldots, \quad k=0,\ldots, 2p.
\]

\end{itemize}
\end{theorem}

The proof of the various parts of the theorem will be given in the
following subsections. It consists, essentially, of an analysis of
the various terms in expressions (\ref{phi:massless}) and
(\ref{phi:massive}) of the massless and massive parts of the Weyl
spinor.

\subsubsection{Proof of the part (i) of theorem \ref{thm:symmetries}  }
Recall the split $\phi^\bullet_{ABCD}=\phi^{\flat \bullet}_{ABCD} +
\phi^{\sharp \bullet}_{ABCD}$ introduced in equation
(\ref{phi:massive}). Consider first the term $\breve{\phi}^{\flat
\bullet}_{ABCD}=\kappa^3\phi^{\flat W}_{ABCD}$. This term coincides,
formally, with the time symmetric $\breve{\phi}^\bullet_{ABCD}$
discussed in reference \cite{Fri98a}. Note however, that in that
refrence, $W$ is the lift of an analytic function on $\mathcal{B}_a$,
while in the case treated here it is the lift of a function belonging
to $E^\infty(\mathcal{B}_a)$. However, by virtue of lemma
\ref{lemma:expansion_type2} one has that $\mbox{type}(W)=p$, and hence
the argument in \cite{Fri98a}, which only requires $W$ having this
expansion type (and not the analyticity) can be reproduced. This will
not be repeated here.  One obtains
\[
\mbox{type}(\breve{\phi}^{\flat \bullet}_{ABCD})=p,
\]
and the  symmetry
\[
\breve{\phi}^{\flat \bullet}_{0,p;2p,k}=\breve{\phi}^{\flat
\bullet}_{4,p;2p,k}, \quad p=0, 1, 2, \ldots, \quad k=0,\ldots,
2p.
\]

Now, consider the term $\phi^{\sharp \bullet}_{ABCD}$. From lemma \ref{lemma:mathringpsi_type} and theorem \ref{thm:v} it follows that the lift, $\psi_{ABCD}$ to $\mathcal{C}_a$ of $\psi_{ij}= \mathring{\psi}_{ij}[A,Q,J,\lambda] +(\mathcal{L}v)_{ij}$  satisfies $\psi_{ABCD}=\rho^{-3} \Phi_{ABCD}$, with
$\Phi_{ABCD}$ a spinorial field of expansion type $p+1$. Hence, it follows
that $\updn{\psi}{EF}{(AB} \psi_{CD)EF} = \rho^{-6}
\updn{\Phi}{EF}{(AB} \Phi_{CD)EF}$, where the spinor
$\updn{\Phi}{EF}{(AB} \Phi_{CD)EF}$ has expansion type $p+2$. Recall
that multiplication by a scalar function of the form
$1+\mathcal{O}(\rho)$ does not change the expansion type or symmetries
of the various terms.

Using the observations in the previous paragraph  one has that
\[
\frac{\rho^9}{U^6 (U+\rho W)^6} \big( 6U^5 \rho W + 15 U^4 \rho^2 W^2 +20 U^3 \rho^3 W^3
+15 U^2 \rho^2 W^4+6 U\rho^3 W^5 +\rho^3 W^6\big) \updn{\psi}{EF}{(AB} \psi_{CD)EF},
\]
---which is essentially the lift of $|x|^3$ times the terms in the
fifth and sixth lines of formula (\ref{phi:massive})--- has at most
expansion type $p-2$. Similarly,
\[
\frac{|x|^3}{U^2(U+|x|W)^2}\big(2U|x|W + |x|^2 W^2  \big) \updn{D}{F}{(A}|x|^2 \psi_{BCD)F}
\]
lifts to $\mathcal{C}_a$ as
\[
\frac{2\rho^2}{U^2(U+\rho W)^2} \left( 2U\rho W + \rho^2 W^2  \right) \updn{x}{F}{(A}\Phi_{BCD)F},
\]
which can be checked to have expansion type of at most $p-1$. Now,
\[
\frac{|x|^3}{U^2(U+|x|W)^3}(3U|x|^2W^2 + U^2 |x|W +|x|^3 W^3) \updn{D}{F}{(A}\psi_{BCD)F}
\]
lifts to
\[
\frac{\rho^2}{U^2(U+\rho W)^3} \left( 3U\rho^2 W^2 +3U^2 \rho W+\rho^3 W^3  \right)
\updn{D}{F}{(A}\Phi_{BCD)F},
\]
hence, using that in the cn-gauge one has that $\mbox{type}(U-1)=p-1$ one concludes that
the whole term has at most expansion type $p-2$. The term
\[
\frac{|x|^5}{(U+|x|W)^3} \updn{D}{F}{(A}|x| \psi_{BCD)F},
\]
lifts to
\[
\frac{\rho^2}{(U+|x|W)^3} \updn{x}{F}{(A}\Phi_{BCD)F},
\]
which can be seen to have at most expansion type $p-1$. The lift of
\[
\frac{|x|^6}{(U+|x|W)^3} \updn{D}{F}{(A}W \psi_{BCD)F}
\]
is given by
\[
\frac{\rho^3}{(U+\rho W)^3} \updn{D}{F}{(A}W \Phi_{BCD)F}.
\]
Now, $D_{AB}W$ has expansion type $p+1$, and hence $\updn{D}{F}{(A}W
\Phi_{BCD)F}$ has expansion type $p+2$. Accordingly, the overall
expansion type of the term is $p-1$. Finally, the lift of
\[
\frac{|x|^5}{U^2(U+|x|W)^2}(2U|x|W + |x|^4 W^2) \updn{D}{F}{(A}\psi_{BCD)F}
\]
is given by
\[
\frac{\rho^2}{U^2(U+\rho W)^2}(2U\rho W +\rho^4 W^2) \updn{D}{F}{(A}\Phi_{BCD)F}
-3\frac{\rho}{U^2(U+\rho W)^2}(2U\rho W +\rho^4 W^2)\updn{x}{F}{(A}\Phi_{BCD)F}.
\]
Hence, noticing that $\updn{D}{F}{(A}\Phi_{BCD)F}$ and
$\updn{x}{F}{(A}\Phi_{BCD)F}$ have both expansion type $p+1$
and $p+1$ one concludes that the overall expansion type of the term is
at most $p-2$. Thus, one has that
\[
\mbox{type}(\breve{\phi}^{\sharp \bullet}_{ABCD})=p-1,
\]
and hence the symmetry
\[
\breve{\phi}^{\sharp \bullet}_{0,p;2p,k}=\breve{\phi}^{\sharp
\bullet}_{4,p;2p,k}, \quad p=0, 1, 2, \ldots, \quad k=0,\ldots,
2p,
\]
holds trivially. This proves part (i) of theorem \ref{thm:symmetries}.

\subsubsection{Proof of the part (ii) of  theorem \ref{thm:symmetries} }

Recall now the expression (\ref{electric:massless}) for
$w^\prime_{ABCD}$. The lift to $\mathcal{C}_a$ of $\breve{\phi}^{\flat
  \prime}_{ABCD}$ is identical to that of the time symmetric
$\breve{\phi}^\prime_{ABCD}$ discussed in \cite{Fri98a}. The
argument used in that reference to analyse the expansion type and
symmetries of the time symmetric $\phi^\prime_{ABCD}$ uses in an
essential manner the analyticity of the conformal metric $h_{ij}$
to introduce a \emph{complex null cone formalism}. To this end a
3-dimensional complex analytic metric manifold $(\mathcal{B},h)$
was introduced. In this setting $h$ defines a complex valued
non-degenerate scalar product. The complex manifold
$(\mathcal{B},h)$ contains $(\mathcal{B}_a,h)$ as a real
Riemannian subspace. The spinor-dyad bundle $SU(\mathcal{B}_a)$
has a complex analytic extension to a bundle $SL(\mathcal{B})$ of
spin frames on $\mathcal{B}$ with structure group
$SL(2,\Complex)$. In the same way that the metric $h$ is
analytically extended to $\mathcal{B}$, one can also analytically
extend the solder and connection forms. Crucial is now to consider
the complex null cone, $\mathcal{N}$, generated by the geodesics
through $i$. As one is restricted to work with analytic functions,
one considers the analytic conformal factor $\Omega^\prime$
obtained setting $W=0$ in formula (\ref{parametrisation_theta}).
Because $\Omega^\prime = \rho^2/U$, one then has that the null
cone $\mathcal{N}$ corresponds to the locus of points on
$\mathcal{B}$ such that $\Omega^\prime$. A suitable frame (i.e.
coordinates and a frame) adapted to the geometry of $\mathcal{N}$
can be introduced on $\mathcal{B}$. This results in a formalism
which allows \emph{to calculate at the point} $i$. A similar
formalism has been used with great effect to discuss the
convergence of multipole expansions of static spacetimes ---see
\cite{Fri07}. The complex null cone formalism is a powerful
machinery to calculate the properties of the expansions of
$\breve{\phi}^\prime_{ABCD}$. Still the required calculations
extend over 10 pages; it will be omitted. The key results are 
one obtains
\[
\mbox{type}(\breve{\phi}^{\flat \prime}_{ABCD}) = p,
\]
and
\[
\breve{\phi}^{\flat \prime}_{0,p;2p,k} = -\breve{\phi}^{\flat \prime}_{4,p;2p,k}, \quad p=0,1,2,\ldots \quad k=0,\ldots,2p.
\]

\textbf{Remark.} This is the only part of the proof of theorem
\ref{thm:symmetries} where the analyticity of $h_{ij}$ is used in
an essential way. Still, it is expected that a similar result
would follow in the smooth setting. This however, would involve
lengthy induction arguments which will not be considered in this
article.

\bigskip
To conclude the analysis of $\breve{w}^\prime_{ABCD}$ consider the
lift to $\mathcal{C}_a$ of
\[
\frac{\kappa^3 |x|^6}{U^6} \updn{\psi}{EF}{(AB} \psi_{CD)EF}.
\]
The lift is given by
\[
\frac{\kappa^\prime 3\rho^3}{U^6} \updn{\Phi}{EF}{(AB} \Phi_{CD)EF}.
\]
As already discussed, $\Phi_{ABCD}$ has expansion
type $p+1$, and hence $\updn{\Phi}{EF}{(AB} \Phi_{CD)EF}$ has
expansion type $p+2$. Accordingly, the expansion type of the whole
term is $p-1$.

Hence, the part of $\breve{w}^\prime_{j,p;2p,k}$ coming from
$\psi_{ABCD}$ satisfies trivially the required antisymmetry condition.
This concludes the proof of part (ii) of theorem \ref{thm:symmetries}.

\subsubsection{Proof of part (iii)  and (iv) of theorem \ref{thm:symmetries}}
Notice that using equation (\ref{magnetic:massless}) one has 
\[
w^{*\prime}_{ABCD} = \frac{2\sqrt{2}}{U^2}\updn{D}{F}{(A}|x|^2 \psi_{BCD)F}
-4\frac{\sqrt{2}|x|^2}{U^3} \updn{D}{F}{(A}U \psi_{BCD)F}
+\frac{\sqrt{2}|x|^2}{U^2}\updn{D}{F}{(A}\psi_{BCD)F}.
\]
As in the proofs of parts (i) and (ii) consider, one by one,  the
lifts of the various terms on $\breve{w}^{*\prime}_{ABCD}=\kappa^3
w^{*\prime}_{ABCD} $. One notes that the lift to $\mathcal{C}_a$
of
\[
\frac{|x|^3}{U^2} \updn{D}{F}{(A}|x|^2 \psi_{BCD)F}
\]
is given by
\begin{equation}
\frac{2\rho}{U^2} \updn{x}{F}{(A}\Phi_{BCD)F}, \label{lift:part2a}
\end{equation}
which has expansion type $p$ as $\Phi_{ABCD}$ has expansion type
$p+1$. Multiplication by the lift of $\kappa^{\prime 3}$ will not
alter the expansion type. The lift to $\mathcal{C}_a$ of
\[
\frac{|x|^5}{U^3} \updn{D}{F}{(A} U \psi_{BCD)F}
\]
is given by
\[
\frac{\rho^2}{U^3} \updn{D}{F}{(A} U \Phi_{BCD)F}.
\]
Now, using lemma \ref{lemma:expansion_type1} one can conclude that in
the cn-gauge $\mbox{type}(D_{AB}U)=p$. Hence
\[
\mbox{type}(\updn{D}{F}{(A} U \Phi_{BCD)F})=p+1.
\]
Accordingly, the overall expansion type of the term is $p-1$. Finally,
the lift of
\[
\frac{|x|^5}{U^2}\updn{D}{F}{(A}\psi_{BCD)F}
\]
is given by
\begin{equation}
-3\frac{\rho}{U^2} \updn{x}{F}{(A}\Phi_{BCD)F} + \frac{\rho^2}{U^2} \updn{D}{F}{(A}\Phi_{BCD)F}. \label{lift:part2b}
\end{equation}
The expansion type of the first term of the latter
expression has already been discussed ---it is, modulo a constant, the same as the term (\ref{lift:part2a}). To analyse the second term,
note that although $D_{AB}\Phi_{CDEF}$ has expansion type $p+2$ if
$\Phi_{CDEF}$ has expansion type $p+1$, one finds that
$\updn{D}{F}{(A}\Phi_{BCD)F}$ has expansion type $p+1$ ---for this,
expand $D_{AB}\Phi_{CDEF}$ in terms of symmetric irreducible terms and
then contract indices. Multiplication by $\rho^2$ renders a term with expansion type $p-1$.

Summarising, it has been found that the only terms in
$\breve{w}^{*\prime}_{ABCD}$ contributing towards an expansion of
type $p$ are proportional to $\rho
U^{-2}\updn{x}{F}{(A}\Phi_{BCD)F}$. The spinor $\Phi_{ABCD}$ is
totally symmetric. Hence one can write it as
\[
\Phi_{ABCD}=\Phi_0 \epsilon^0_{ABCD} + \Phi_1 \epsilon^1_{ABCD} + \Phi_2 \epsilon^2_{ABCD}
+ \Phi_3 \epsilon^3_{ABCD} + \Phi_4 \epsilon^4_{ABCD}.
\]
Furthermore, noting that the following relations hold
\begin{subequations}
\begin{eqnarray}
&& \updn{x}{F}{(A} \epsilon^0_{BCD)F} =\frac{1}{\sqrt{2}} \epsilon^0_{ABCD}, \quad  \updn{x}{F}{(A} \epsilon^1_{BCD)F} =\frac{1}{2\sqrt{2}} \epsilon^1_{ABCD},  \label{xepsilon1}\\
&& \updn{x}{F}{(A} \epsilon^2_{BCD)F} =0, \label{xepsilon2}\\
&& \updn{x}{F}{(A} \epsilon^3_{BCD)F} =-\frac{1}{2\sqrt{2}} \epsilon^3_{ABCD}, \quad
\updn{x}{F}{(A} \epsilon^4_{BCD)F} =-\frac{1}{\sqrt{2}} \epsilon^4_{ABCD}, \label{xepsilon3}
\end{eqnarray}
\end{subequations}
one finds that
\[
\updn{x}{F}{(A}\Phi_{BCD)F} = \frac{1}{\sqrt{2}} \Phi_0 \epsilon^0_{ABCD} +\frac{1}{2\sqrt{2}} \Phi_1 \epsilon^1_{ABCD} -\frac{1}{2\sqrt{2}} \Phi_3 \epsilon^3_{ABCD} -\frac{1}{\sqrt{2}} \Phi_4 \epsilon^4_{ABCD}.
\]
Now, $\Phi_{ABCD}$ has expansion type $p+1$, that is
\[
\Phi_j \sim \sum_{p=0}^\infty \sum_{q=\max\{|1-j|,2-p\}}^{p+1} \sum_{k=0}^{2q} \frac{1}{p!} \Phi_{j,p;2q,k} \TT{2q}{k}{q-2+j} \rho^p.
\]

A quick comparison with formulae
(\ref{soln_lambda_1})-(\ref{soln_lambda_2}) shows that the expansion
coefficients of $\Phi_0$ and $\Phi_4$ are essentially those of $\mu_2$
and $\overline{\mu}_2$. Now, from lemma \ref{lemma:mathringpsi_type}
and theorem \ref{thm:v} it follows that all the contributions of type
$p$ in $\Phi_{ABCD}$ come from
$\mathring{\psi}_{ABCD}[J,\lambda^{(2)}/r]$. Recall that
$\lambda^{(2)}=\tilde{\lambda}^{(2)}$ where $\tilde{\lambda}^{(2)}$ is
a complex function with real and imaginary parts in
$C^\infty(\mathcal{B}_a)$. Accordingly, write
\begin{equation}
\label{lambda2_expansion} \lambda^{(2)}/\rho \sim
\sum_{p=1}^\infty \sum_{q=2}^{p+1} \sum_{k=0}^{2q} \left(
f_{p;2q,k} + g_{p;2q,k} \right) \TT{2q}{k}{q-2}\rho^p,
\end{equation}
where $f_{p;2q,k}, \; g_{p;2q,k} \in \Complex$. The coefficients $f_{p;2q,k}$ satisfy the conditions
\[
f_{p;2q,k}=(-1)^{k+q} \overline{f}_{p;2q,2q-k}, \quad p=1,\ldots, \quad q=2,\ldots, p+1, \quad k=0,\ldots p,
\]
so that they are associated with the real part of $\tilde{\lambda}^{(2)}$
while the coefficients $g_{p;2q,k}$ satisfy the conditions
\[
g_{p;2q,k}=(-1)^{k+q+1} \overline{g}_{p;2q,2q-k},\quad p=1,\ldots, \quad q=2,\ldots, p+1, \quad k=0,\ldots p
\]
so that they are associated with the imaginary part of $\tilde{\lambda}^{(2)}$. A direct calculation reveals that
\begin{eqnarray*}
&&\mu_2[\mbox{Re}(\lambda^{(2)})/\rho] \sim \sum_{p=1}^\infty \sum_{q=2}^{p+1} \sum_{k=0}^{2q} \frac{1}{p!} \left( 2p(p-1)\beta_q -(q-1)(q-2)-2 \right) f_{p;2q,k} \TT{2q}{k}{q-2} \rho^p, \\
&& \mu_2[\mbox{Im}(\lambda^{(2)})/\rho] \sim -\sum_{p=1}^\infty
\sum_{q=2}^{p+1} \sum_{k=0}^{2q} \frac{1}{(p-1)!} g_{p;2q,k}
\TT{2q}{k}{q-2} \rho^p,
\end{eqnarray*}
with
\[
\beta_q = \sqrt{(q-1)q(q+1)(q+2)}.
\]
From here another short computation using the rules (\ref{xepsilon1})-(\ref{xepsilon3}) shows that
\begin{eqnarray*}
&& \Phi_{0,p;2p+2,k}[\mbox{Re}(\lambda^{(2)})/\rho)] =-\Phi_{4,p;2p+2,k}[\mbox{Re}(\lambda^{(2)})/\rho], \\
&& \Phi_{0,p;2p+2,k}[\mbox{Im}(\lambda^{(2)})/\rho] =\Phi_{4,p;2p+2,k}[\mbox{Im}(\lambda^{(2)})/\rho].
\end{eqnarray*}
The latter relations imply the antisymmetry and symmetry conditions,
\begin{eqnarray*}
&& \breve{w}_{0,p;2p,k}[\mbox{Re}(\lambda^{(2)})/\rho]=-\breve{w}_{4,p;2p,k}[\mbox{Re}(\lambda^{(2)})/\rho], \\
&& \breve{w}_{0,p;2p,k}[\mbox{Im}(\lambda^{(2)}/\rho)]=-\breve{w}_{4,p;2p,k}[\mbox{Im}(\lambda^{(2)})/\rho].
\end{eqnarray*}
Finally, a similar argument with $\rho
U^{-2}\updn{x}{F}{(A}\mathring{\psi}_{BCD)F}$ shows that
$\Phi_0[J]=\Phi_4[J]$ so that the symmetry conditions are trivially
satisfied. This proves points (iii) and (iv) of theorem \ref{thm:v}.  \hfill $\Box$

\subsection{Some further results}
Besides the information provided by theorem \ref{thm:symmetries} about
the properties of the expansions of $\breve{\phi}_{ABCD}$ and derived
spinors, a couple of further results will be required. The following
spinor will be used in the sequel
\begin{equation}
\label{c_spinor}
c_{ABCD}\equiv \mbox{i}\sqrt{2} \updn{D}{F}{(A} \chi_{BCD)F}.
\end{equation}

The following lemma is a direct consequence of the proof of theorem \ref{thm:symmetries}.
\begin{lemma}
\label{lemma:c}
For the class of data under consideration
\[
\mbox{type}(c_{ABCD})=p+1.
\]
\end{lemma}

In accordance with the previous result one has that the essential
components $c_j$ of $c_{ABCD}$ are of the form
\[
c_{j} \sim \sum_{p=0} ^\infty \sum_{q=\max\{|1-j|,2-p\}}^{2p+2} \sum_{k=0}^{2q} \frac{1}{p!} c_{j,p;2q,k} \TT{2q}{k}{q-2+j}\rho^p.
\]
Now, recall the expansion (\ref{lambda2_expansion}) of
$\lambda^{(2)}/\rho$. The coefficients $f_{p;2q,k}$ associated to the
real part of $\tilde{\lambda}^{(2)}$ satisfy the following lemma.
\begin{lemma}
\label{lemma:vanishing}
For the class of initial data under consideration:
\begin{itemize}
\item[(i)]
\[
\breve{w}^{*\prime}_{j,p;2p,k}[\mbox{Re}(\lambda^{(2)}/\rho)]=0 \Leftrightarrow f_{p;2p+2,k}=0.
\]
\item[(ii)]
\[
c_{j,p;2p+2,k}[\mbox{Re}(\lambda^{(2)}/\rho)]=0 \Leftrightarrow f_{p;2p+2,k}=0.
\]
\end{itemize}
\end{lemma}

\textbf{Proof.} The proof of (i) follows directly from the analysis
carried in the proof of parts (iii) and (iv) of theorem
\ref{thm:symmetries}. The proof of part (ii) follows directly from (i) by
noticing that $w_{ABCD}^*=-\Omega^{-1}c_{ABCD}$. \hfill $\Box$

\bigskip
Finally, to state the main result of this article one needs
the following theorem, which for the class of data under consideration,
generalises part (iii) of theorem 4.1 in \cite{Fri98a}.

\begin{theorem} \label{thm:vanishing}
For the class of initial data under consideration one has that:
\begin{itemize}
\item[(i)]
\[
\breve{w}^\prime_{0,p;2p,k}=0, \quad p=2,3,\ldots, \quad k=0,\dots,2p,
\]
if and only if
\[
D_{(A_pB_p} \cdots D_{A_1B_1} b_{ABCD)}(i)=0, \quad p=0,1,2,\ldots.
\]

\item[(ii)]
\[
\breve{w}^{*\prime}_{0,p;2p,k}[\mbox{Re}(\lambda^{(2)}/\rho)]=0,
\quad p=2,3,\ldots, \quad k=0,\dots,2p
\]
if and only if
\[
D_{(A_p B_p} \cdots D_{A_1|E|} \dnup{c}{ABC)}{E}[\mbox{Re}(\lambda^{(2)}/r)](i)=0, \quad p=1,2,3, \ldots.
\]

\end{itemize}
\end{theorem}

\textbf{Remark 1.} Note that the class of data under consideration
automatically satisfies
\[
c_{ABCD}(i)=0.
\]

\textbf{Proof of theorem \ref{thm:vanishing}.} From the proof of
part (ii) of theorem \ref{thm:symmetries} one has that
\[
\mbox{type}(\breve{\phi}_{ABCD}^{\flat \prime})=p, \quad \mbox{type}(\breve{\phi}_{ABCD}^{\sharp \prime})=p-1.
\]
Hence, itn order to prove part (i) of theorem \ref{thm:vanishing}
one has to consider only the expansion of the term
$\breve{\phi}_{ABCD}^{\flat \prime}$ which is formally identical
to the spinor $\breve{\phi}^\prime_{ABCD}$ considered in
\cite{Fri98a} under the assumption of an analytic conformal
metric. Thus, (i) follows from the analysis based on the complex
null cone formalism given in the aforementioned reference and
which will be omitted here.

In order to prove part (ii) notice that consistent with lemma \ref{lemma:c}
\[
c_j \sim^\infty \sum_{p=0} \frac{1}{p!} c_j^{(p)} \rho^p,
\]
with
\[
c_j^{(p)} =\sum^{2p+2}_{k=0} c_{j,p;2p+2,k} \TT{2p+2}{k}{p-1+j} + \sum^{p}_{q=\max\{|1-j|,2-p\}} \sum_{k=0}^{2q} c_{j,p;2q,k} \TT{2q}{k}{q+j}, \quad p=1,2,\ldots,
\]
where from general considerations about normal expansions at
$\mathcal{I}$ ---see section \ref{subsection:normal_exp_I0} and in
particular the expression (\ref{expansion:Ca3})--- one has that
\[
c_{j,p;2p+2,k} = K_{p,j,k} D_{(A_pB_p} \cdots D_{A_1|E|} \dnup{c}{ABC)_j}{|E|}(i), \quad j=0,\ldots, 4, \quad k=0,\ldots, 2p,
\]
and $K_{p,j,k}$ are some constants. Part (ii) now follows from
lemma \ref{lemma:vanishing}. \hfill $\Box$

\bigskip
\textbf{Remark 2.} It is perhaps worth noticing that $c_{ABCD}$ being 
of expansion type $p+1$, then $c_{j,p;2p+4,k}=0$ and accordingly,
\[
D_{(A_pB_p} \cdots D_{A_1B_1} c_{ABCD)}(i)=0.
\]

\section{The spacetime Friedrich gauge}
\label{section:Fgauge}
The final aim of the framework developed in \cite{Fri98a} is to gain
control over the evolution of the gravitational field in a neighbourhood
of spacelike infinity which extends to null infinity. This problem is
generically known as the \emph{initial value problem near spatial
  infinity}. In slight contrast to the analysis of the non-linear stability
of the Minkowski spacetime ---see e.g. \cite{ChrKla93,KlaNic03},
the aim of the initial value problem near spatial infinity as
discussed in \cite{Fri98a} is not only that of showing that the
outgoing null geodesics starting close to spatial infinity are
complete, but also to analyse under which conditions on the initial data,
the resulting spacetime will admit a smooth conformal extension
through null infinity ---and hence giving rise to an
asymptotically simple spacetime.

In the standard representation of spatial infinity as a point, the
direct formulation of an initial value problem for the conformal field
equations with data prescribed in a neighbourhood, say $\mathcal{B}_a$
of infinity, renders a problem which although local is also singular
---this can be seen in a very poignant way by considering the
expressions for the massless and massive parts of the Weyl spinor
discussed in section \ref{section:weyltensor}. 

The formulation of
the initial value problem near spatial infinity presented in
\cite{Fri98a} employs gauge conditions based on timelike
conformal geodesics. The conformal geodesics are autoparallel with
respect to a Weyl connection
---i.e. a torsion-free connection which is not necessarily the
Levi-Civita connection of a metric. An analysis of Weyl
connections in the context of the conformal field equations has
been given in \cite{Fri95}. In terms of this gauge based on
conformal geodesics
---which shall be called the \emph{Friedrich gauge} or \emph{F-gauge}
for short--- the conformal factor of the spacetime can be
determined explicitly in terms of the initial data for the
Einstein vacuum equations. Hence, provided that the congruence of
conformal geodesics and the fields describing the gravitational
field extend in a regular manner to null infinity, one has
complete control on the location of null infinity. In addition,
the F-gauge renders a particularly simple representation of the
propagation equations. Using this framework, the singular initial
value problem at spatial infinity can be reformulated into another
problem where null infinity is represented by an explicitly known
hypersurface and where the data are regular at spacelike infinity.
The construction of the bundle manifold $\mathcal{C}_a$ and the
blowing up of the point $i\in \mathcal{B}_a$ to the set
$\mathcal{I}^0\subset \mathcal{C}_a$ briefly described in section
\ref{section:manifold_Ca} is the first step in the construction of
the regular setting. The next step in the construction is to
introduce a rescaling in the frame bundle so that fields that are
singular at $\mathcal{I}^0$ become regular. The required rescaling
has already been hinted in theorem \ref{thm:symmetries} where the
regular spinorial field $\breve{\phi}_{ABCD}=\kappa^3 \phi_{ABCD}$
has been considered instead of the singular $\phi_{ABCD}$.

\subsection{The  manifold $\mathcal{M}_{a,\kappa}$}
Following the discussion of \cite{Fri98a} assume that in the
development of data prescribed on $\mathcal{B}_a$ the timelike
spinor $\tau^{AA'}$ introduced in section
\ref{section:manifold_Ca} is tangent to a congruence of timelike
conformal geodesics which are orthogonal to $\mathcal{B}_a$. The
canonical factor rendered by the consideration of this congruence
of conformal geodesics is given in terms of an affine parameter
$\tau$ of the conformal geodesics by
\begin{equation}
\Theta=\kappa^{-1}\Omega\left(1-\frac{\kappa^2\tau^2}{\omega^2}\right), \quad \mbox{ with }  \omega=\frac{2\Omega}{\sqrt{|D_\alpha\Omega D^\alpha \Omega|}},\label{Theta}
\end{equation}
where $\Omega=\vartheta^{-2}$ and $\vartheta$ solves the Licnerowicz
equation (\ref{Licnerowicz:eqn}). The function $\kappa>0$ ---which will be
taken to be of the form $\kappa=\kappa^\prime \rho$, with
$\kappa^\prime$ smooth, $\kappa'(i)=1$--- expresses the remaining
conformal freedom in the construction. Consistent with the scalings $\delta_A \mapsto \kappa^{1/2} \delta_A$ induced by the function $\kappa$ one considers the set $\mathcal{C}_{a,\kappa}=\kappa^{1/2}\mathcal{C}_a$ of scaled spinor
dyads. Furthermore, define the bundle manifold
\[
\mathcal{M}_{a,\kappa}=\left\{ (\tau,q) \big |  q\in
\mathcal{C}_{a,\kappa}, -\frac{\omega(q)}{\kappa(q)} \leq \tau \leq \frac{\omega(q)}{\kappa(q)} \right\},
\]
which, assuming that the congruence of null
geodesics and the relevant fields extend adequately, can be identified with the
development of $\mathcal{B}_a$ up to null infinity ---that is, the region of
spacetime near null and spatial infinity. In addition to $\mathcal{C}_{a,\kappa}$ one defines the sets:
\begin{subequations}
\begin{eqnarray*}
&& \mathcal{I}=\big \{(\tau,q)\in \mathcal{M}_{a,\kappa} \;\big|\; \rho(q)=0, \;|\tau|<1\big\}, \\
&& \mathcal{I}^\pm= \big \{ (\tau,q)\in \mathcal{M}_{a,\kappa} \;\big |\; \rho(q)=0, \;\tau=\pm1\big \}, \\
&& \mathscr{I}^\pm=\left\{ (\tau,q)\in \mathcal{M}_{a,\omega} \;\big |  \; q\in \mathcal{C}_{a,\omega}, \;\; \rho>0, \;\; \tau=\pm \frac{\omega(q)}{\kappa(q)} \right\},
\end{eqnarray*}
\end{subequations}
which will be referred to as, respectively, the \emph{cylinder at
  spatial infinity}, the \emph{critical sets} (where null infinity
touches spatial infinity) and \emph{future and past null infinity}. In
order to coordinatise the hypersurfaces of constant parameter $\tau$, one extends
the coordinates $(\rho,\updn{t}{A}{B})$ off $\mathcal{C}_{a,\kappa}$
by requiring them to be constant along the conformal geodesics ---i.e.
one has a system of conformal Gaussian coordinates.

\subsection{The conformal propagation equations}
On the manifold $\mathcal{M}_{a,\kappa}$ it is possible to introduce a
calculus based on the derivatives $\partial_\tau$ and $\partial_\rho$
and on the operators $X_+$, $X_-$ and $X$. The operators
$\partial_\rho$, $X_+$, $X_-$ and $X$ originally defined on
$\mathcal{C}_{a}$ can be suitably extended to the rest of the manifold
in a standard way. A frame $c_{AA'}$ and the associated
spin connection coefficients $\Gamma_{AA'BC}$ of the Weyl
connection $\nabla$ will be used. The gravitational field is, in addition,
described by the spinorial counterparts of the Ricci tensor of the
Weyl connection, $\Theta_{AA'BB'}$, and of the rescaled Weyl
tensor, $\phi_{ABCD}$. In order to describe the conformal
propagation equations consider the vector
\[
\upsilon=(c_{AB},\Gamma_{ABCD},\Theta_{ABCD}), \quad
\phi=(\phi_{ABCD}),
\]
where $c_{AB}$, $\Gamma_{ABCD}$, $\Theta_{ABCD}$ are the
space-spinor versions of the spacetime spinors $c_{AA'}$,
$\Gamma_{AA'BC}$, $\Theta_{AA'BB'}$. The relation between
space-spinors and spacetime spinors is implemented by means of
suitable contractions with the spinor $\tau^{AA'}$. This will not
be elaborated further ---the interested reader is referred to
\cite{Som80,Fra98a,Fri98a}. The explicit form of the propagation
equations will not be required. Only general properties will be
used. Schematically one has
\begin{subequations}
\begin{eqnarray}
&&\partial_\tau \upsilon=K \upsilon +Q(\upsilon,\upsilon)+L\phi, \label{v_eqns}\\
&& \sqrt{2} E \partial_\tau \phi + A^{AB}c^\mu_{AB}\partial_\mu\phi=B(\Gamma_{ABCD})\phi, \label{phi_eqns}
\end{eqnarray}
\end{subequations}
where $E$ denotes the $(5\times 5)$ unit matrix, $A^{AB}c^\mu_{AB}$
are $(5\times 5)$ matrices depending on the coordinates, and
$B(\Gamma_{ABCD})$ is a linear $(5\times 5)$ matrix valued function
with constant entries of the connection coefficients $\Gamma_{ABCD}$.
In addition to the above propagation equations it is essential to
consider the following constraint equations derived from the Bianchi
identities:
\[
F^{AB}c^\mu_{AB} \partial_\mu\phi = H(\Gamma_{ABCD}),
\]
where now $F^{AB}c^\mu_{AB}$ denote $(3\times 5)$ matrices, and
$H(\Gamma_{ABCD})$ is a $(3\times5)$ matrix valued function of the
connection with constant entries.

Equations (\ref{v_eqns}) and (\ref{phi_eqns}) can be casted as a
symmetric hyperbolic system on a neighbourhood $\mathcal{N}\subset
\mathcal{M}_{a,\kappa}$ of the initial hypersurface. Hence, given data
that extends smoothly to $\mathcal{I}^0$, one obtains a unique smooth
solution on $\mathcal{N}$. The metric can be seen to degenerate as
$\rho\rightarrow 0$. A consequence of this is the fact that
$A^{AB}c_{AB}^1=0$ on $\mathcal{I}\cap \mathcal{N}$ ---i.e. the
coefficient associated with the $\rho$ derivatives in the Bianchi
propagation equations (\ref{phi_eqns}). The restriction of the
propagation equations to $\mathcal{I}$ implies an interior system on
$\mathcal{I}$ which determines $\upsilon$ and $\phi$ on $\mathcal{I}$
uniquely from the restriction of the initial data to $\mathcal{I}^0$.
Differentiating the propagation equations repeatedly with respect to
$\rho$ and restricting the result to $\mathcal{I}$ one obtains a
hierarchy of interior equations for $\upsilon^{(p)}=\partial^p_\rho
\upsilon|_\mathcal{I}$ from where it is possible to determine a formal
expansions
\[
\label{formal_expansion}
\upsilon=\sum_{p\geq 0} \frac{1}{p!} \upsilon^{(p)} \rho^p,  \quad \phi=\sum_{p\geq 0} \frac{1}{p!} \phi^{(p)} \rho^p,
\]
of the solution at $\mathcal{I}$. The calculation of the lowest order
terms $\upsilon^{(0)}$ shows that the matrix $A^{AB}c^1_{AB}$ is
positive definite on $\mathcal{I}$ and extends smoothly to
$\mathcal{I}^\pm$, but it looses rank there. The entries of the vectors $\upsilon^{(p)}$ and $\phi^{(p)}$ on $\mathcal{I}$ can
be seen to have definite spin-weights and hence admit very particular
expansions in terms of the functions $\TT{j}{k}{l}$ ---as described in
section \ref{subsection:normal_exp_I0}. The use of these \emph{time
  dependent} normal expansions reformulates the problem of calculating
the vector $\upsilon^{(p)}$ into a problem of linear ordinary
differential equations. Once the entries of the vectors $\upsilon^{(p)}$ and $\phi^{(p)}$ have been expanded using the
functions $\TT{j}{k}{l}$, the interior equations are reduced to
systems of ordinary differential equations for the $\tau$-dependent
expansion coefficients. The task of solving these equations reduces, in
the end, to solving a hierarchy of ordinary differential equations of the form
\begin{equation}
\label{reduced}
y_\alpha'=C_\alpha y_\alpha +b_\alpha,
\end{equation}
where $C_\alpha$ is a $2\times 2$ matrix and $y_\alpha$ and $b_\alpha$
are $2\times 1$ column vectors. The components of $y_\alpha$ consist
of $p$th-order $\rho$ derivatives of certain components of the Weyl
spinor $\phi_{ABCD}$. In equation (\ref{reduced}) note the presence of a
multi-index $\alpha=(p,q,k)$ indicating the order, $\rho^p$, in the
expansions and to which harmonic $\TT{2q}{k}{l}$ the term is
associated. In what follows, in order to ease the discussion, the
multi-index $\alpha$ will be, sometimes, suppressed. For a given
multi-index $\alpha=(p,q,k)$, the components of the vector $b$ are
calculated from the lower order solutions $\upsilon^{(q)}$ and
$\phi^{(q)}$, $0\leq q\leq p-1$.  The solutions to these equations can
be written in the form
\begin{equation} \label{integral:formula}
y(\tau)=X(\tau)X^{-1}(0) y_0 +X(\tau) \int_0^\tau X^{-1}(s) b(s) \mbox{d}s,
\end{equation}
where $y_0=y(0)$ and $X(\tau)$ ---again suppressing the relevant
multi-index--- denotes the fundamental matrix of the system of
ordinary differential equations.  The matrices $X$ have been
explicitly calculated in \cite{Fri98a}. For increasing $p$ the
explicit expressions for $b$ become more complicated. One can,
nevertheless, implement the aforediscussed procedure in a computer
algebra system ---see e.g.  \cite{Val04a,Val05a}. Now, due to the
degeneracy of $A^{AB}c^1_{AB}$ on $\mathcal{I}$, the ordinary
differential equations for the vector $y$ are singular at
$\tau=\pm 1$. As a consequence, for a given $p$, there are certain
choices of the multi-index $\alpha_*=(p,p,k)$, $k=0,\ldots,2p$, of
for which the fundamental system develops logarithmic singularities
at $\tau=\pm 1$. For all other allowed values of the multi-index,
the fundamental matrices can be written explicitly in terms of
Jacobi polynomials. It can be seen that the vector $b$ for the
corresponding multi-index $\alpha_*$ vanishes. Hence the
logarithmic divergences cannot be cancelled out by the integral in
(\ref{integral:formula}). Thus, in general the functions
$\upsilon^{(p)}$ and $\phi^{(p)}$ develop logarithmic
singularities for all $p\geq 2$ on $\mathcal{I}$.

\subsection{Regularity conditions}
\label{subsection:regularityconds}

Given the aforementioned state of affairs, can one find conditions on
the initial data so that the functions $y_\alpha$ extend smoothly to
$\mathcal{I}^\pm$? An inspection of the term $X(\tau)X^{-1}(0)y_0$ for
the singular multi-index values $\alpha_*$ shows that there are
conditions on the initial data for which the observed singularities at
$\mathcal{I}^\pm$ do not arise. This analysis is completely general as
long as the initial data is expandable in powers of $\rho$ near
$\mathcal{I}^0$.

With the aim of formulating the main result of this article, it
will be necessary to provide a more precise description of the
conditions oo the data mentioned in the previous paragraph.
Following the conventions of section \ref{section:weyltensor}
consider the spinor $\breve{\phi}_{ABCD}=\kappa^3 \phi_{ABCD}$,
and let
\[
\breve{\phi}_j^{(p)}= \partial^p_\rho \breve{\phi}_j |_{\rho=0}.
\]
Observing theorem \ref{thm:symmetries} one expands
\[
\breve{\phi}_j^{(p)}
=\sum_{q=|2-j|}^p\sum_{k=0}^{2q}a_{j,p;2q,k}\TT{2q}{k}{q-2+j},
\]
with $\tau$-dependent complex coefficients $a_{j,p;2q,k}$. The
coefficients $a_{j,p;2q,k}$ can be determined along $\mathcal{I}$ from
their value at $\mathcal{I}^0$ by solving a set of transport equations
implied by the conformal field equations as described in section
\ref{section:Fgauge}. The analysis of \cite{Fri98a} implies the
following result.

\begin{theorem}
\label{thm:helmut}
For the class of data under consideration one has that
\begin{eqnarray*}
&& a_{0,p;2p,k}(\tau) = (1-\tau)^{p+2} (1+\tau)^{p-2} \left(  C_{0,k} + C_{1,k} \int_0^\tau \frac{\mbox{d}s}{(1+s)^{p-1}(1-s)^{p+3}} \right), \\
&& a_{4,p;2p,k}(\tau) = (1+\tau)^{p+2} (1-\tau)^{p-2} \left(  C_{0,k} + C_{1,k} \int_0^\tau \frac{\mbox{d}s}{(1-s)^{p-1}(1+s)^{p+3}} \right),
\end{eqnarray*}
with $C_{0,k}$ and $C_{1,k}$ constants. In particular, $a_{0,p;2p,k}$
and $a_{4,p;2p,k}$ extend analytically through $\tau=\pm 1$ if and
only if
\[
a_{0,p;2p,k}(0)=a_{4,p;2p,k}(0),
\]
with $k=0,\ldots,2p$.
\end{theorem}

\section{The main result}
\label{section:main}

In \cite{Fri98a} it was shown that if one restricts
the attention to the class of time symmetric data with a conformal
metric that is analytic in a neighbourhood of infinity these
conditions can be reformulated in terms of the vanishing of the Coton
tensor and its symmetrised higher order derivatives at infinity. This
condition happens to be a purely asymptotic condition on the freely
specifiable data.

The main result in this article is a generalisation of the result
described in the previous paragraphs to the class of data with a
non-vanishing second fundamental form. This main result brings
together the discussion in section \ref{section:weyltensor} (on the
structure of initial data for the Weyl tensor near infinity), in
particular of theorems \ref{thm:symmetries} and \ref{thm:vanishing},
with the discussion of section \ref{section:Fgauge} (on the regular
finite initial value problem at spatial infinity) leading to theorem
\ref{thm:helmut}.

From theorem \ref{thm:symmetries} it follows readily that
\begin{eqnarray*}
&& a_{0,p;2p,k}(0)= \breve{\phi}^W_{0,p;2p,k} + \breve{w}^\prime_{0,p;2p,k} + \breve{w}^{*\prime}_{0,p;2p,k}[\mbox{Re}(\lambda^{(2)}/\rho)] + \breve{w}^{*\prime}_{0,p;2p,k}[\mbox{Im}(\lambda^{(2)}/\rho)],\\
&&  a_{4,p;2p,k}(0)= \breve{\phi}^W_{0,p;2p,k}- \breve{w}^\prime_{0,p;2p,k} - \breve{w}^{*\prime}_{0,p;2p,k}[\mbox{Re}(\lambda^{(2)}/\rho)] + \breve{w}^{*\prime}_{0,p;2p,k}[\mbox{Im}(\lambda^{(2)}/\rho)].
\end{eqnarray*}
Hence the condition
\[
a_{0,p;2p,k}(0)=a_{4,p;2p,k}(0), \quad k=0,\ldots,2p,
\]
of theorem \ref{thm:helmut}, using that $w_{ABCD}$ is associated to a
real spatial tensor whereas $w^*_{ABCD}$ is associated to an imaginary
one, implies
\[
\breve{w}^\prime_{0,p;2p,k}=0, \quad \breve{w}^{*\prime}_{0,p;2p,k}[\mbox{Re}(\lambda^{(2)}/\rho)]=0,
\]
on $\mathcal{C}_a$ independently of the choice of $\kappa$. These last two
conditions can be readily reformulated in terms of the tensors
$b_{ABCD}$ and $c_{ABCD}$ using theorem \ref{thm:vanishing}. In this
way one obtains the main result of the article.

\begin{theorem}\label{thm:main}
  For the class of data under consideration, the solution to the
  regular finite initial value problem at spatial infinity is smooth
  through $\mathcal{I}^\pm$ only if the conditions
\begin{subequations}
\begin{eqnarray}
&& D_{(A_pB_p} \cdots D_{A_1B_1} b_{ABCD)}(i)=0, \quad p=0,1,2,\ldots \label{condition_b}\\
&& D_{(A_qB_q} \cdots D_{A_1|E|} \dnup{c}{ABC)}{|E|}[\lambda^{(R)}](i)=0, \quad q=0,1,2,\ldots \label{condition_c}
\end{eqnarray}
\end{subequations}
are satisfied by the free initial data. If the above conditions are
violated at some order $p$ or $q$, then the solution will develop
logarithmic singularities at $\mathcal{I}^\pm$.
\end{theorem}

\textbf{Remark 1.} This result is a non-time symmetric generalisation
of theorem 8.2 in \cite{Fri98a}.

\textbf{Remark 2.} The result in theorem \ref{thm:main} which is
expressed in the language of space-spinors can be readily reformulated
in terms of spatial tensors. Using standard transcription rules one
finds the tensorial version of the main theorem given in the
introductory section.

\section{Extensions}
\label{section:extensions}
In view of the main result \ref{thm:main} it is natural to ask whether
the assertions are true for more general classes of initial data. In
particular, it is important to see how stationary initial data
fits into this picture. In \cite{Fri88} it has been shown that the
condition (\ref{condition_b}) for the spinor $b_{ABCD}$ is satisfied
by static data ---the condition (\ref{condition_c}) is in this case
satisfied trivially.

In \cite{Dai01b} it has been shown that there is a gauge for which in
a neighbourhood $\mathcal{B}_a$ of infinity, the conformal metric of
stationary data is of the form
\begin{equation}
h_{ij} = h^{(1)}_{ij} + r^3 h^{(2)}_{ij}, \label{stationary:conformal_metric}
\end{equation}
with $h^{(1)}_{ij}$ and $h^{(2)}_{ij}$ analytic. This type of
conformal metrics is not smooth analytic at $i$ ---i.e. at $r=0$.
In fact, it follows directly that it is $C^{2,\alpha}$. For the
tensor $\psi_{ij}$ it was shown that in that gauge stationary data
are such that
\begin{equation}
\psi_{ij}= r^{-5} \psi^{(1)}_{ij} + r^{-4} \psi^{(2)}_{ij}, \label{stationary:conformal_psi}
\end{equation}
with $\psi^{(1)}_{ij}=\mathcal{O}(r^2)$ and
$\psi^{(2)}_{ij}=\mathcal{O}(r^2)$ analytic. A choice of $h_{ij}$ and
$\psi_{ij}$ consistent with (\ref{stationary:conformal_metric}) and
(\ref{stationary:conformal_psi}) provide a natural ground for
generalising the results in this article. Unfortunately, this analysis
is complicated by the fact that up to date there is no generalisation
of the analysis in \cite{DaiFri01} for conformal metrics which are
non-smooth.

\section{Conclusions}
\label{section:conclusions}

In order to bring further context to the content of the main
theorem \ref{thm:main}, some connections with the construction of
data for \emph{purely radiative spacetimes} are raised: in
reference \cite{Fri88} it has been shown how time symmetric Cauchy
initial data with vanishing mass ---so that $\Omega=\Omega^\prime$
in the notation of section \ref{section:weyltensor}--- can be used to
obtain analytic data at past null infinity. The crucial point in
that analysis was to obtain conditions on the Cauchy data so that
the spinor $\phi_{ABCD}=w_{ABCD}$ is analytic at $i$. The
Cauchy-Kowalevskaya theorem can then be used to obtain analytic
data on past null infinity close to $i^-$ (purely radiative data).
The conditions are given by
\[
D_{(A_qB_q} \cdots D_{A_1B_1} b_{ABCD)}(i)=0, \quad q=0,1,2,\ldots,
\]
that is, the same as condition (\ref{condition_b}) in theorem
\ref{thm:main}. The way the analysis in reference \cite{Fri88} can be
generalised to the case of non-time symmetric Cauchy data has been
briefly discussed in the concluding remarks of \cite{Fri98a}. That
analysis will not be repeated here, but under the extra assumption
that the Cauchy data has no linear momentum ---so that
$\chi_{ij}=\mathcal{O}(r)$---, necessary and sufficient
conditions for $\phi_{ABCD}= w_{ABCD}+\mbox{i}w^*_{ABCD}$ to be
analytic at $i$ are in addition to (\ref{condition_b}) that
\begin{equation} D_{(A_pB_p} \cdots D_{A_1B_1)} c_{ABCD}(i)=0, \quad
p=0,1,2, \ldots . \label{guess:helmut}
\end{equation} Note that condition (\ref{guess:helmut}) is much
stronger than condition (\ref{condition_c}). Indeed, from the
discussion in subsection \ref{subsection:normal_exp_i} one has that
\[
D_{(A_pB_p} \cdots D_{A_1B_1)} c_{ABCD}=0 \Rightarrow D_{(A_pB_p} \cdots D_{A_1|E|} \dnup{c}{ABC)}{E}=0,
\]
but not conversely. Furthermore, condition (\ref{guess:helmut})
involves the whole of $\mathring{\psi}_{ij}$ and not only the part
depending on $\mbox{Re}(\lambda^{(2)}/\rho)$.

\section*{Acknowledgements}
This research is funded by an EPSRC Advanced Research Fellowship. I
thank H. Friedrich and S. Dain for valuable discussions and insights.
I thank CM Losert-VK for a careful reading of the manuscript.

\appendix

\section{Spinorial identities}
\label{appendix:spinors}
The following identities will be used throughout the main text:
\begin{eqnarray*}
&& x_{(AB}x_{CD)} = 2 \epsilon^2_{ABCD}, \quad y_{(AB}y_{CD)} = \frac{1}{2} \epsilon^4_{ABCD}, \\
&& x_{(AB} y_{CD)} = -\epsilon^3_{ABCD}, \quad y_{(AB} z_{CD)} =-\frac{1}{2} \epsilon^2_{ABCD}, \\
&& x_{(AB}z_{CD)} = \epsilon^1_{ABCD}, \quad  z_{(AB} z_{CD)} =\frac{1}{2} \epsilon^0_{ABCD}.
\end{eqnarray*}
Also,
\[
\dnup{x}{A}{F}x_{BF}= \frac{1}{2}\epsilon_{AB} \quad \dnup{x}{A}{F}y_{BF} =-\frac{1}{\sqrt{2}} y_{AB}, \quad \dnup{x}{A}{F}z_{BF} =\frac{1}{\sqrt{2}} z_{AB},
\]
and
\begin{eqnarray*}
&& \updn{x}{F}{(A} \epsilon^0_{BCD)F} = \frac{1}{\sqrt{2}} \epsilon^0_{ABCD}, \quad \updn{x}{F}{(A} \epsilon^1_{BCD)F} = \frac{1}{\sqrt{2}} \epsilon^1_{ABCD}, \\
&& \updn{x}{F}{(A} \epsilon^2_{BCD)F} = 0, \\
&& \updn{x}{F}{(A} \epsilon^3_{BCD)F} = -\frac{1}{\sqrt{2}} \epsilon^3_{ABCD}, \quad \updn{x}{F}{(A} \epsilon^4_{BCD)F} = -\frac{1}{\sqrt{2}} \epsilon^4_{ABCD}.
\end{eqnarray*}

\section{Technical results concerning solutions of the constraints}
\label{appendix:momentum}

A detailed analysis of solutions to the vacuum Einstein constraint
equations such that near
spacelike infinity they admit asymptotic expansions of the form
\begin{eqnarray*}
&& \tilde{h}_{ij} \sim \left( 1 + \frac{2m}{\tilde{r}} \right) \delta_{ij} + \sum_{k\geq 2} \frac{\tilde{h}^k_{ij}}{\tilde{r}}, \\
&& \tilde{\chi}_{ij} \sim \sum_{k\geq 2} \frac{\tilde{\chi}^k_{ij}}{\tilde{r}^k}
\end{eqnarray*}
where $\tilde{h}^k_{ij}$ and $\tilde{\chi}^k_{ij}$ are smooth
functions on $\Sphere^2$ has been given in reference \cite{DaiFri01}. The
results given in the latter reference will constitute fundamental
building blocks of the analysis carried out in the present article.
This appendix offers a summary of the results of reference \cite{DaiFri01}
relevant for the analysis together with some extensions of their work
as the class of initial data considered in the present article turns
out to be more general than the one in reference \cite{DaiFri01}.

\subsection{Solutions to the Hamiltonian constraint}

A function $f\in C^\infty(\tilde{S})$ is
said to be in $E^\infty(\mathcal{B}_a)$ if on $\mathcal{B}_a$ one can
write $f=f_1 + rf_2$ with $f_1, \;f_2\in C^\infty(\mathcal{B}_a)$,
with $r^2=\delta_{ij}x^ix^j$. One could define in an
analogous fashion the spaces $E^k(\mathcal{B}_a)$ and
$E^\omega(\mathcal{B}_a)$.  Theorem 1 in \cite{DaiFri01} then states
that:

\begin{theorem}
\label{thm:DaiFri}
  Let $h_{ij}$ be a smooth metric on $\mathcal{S}$ with
  positive Ricci scalar. Assume that $\psi_{ij}$ is smooth in
  $\tilde{\mathcal{S}}$ and satisfies on $\mathcal{B}_a$
\[
r^8 \psi_{ij}\psi^{ij} \in E^\infty(\mathcal{B}_a).
\]
Then there exists on $\tilde{\mathcal{S}}=\mathcal{S}\setminus \{ i\}$ a unique solution
$\vartheta$ to the Licnerowicz equation (\ref{Licnerowicz:eqn}), which
is positive, satisfies $\lim_{r\rightarrow 0} r\vartheta =1$ and has
in $\mathcal{B}_a$ the form
\[
\vartheta = \frac{1}{r}(u_1+ru_2), \quad u_1\in\mathcal{B}_a, \quad u_2\in E^\infty(\mathcal{B}_a), \quad u_1=1+\mathcal{O}(r^2).
\]
\end{theorem}

A class of symmetric trace-free tensors $\psi_{ij}$ solving the
momentum constraint and satisfying the condition $r^8
\psi_{ij}\psi^{ij} \in E^\infty(\mathcal{B}_a)$ has been discussed in
section 4.3 of \cite{DaiFri01} and their asymptotic properties
in corollary 5. Unfortunately this class of tensors
$\psi_{ij}$ is not general enough for our purposes. Accordingly,
further analysis is required.

\subsection{Solutions to the momentum constraint}
\subsubsection{On the solutions to the flat space momentum constraint}

As in the main text, denote by $\mathring{\psi}_{ij}$ a solution to the
Euclidean momentum constraint and write
\[
\mathring{\psi}_{ij} = \mathring{\psi}_{ij}[P] + \mathring{\psi}_{ij}[A] + \mathring{\psi}_{ij}[J] + \mathring{\psi}_{ij}[Q] + \mathring{\psi}_{ij}[\lambda],
\]
where the $[P]$ denotes the part of $\mathring{\psi}_{ij}$ depending
on $P^i$, etc. It can be seen that ---see theorem 15 in reference 
\cite{DaiFri01}--- that if $r\lambda \in E^\infty(\mathcal{B}_a)$ and
$P^i=0$ then $r^8 \mathring{\psi}_{ij}\mathring{\psi}^{ij} \in
E^\infty(\mathcal{B}_a)$. This result justifies the choice of
$\lambda$ made in the main text ---cfr. Ansatz (\ref{Ansatz:lambda}).
The $h$-trace-free part
\[
\mathring{\Phi}_{ij}=\mathring{\psi}_{ij} -\frac{1}{3}h_{ij} h^{kl}\mathring{\psi}_{kl}
\]
of $\mathring{\psi}_{ij}$ will be used to specify the
freely-specifiable data in the solutions to the momentum constraint.
As mentioned in the main text, if $r\lambda \in
E^\infty(\mathcal{B}_a)$ then $\tilde{\lambda} =\tilde{\lambda}^{(1)}
+ \tilde{\lambda}^{(2)}/r$, with $\tilde{\lambda}^{(1)}, \;
\tilde{\lambda}^{(2)} \in C^\infty(\mathcal{B}_a)$. Accordingly, one
can write without loss of generality
\[
\tilde{\lambda}^{(1)} \sim \sum_{k\geq 2} \alpha_{i_1\cdots i_k} x^{i_1}\cdots x^{i_k}, \quad \tilde{\lambda}^{(2)} \sim \sum_{k\geq 2} \beta_{i_1\cdots i_k} x^{i_1}\cdots x^{i_k}
\]
with $\alpha_{i_1\cdots i_k}, \; \beta_{i_1\cdots i_k}\in \Complex$,
symmetric. It is recalled that the above expressions are to be
interpreted as
\[
\tilde{\lambda}^{(1)} = \sum_{k\geq 2}^m \alpha_{i_1\cdots i_k} x^{i_1}\cdots x^{i_k}+\tilde{\lambda}^{(1)}_R, \quad \tilde{\lambda}^{(2)} = \sum_{k\geq 2}^m \beta_{i_1\cdots i_k} x^{i_1}\cdots x^{i_k}+\tilde{\lambda}^{(2)}_R,
\]
with $\tilde{\lambda}^{(1)}_R=o(r^m)$ and
$\tilde{\lambda}^{(2)}_R=o(r^m)$ for all $m$.

\bigskip
The expressions given in section \ref{section:tensor_mc}
for the various possible solutions of the momentum constraint
are given in terms of its components with respect to a frame $\{ n^i,
m^i, \overline{m}^i \}$. This choice is not ideal for the subsequent discussion
of the solutions of equation (\ref{elliptic:eqn:v}). Therefore, the
expressions involving the parts of the solution arising from the
complex function $\tilde{\lambda}$ are reformulated in terms of a
Cartesian basis related to the normal coordinates $\{x^i\}$.  The
lengthy calculations, which are omitted here, reveal that the tensor
$\mathring{\psi}_{ij}[\mbox{Re}(\lambda^{(1)})]$ is a sum of terms of
the form
\[
\frac{A^{(1)}_k}{r^5}\mbox{Re}(\alpha_{i_1\cdots i_k})x^{i_1}\cdots x^{i_k} x_i x_j
+ \frac{B^{(1)}_k}{r^3}\mbox{Re}(\alpha_{i_1\cdots i_k})x^{i_1}\cdots x^{i_k} \delta_{ij} + \frac{C^{(1)}_k}{r^3} x_{(i} \mbox{Re}(\alpha_{j)i_1\cdots i_{p-1}}) x^{i_1}\cdots x^{p-1},
\]
where $A^{(1)}_k$, $B^{(1)}_k$ and $C^{(1)}_k$ are constants depending
on $k\geq 2$ such that the whole term is $\delta$-trace free and
$\delta$-divergence free. The contributions due to the imaginary part
of $\lambda^{(1)}$ render a tensor
$\mathring{\psi}_{ij}[\mbox{Im}(\lambda^{(1)})]$ which is a sum of
terms of the form
\begin{eqnarray*}
&&
\bigg(\frac{D^{(1,0)}_m}{r^4} x_{(i} \dnup{\epsilon}{j)}{kl}x_k \mbox{Im}(\alpha_{li_1 \cdots i_{m-1}}) x^{i_1}\cdots x^{i_{m-1}} +  \frac{D^{(1,2)}_m}{r^2} x_{(i} \dnup{\epsilon}{j)}{kl}x_k \mbox{Im}(\dnup{\alpha}{li_1 \cdots i_{m-3}j_1j_2}{j_1j_2})x^{i_1}\cdots x^{i_{m-3}} + \cdots \bigg) \\
&&\hspace{1cm}+ \bigg(  \frac{E^{(1,0)}_m}{r^2} x_k \updn{\epsilon}{kl}{(i} \mbox{Im}(\alpha_{j)li_1\cdots i_{m-2}})x^{i_1}\cdots x^{i_{m-2}}+ E^{(1,2)}_m x_k \updn{\epsilon}{kl}{(i} \mbox{Im}(\dnup{\alpha}{j)li_1\cdots i_{m-4}j_1j_2}{j_1j_2})x^{i_1}\cdots x^{i_{m-4}} + \cdots \bigg),
\end{eqnarray*}
where again $D^{(1,0)}_k,D^{(1,2)}_k,\cdots$ and
$E^{(1,0)}_k,E^{(1,2)}_k,\cdots$ are constants making the whole term
$\delta$-divergence free ---note that the terms are $\delta$-trace
free by construction. Similarly, the tensor
$\mathring{\psi}_{ij}[\mbox{Re}(\lambda^{(2)})/r]$ is a sum of terms of the form
\[
\frac{A^{(2)}_k}{r^6}\mbox{Re}(\beta_{i_1\cdots i_k})x^{i_1}\cdots x^{i_k} x_i x_j
+ \frac{B^{(2)}_k}{r^4}\mbox{Re}(\beta_{i_1\cdots i_k})x^{i_1}\cdots x^{i_k} \delta_{ij} + \frac{C^{(2)}_k}{r^4} x_{(i} \mbox{Re}(\beta_{j)i_1\cdots i_{p-1}}) x^{i_1}\cdots x^{p-1},
\]
where $A^{(2)}_k$, $B^{(2)}_k$ and $C^{(2)}_k$ are constants depending
on $k\geq 2$ such that the term is $\delta$-trace free and
$\delta$-divergence free. In addition, one has that
$\mathring{\psi}_{ij}[\mbox{Im}(\lambda^{(2)})/r]$ is a sum of terms
of the form
\begin{eqnarray*}
&&
\bigg(\frac{D^{(2,0)}_m}{r^5} x_{(i} \dnup{\epsilon}{j)}{kl}x_k \mbox{Im}(\beta_{li_1 \cdots i_{m-1}}) x^{i_1}\cdots x^{i_{m-1}} +  \frac{D^{(2,2)}_m}{r^3} x_{(i} \dnup{\epsilon}{j)}{kl}x_k \mbox{Im}(\dnup{\beta}{li_1 \cdots i_{m-3}j_1j_2}{j_1j_2})x^{i_1}\cdots x^{i_{m-3}} + \cdots \bigg) \\
&&\hspace{1cm}+ \bigg(  \frac{E^{(2,0)}_m}{r^3} x_k \updn{\epsilon}{kl}{(i} \mbox{Im}(\beta_{j)li_1\cdots i_{m-2}})x^{i_1}\cdots x^{i_{m-2}}+ \frac{E^{(2,2)}_m}{r} x_k \updn{\epsilon}{kl}{(i} \mbox{Im}(\dnup{\beta}{j)li_1\cdots i_{m-4}j_1j_2}{j_1j_2})x^{i_1}\cdots x^{i_{m-4}} + \cdots \bigg)
\end{eqnarray*}
where $D^{(2,0)}_k,D^{(2,2)}_k,\ldots$ and
$E^{(2,0)}_k,E^{(2,2)}_k,\ldots$ are constants making the term
$\delta$-divergence free ---note that the individual terms are $\delta$-trace
free by construction.

For future reference it will be convenient to write
\begin{eqnarray*}
&& \mathring{\psi}_{ij}[\mbox{Re}(\lambda^{(1)})] = r^{-5} \mathring{\Xi}_{ij}[\mbox{Re}(\lambda^{(1)}], \quad \mathring{\psi}_{ij}[\mbox{Im}(\lambda^{(1)})] = r^{-4} \mathring{\Xi}_{ij}[\mbox{Im}(\lambda^{(1)})], \\
&& \mathring{\psi}_{ij}[\mbox{Re}(\lambda^{(2)})/r] = r^{-6} \mathring{\Xi}_{ij}[\mbox{Re}(\lambda^{(2)})/r], \quad \mathring{\psi}_{ij}[\mbox{Im}(\lambda^{(2)})/r] = r^{-5} \mathring{\Xi}_{ij}[\mbox{Im}(\lambda^{(2)})/r],
\end{eqnarray*}
with $\mathring{\Xi}_{ij}[\mbox{Re}(\lambda^{(1)})], \; \mathring{\Xi}_{ij}[\mbox{Im}(\lambda^{(1)})],\; \mathring{\Xi}_{ij}[\mbox{Re}(\lambda^{(2)})/r],\;\mathring{\Xi}_{ij}[\mbox{Im}(\lambda^{(2)})/r]\in C^\infty(\mathcal{B}_a)$ ---and actually polynomial. It is noticed that
\begin{eqnarray*}
&& x^ix^j\mathring{\Xi}_{ij}[\mbox{Re}(\lambda^{(1)})]=r^2 \mathring{\Xi}[\mbox{Re}(\lambda^{(1)})], \quad x^ix^j\mathring{\Xi}_{ij}[\mbox{Im}(\lambda^{(1)})]=0, \\
&& x^ix^j\mathring{\Xi}_{ij}[\mbox{Re}(\lambda^{(2)})/r]=r^2 \mathring{\Xi}[\mbox{Re}(\lambda^{(2)})/r], \quad x^ix^j\mathring{\Xi}_{ij}[\mbox{Im}(\lambda^{(2)})/r]=0,
\end{eqnarray*}
for
$\mathring{\Xi}[\mbox{Re}(\lambda^{(1)})],\;\mathring{\Xi}[\mbox{Re}(\lambda^{(2)})/r]$
some scalar functions.

\subsubsection{General issues concerning the solutions to the non-flat momentum constraint}

An analysis of the solutions to the elliptic equation for the
vector $v^i$ ---see equation (\ref{elliptic:eqn:v})--- for free data of
the form
$\mathring{\Phi}_{ij}[A,J,Q,\mbox{Re}(\lambda^{(1)}),\mbox{Im}(\lambda^{(2)})/r]$,
has been given in theorem 16 and of reference \cite{DaiFri01}.

If the conformal metric $h_{ij}$ admits no conformal Killing vectors
on $\mathcal{S}$, then there exists a unique vector $v^i\in W^{2,q}$,
$q>1$ solving equation (\ref{elliptic:eqn:v}). If the conformal metric
admits conformal Killing vectors one can guarantee the existence of a solution
only if the constants $A$, $J^i$ and $Q^i$ satisfy a particular
relation. In order to discuss the asymptotic expansions of the vector
$v^i$ recall the definition of $\mathcal{Q}_\infty(\mathcal{B}_a)$ given in section
\ref{consequences:freedata}. Furthermore, introduce for $m\in
\Natural$, $m\geq 1$ the following real spaces of functions:
\[
\mathcal{Q}_m = \left \{ v^i\in C^\infty(\Real^3,\Real^3) \;|\; v^i\in \mathcal{P}_m, \; v^ix_i = r^2 v \mbox{ with }v\in\mathcal{P}_{m-1} \right\}, \\
\]
where $\mathcal{P}_m$ denotes the space of homogeneous polynomials of
degree $m$ ---that is, $v \in \mathcal{P}_m$ if and only if $v=
v_{i_1\cdots i_m} x^{i_1}\cdots x^{i_m}$, with $v_{i_1\cdots i_m}\in
\Real$ totally symmetric.

It turns out that a convenient way of grouping the different terms in
the free specifiable data is the following:
\[
\mathring{\psi}_{ij}= \mathring{\psi}_{ij}[A,J,Q] + \mathring{\psi}_{ij}[\mbox{Re}(\lambda^{(1)}),\mbox{Im}(\lambda^{(2)})/r]  + \mathring{\psi}_{ij}[\mbox{Re}(\lambda^{(2)})/r,\mbox{Im}(\lambda^{(1)})].
\]
We shall proceed to analyse the asymptotic expansions of the solutions
to (\ref{elliptic:eqn:v}) implied by each of these terms. However,
first an analysis of the source term $D^i\mathring{\Phi}_{ij}$ will be
required.

\subsubsection{Analysis of $D^i\mathring{\Phi}_{ij}$ }
As seen in sections \ref{section:freedata} and \ref{section:cngauge} ,
if the conformal metric $h_{ij}$ satisfies the cn-gauge, then one can
write
\[
h_{ij}= -\delta_{ij} + r^2 \check{h}_{ij}, \quad h^{ij}=-\delta_{ij} + r^2 \check{h}^{ij}.
\]
In addition, the technical assumption (\ref{technical_condition}) is recalled:
$\delta^{ij} \check{h}_{ij}=0, \quad \delta_{ij} \check{h}^{ij}=0$. A calculation renders
\begin{eqnarray*}
&& \Gamma_{kij}= \left( x_i \check{h}_{jk} + x_j \check{h}_{ik} -x_k\check{h}_{ij}\right) + \frac{1}{2} r^2 \left( \partial_i \check{h}_{kj} + \partial_j \check{h}_{ik} - \partial_k \check{h}_{ij}  \right) \\
&& \phantom{\Gamma_{kij}}= \tilde{\Gamma}_{kij} + r^2 \check{\Gamma}_{kij}.
\end{eqnarray*}
In addition, let
\[
\updn{\tilde{\Gamma}}{l}{ij} = h^{lk} \tilde{\Gamma}_{kij}, \quad \updn{\check{\Gamma}}{l}{ij} = h^{lk} \check{\Gamma}_{kij}.
\]
One readily verifies that
\[
x^i \updn{\tilde{\Gamma}}{l}{ij}= r^2 \updn{\check{h}}{l}{j}, \quad x^j \updn{\tilde{\Gamma}}{l}{ij}= r^2 \updn{\check{h}}{l}{i} \quad x_l \updn{\tilde{\Gamma}}{l}{ij}= r^2 \check{h}_{ij},
\]
and that
\[
h^{ij} \updn{\tilde{\Gamma}}{l}{ij}=-x^l h^{ij} \check{h}_{ij}.
\]
Now, recall that
\[
D^i \mathring{\Phi}_{ij} = D^i\mathring{\psi}_{ij} -\frac{1}{3} \updn{h}{i}{j} h^{kl} D_i \mathring{\psi}_{kl},
\]
and that
\[
D_k \mathring{\psi}_{ij} = \partial_k \mathring{\psi}_{ij} - \updn{\Gamma}{l}{ki} \mathring{\psi}_{lj} - \updn{\Gamma}{l}{kj}\mathring{\psi}_{il},
\]
so that
\begin{equation}
D^i \mathring{\psi}_{ij} = r^2 \check{h}^{ik} \partial_k \mathring{\psi}_{ij}
- h^{ki} \updn{\Gamma}{l}{ki}\mathring{\psi}_{lj} - h^{ki} \updn{\Gamma}{l}{kj}\mathring{\psi}_{il}. \label{Dpsi}
\end{equation}
In the last equation it has been used that by construction
\[
\delta^{ik}\partial_k \mathring{\psi}_{ij} =0.
\]

\bigskip
Now, consider
$\mathring{\Phi}_{ij}[\mbox{Re}(\lambda^{(2)})/r]$, which is the most singular
contribution of $\lambda$ to the seed tensor $\mathring{\Phi}_{ij}$. A direct calculation shows that $r^2 \check{h}^{ik} \partial_k \mathring{\psi}_{ij}$ is given by a sum of terms of the form
\begin{eqnarray*}
&&\hspace{-1.5cm}\frac{A^{(2)}_n}{r^4} \check{h}^{ik}\delta_{ik} \mbox{Re}(\beta_{i_1\cdots i_n}) x^{i_1}\cdots x^{i_n} x_j + \frac{nB^{(2)}_n}{r^2} \check{h}^{ik}\delta_{ij}  \mbox{Re}(\beta_{i_1\cdots i_{n-1}k}) x^{i_1}\cdots x^{i_{n-1}} + \frac{C^{(2)}_n}{2r^2} \check{h}^{ki}\delta_{ki}\mbox{Re}(\beta_{i_1\cdots i_{n-1}j})x^{i_1}\cdots x^{i_{n-1}} \\
&& \hspace{2cm}+ \frac{C^{(2)}_n}{2r^2} \check{h}^{ki}\delta_{kj}  \mbox{Re}(\beta_{i_1\cdots i_{n-1}i})x^{i_1}\cdots x^{i_{n-1}} + \frac{C^{(2)}_n(n-1)}{r^2} x_j\check{h}^{ki}  \mbox{Re}(\beta_{i_1\cdots i_{n-2}ki})x^{i_1}\cdots x^{i_{n-2}}.
\end{eqnarray*}
Hence, if the technical condition (\ref{technical_condition}) holds,
one can conclude that
\[
D^i \mathring{\Phi}_{ij}[\mbox{Re}(\lambda^{(2)})/r]= \frac{1}{r^2} S_j[\mbox{Re}(\lambda^{(2)})/r],
\]
with
\[
S_j[\mbox{Re}(\lambda^{(2)})/r] \in \mathcal{Q}_\infty(\mathcal{B}_a), \quad S_j[\mbox{Re}(\lambda^{(2)})/r]=O(r^4).
\]
A similar analysis can be carried out with the contributions due to $\mbox{Re}(\lambda^{(1)})$, $\mbox{Im}(\lambda^{(1)})$ and $\mbox{Im}(\lambda^{(2)})/r$. One obtains the following lemma.

\begin{lemma} \label{source:momentum}
For the class of initial data under consideration expressed in the cn-gauge, and assuming that condition (\ref{technical_condition}) holds, one has that:
\begin{eqnarray*}
&& D^i \mathring{\Phi}_{ij}[A]= \frac{1}{r} S_j[A], \quad S_j[A]\in \mathcal{Q}_\infty(\mathcal{B}_a), \quad S_j[A]=O(r^0), \\
&& D^i \mathring{\Phi}_{ij}[J]= \frac{1}{r^3} S_j[J], \quad S_j[J]\in \mathcal{Q}_\infty(\mathcal{B}_a), \quad S_j[J]=O(r^2), \\
&& D^i \mathring{\Phi}_{ij}[Q]= \frac{1}{r} S_j[Q], \quad S_j[Q]\in \mathcal{Q}_\infty(\mathcal{B}_a), \quad S_j[Q]=O(r), \\
&& D^i \mathring{\Phi}_{ij}[\mbox{\em Re}(\lambda^{(1)})]= \frac{1}{r} S_j[\mbox{\em Re}(\lambda^{(1)})], \quad  S_j[\mbox{\em Re}(\lambda^{(1)})] \in \mathcal{Q}_\infty(\mathcal{B}_a), \quad S_j[\mbox{\em Re}(\lambda^{(1)})]=O(r^4), \\
&& D^i \mathring{\Phi}_{ij}[\mbox{\em Im}(\lambda^{(1)})]= \frac{1}{r^2} S_j[\mbox{\em Im}(\lambda^{(1)})], \quad  S_j[\mbox{\em Im}(\lambda^{(1)})] \in \mathcal{Q}_\infty(\mathcal{B}_a), \quad S_j[\mbox{\em Im}(\lambda^{(1)})]=O(r^3),\\
&& D^i \mathring{\Phi}_{ij}[\mbox{\em Re}(\lambda^{(2)})/r]= \frac{1}{r^2} S_j[\mbox{\em Re}(\lambda^{(2)})/r],  \quad S_j[\mbox{\em Re}(\lambda^{(2)})/r] \in \mathcal{Q}_\infty(\mathcal{B}_a), \quad S_j[\mbox{\em Re}(\lambda^{(2)})/r]=O(r^4), \\
&& D^i \mathring{\Phi}_{ij}[\mbox{\em Im}(\lambda^{(2)})/r]=\frac{1}{r^3} S_j[\mbox{\em Im}(\lambda^{(2)})/r], \quad  S_j[\mbox{\em Im}(\lambda^{(2)})/r] \in \mathcal{Q}_\infty(\mathcal{B}_a), \quad S_j[\mbox{\em Im}(\lambda^{(2)})/r]=O(r^3).
\end{eqnarray*}
\end{lemma}

\subsubsection{Free data $\mathring{\psi}_{ij}[A,J,Q]$}
The case has been discussed in corollary 5 of \cite{DaiFri01}. Using
the results of lemma \ref{source:momentum} which hold for the cn-gauge
and assuming condition (\ref{technical_condition}), a direct
application of the techniques of section 4.2 of reference
\cite{DaiFri01} render
\begin{eqnarray*}
&& v^i[A]= r v^i_1[A] + v^i_2[A], \\
&& v^i[J]= \frac{1}{r} v^i_1[J] + v^i_2[J] \\
&& v^i[Q]=  \frac{1}{r} v^i_1[Q] + v^i_2[Q],
\end{eqnarray*}
with
\begin{eqnarray*}
&& v_1^i[A]\in Q_\infty(\mathcal{B}_a), \quad  v_1^i[A]=O(1), \quad  v_2^i[A]\in C^\infty(\mathcal{B}_a), \\
&& v_1^i[J]\in Q_\infty(\mathcal{B}_a), \quad  v_1^i[J]=O(r^2), \quad  v_2^i[J]\in C^\infty(\mathcal{B}_a), \\
&&  v_1^i[Q]\in Q_\infty(\mathcal{B}_a), \quad  v_1^i[Q]=O(r), \quad  v_2^i[Q]\in C^\infty(\mathcal{B}_a).
\end{eqnarray*}

\subsubsection{Free data $\mathring{\psi}_{ij}[\mbox{Re}(\lambda^{(1)}),\mbox{Im}(\lambda^{(2)})/r]$}
This case has not been discussed in complete generality in
\cite{DaiFri01}, but in view of the results of lemma
\ref{source:momentum} it is essentially a direct application of
theorem 17 there. One obtains that
\begin{eqnarray*}
&& v^i[\mbox{Re}(\lambda^{(1)})]=rv^i_1[\mbox{Re}(\lambda^{(1)})] + v^i_2[\mbox{Re}(\lambda^{(1)})], \\
&&  v^i[\mbox{Im}(\lambda^{(2)})/r]=\frac{1}{r}v^i_1[\mbox{Im}(\lambda^{(2)})/r] + v^i_2[\mbox{Im}(\lambda^{(2)})/r],
\end{eqnarray*}
with
\begin{eqnarray*}
&& v^i_1[\mbox{Re}(\lambda^{(1)})]\in Q_\infty(\mathcal{B}_a), \quad v^i_1[\mbox{Re}(\lambda^{(1)})]=O(r^4), \quad  v^i_2[\mbox{Re}(\lambda^{(1)})]\in C^\infty(\mathcal{B}_a), \\
&& v^i_1[\mbox{Im}(\lambda^{(2)})/r]\in Q_\infty(\mathcal{B}_a), \quad v^i_1[\mbox{Im}(\lambda^{(2)})/r]=O(r^3), \quad  v^i_2[\mbox{Im}(\lambda^{(2)})/r]\in C^\infty(\mathcal{B}_a)
\end{eqnarray*}

\subsubsection{Free data $\mathring{\psi}_{ij}[\mbox{Re}(\lambda^{(2)})/r,\mbox{Im}(\lambda^{(1)})]$}

The discussion of the asymptotic structure of solutions of equation
(\ref{elliptic:eqn:v}) if the free data is given by
$\mathring{\psi}_{ij}[\mbox{Re}(\lambda^{(2)})/r,\mbox{Im}(\lambda^{(1)})]$
is not directly covered by the techniques of \cite{DaiFri01}. The
reason for this is the appearance of terms with even powers of $1/r$
in the source terms $D^i\mathring{\Phi}_{ij}$. As it will be
shown, these cases ---which are the ones of more relevance for the
present work--- can be analysed under further assumptions on the free data.

The analysis in section \ref{section:weyltensor} shows that free data
of the form
$\mathring{\psi}_{ij}[A,J,Q]+\mathring{\psi}_{ij}[\mbox{Re}(\lambda^{(1)}),\mbox{Im}(\lambda^{(2)})/r]$
does not contribute to the regularity condition discussed in the
present article. In order to have a non-trivial contribution from the
second fundamental form in the regularity conditions at the critical
points $\mathcal{I}^\pm$ one has to consider free data of the form
$\mathring{\psi}_{ij}[\mbox{Re}(\lambda^{(2)})/r,\mbox{Im}(\lambda^{(1)})]$.
The complications concerning this case will be discussed in the
sequel.

In a neighbourhood $\mathcal{B}_a$ of infinity, the form that the conformal metric acquires in normal coordinates can be used to decompose the differential operator appearing in equation (\ref{elliptic:eqn:v}), namely
\[
\mathbf{L}_h v^i = D_k D^k v^i +\frac{1}{3} D^i D_k v^k + \updn{r}{i}{k} v^k
\]
in the form
\[
\mathbf{L}_h = \mathbf{L}_0 + \hat{\mathbf{L}}_h,
\]
where
\begin{eqnarray*}
&& \mathbf{L}_0 v_i = \partial^k \partial_k v_i +\frac{1}{3} \partial_i \partial^j v_j, \\
&& \hat{\mathbf{L}}_h v_i= \hat{h}^{jk} \partial_j \partial_k v_i + \frac{1}{3} \hat{h}^{jk} \partial_i \partial_k v_j + \updn{B}{jk}{i}\partial_j v_k + \updn{A}{j}{i}v_j,
\end{eqnarray*}
with $\updn{A}{j}{i}$ and $\updn{B}{kj}{i}$ functions of the metric coefficients and their first and second derivatives. They are smooth functions and satisfy
\[
\updn{A}{j}{i}=O(r), \quad \updn{B}{kj}{i}=O(r^2).
\]

\paragraph{Properties of the operators $\Delta_0$ and $\mathbf{L}_h$.} 
As before, let $\mathcal{P}_m$ denote the real linear space of
homogeneous polynomials of degree $m$. Likewise, let $\mathcal{H}_m$
denote the space harmonic polynomials of degree $m$. If
$\alpha_{i_1\cdots i_m}x^{i_1}\cdots x^{i_m} \in \mathcal{P}_m$,
$\alpha_{i_1\cdots i_m}=\alpha_{(i_1\cdots i_m)}$, then it is an
harmonic polynomial if and only if $ \alpha_{i_1\cdots i_m}$ is trace
free. It is well known that the space $\mathcal{P}_m$ can be written
as a direct sum
\[
\mathcal{P}_m = \mathcal{H}_m \oplus r^2 \mathcal{H}_{m-2} \oplus r^4 \mathcal{H}_{m-4} \oplus \cdots.
\]
Let $s\in \Integer$,
\[
\quad \Delta_0 =\partial^k \partial_k, \quad \Delta_0: r^s \mathcal{P}_m \rightarrow r^{s-2} \mathcal{P}_m,  
\]
defines a bijective linear map
if either $s>0$ or $s<0$, $|s|$ is odd and $m+s \geq 0$ ---see
\cite{DaiFri01}.  If one wants to discuss the bijectivity of the
Laplacian for functions of the form $r^s p_m$ with $s<0$, $|s|$ even
and $p_m \in \mathcal{P}_m$, one has to restrict the domain and the
range sets. To this end define
\[
\mathcal{P}_{m,s} \equiv \mathcal{H}_m \oplus r^2 \mathcal{H}_{m-2} \cdots \oplus r^{m-|s|-1}\mathcal{H}_{|s|+1}
\]
Elaborating from the proof of lemma 3 in \cite{DaiFri01} one finds the
following lemma.

\begin{lemma}
Let $s \in \Integer$, with $s<0$, $|s|$ even, then
\[
\Delta_0: r^s \mathcal{P}_{m,s} \rightarrow r^{s-2} \mathcal{P}_{m,s}
\]
is a bijective mapping if $m+s\geq 0$.
\end{lemma}

The operator $\mathbf{L}_0$ has nice properties with regard to the
spaces $\mathcal{Q}_m$. Indeed, if $s\in \Integer$, then
\[
\mathbf{L}_0: r^s \mathcal{Q}_m \rightarrow r^{s-2} \mathcal{Q}_m
\]
is a bijective linear mapping again if $s>0$ or $s<0$, $|s|$ is odd
and $m+s \geq 0$. As in the case of the Laplacian, in order to obtain bijectivity for $s<0$, $|s|$ even, one has to restrict both domain and range. Let
\[
\mathcal{Q}_{m,s}\equiv \left \{ v^i \in C^\infty(\Real^3,\Real^3) \;|\; v^i\in \mathcal{P}_{m,s}, \; v^ix_i=r^2 v, \; v\in \mathcal \;{P}_{m,s} \right\},
\]
then again, following closely the arguments of \cite{DaiFri01} one can
prove the following lemma ---cfr. lemma 11 in \cite{DaiFri01}.

\begin{lemma}
\label{lemma:invert_L}
Let $s \in \Integer$, with $s<0$, $|s|$ even, then
\[
\mathcal{L}_0: r^s \mathcal{Q}_{m,s} \rightarrow r^{s-2} \mathcal{Q}_{m,s}
\]
is a bijective mapping if $m+s\geq 0$.
\end{lemma}

\paragraph{Concluding the analysis.} Returning now to the behaviour of
solutions to equation (\ref{elliptic:eqn:v}) with seeds
$\mathring{\psi}_{ij}[\mbox{Re}(\lambda^{(2)})/r,\mbox{Im}(\lambda^{(1)})]$,
lemma \ref{source:momentum} renders that
\[
D^i\mathring{\Phi}_{ij}[\mbox{Re}(\lambda^{(2)})/r,\mbox{Im}(\lambda^{(1)})] =r^{-2} S_j[\mbox{Re}(\lambda^{(2)})/r,\mbox{Im}(\lambda^{(1)})], \quad S_j[\mbox{Re}(\lambda^{(2)})/r,\mbox{Im}(\lambda^{(1)})]=O(r^3),
\]
with $S_j[\mbox{Re}(\lambda^{(2)})/r,\mbox{Im}(\lambda^{(1)})]\in
C^\infty(\mathcal{S})$. For the ease of notation, in what follows the
affix $[\mbox{Re}(\lambda^{(2)})/r,\mbox{Im}(\lambda^{(1)})]$ will be
dropped. One can write for some arbitrary $m\in \Natural$
\[
S^i= \sum_{k=3}^m s^{i}_{(k)} + s^i_\mathcal{R}, \quad s^i_{(k)}\in \mathcal{P}_k,
\]
with $s^i_\mathcal{R}=o(r^m)$. The latter suggests considering
\[
v_i = \sum^m_{k=s} v^i_{(k)} + v^i_\mathcal{R},
\]
so that
\[
\mathbf{L}_h v_i = \mathbf{L}_0 \left(\sum^m_{k=3} v^i_{(k)}  \right) + \hat{\mathbf{L}}_h \left(\sum^m_{k=3} v^i_{(k)}  \right) + \mathbf{L}_h \left( v^i_\mathcal{R} \right).
\]
Using lemma 12 in reference \cite{DaiFri01} one readily finds that
\[
\hat{\mathbf{L}}_h \left( \sum^m_{k=3} v^i_{(k)}  \right) = \sum^m_{k=3} \hat{\mathbf{L}}_h(v^i_{(k)}) = r^{-2} \sum_{k=3}^m u^i_{(k)},
\]
with $u^i_{(k)}= O(r^{k+3})$, $u^i_{(k)} \in \mathcal{Q}_k(\mathcal{B}_a)$. The functions $u^i_{(k)}$ can also be written as
\[
u^i_{(k)} = \sum^m_{j=k+3} u^i_{(k,j)} + u^i_{(k)\mathcal{R}}, \quad u^i_{(k,j)} \in \mathcal{P}_j, \quad u^i_{(k,j)}=o(r^m),
\]
so that
\begin{eqnarray*}
&& \hat{\mathbf{L}}_h \left( \sum^m_{k=3} v^i_{(k)}  \right) = r^{-2} \sum_{k=3}^m \sum_{j=k+3}^m u^i_{(k,j)} + r^{-2} \sum_{k=3}^m u^i_{(k)\mathcal{R}}, \\
&& \phantom{\widehat{\mathbf{L}}_h \left( \sum^m_{k=3} v^i_{(k)}  \right)}=r^{-2} \sum_{j=6}^m \sum_{k=3}^{j-3} u^i_{(k,j)}+ r^{-2} \sum_{k=3}^m u^i_{(k)\mathcal{R}}.
\end{eqnarray*}

The above analysis suggests calculating the polynomials
$v^i_{(k)}$ recursively if one can find solutions to the equations
\begin{eqnarray}
&& \mathbf{L}_0 \left( v^i_{(k)} \right) = r^{-2} s^i_{(k)}, \quad 6 \leq k \leq 8 \label{recursive_1}\\
&& \mathbf{L}_0 \left( v^i_{(k)} \right) = r^{-2} \left( s^i_{(k)} - \sum_{j=3}^{k-3} u^i_{(j,k)} \right), \quad 9 \leq k \leq m. \label{recursive_2}
\end{eqnarray}
From lemma \ref{lemma:invert_L} one sees that one can only find
polynomials $v^i_{(k)}$ solving (\ref{recursive_1}) and
(\ref{recursive_2}) if the right-hand sides are in
$r^{-6}\mathcal{Q}_{k,3}$. Thus, in general, solutions to 
equation (\ref{elliptic:eqn:v}) with seed
$\mathring{\psi}_{ij}[\mbox{Re}(\lambda^{(2)})/r,\mbox{Im}(\lambda^{(1)})]$
cannot be expected to be smooth at $i$, and its asymptotic expansions
may have terms involving the function $\ln r$.

The conditions
\begin{eqnarray*}
&& s^i_{(k)} \in \mathcal{Q}_{k,3}, \quad 6 \leq k \leq 8, \\
&& \left( s^i_{(k)} - \sum_{j=6}^{k-3} u^i_{(j,k)} \right) \in \mathcal{Q}_{k,3}, \quad 9 \leq k \leq m,
\end{eqnarray*}
can, in principle, be reformulated as conditions on the free data
$h_{ij}$ and $\mbox{Re}(\lambda^{(2)})/r$, $\mbox{Im}(\lambda^{(1)})$, so
that they cannot be independent of each other. \emph{We shall neither be
  concerned with a detailed analysis of these conditions nor with an explicit formulation of them as this goes beyond the scope of the present work. Through out it will be assumed that they are satisfied}. Nevertheless, it is important
to remark that there exists a class of free-data,
$(h_{ij},\mathring{\psi}_{ij})$, with $h_{ij}$ smooth for which this
is the case: axial symmetric data ---see \cite{Dai04a}. The proof that
this is the case is done by methods different to the ones discussed
here and exploits that in axial symmetry, one has an explicit
solution, $\psi_{ij}$, to the momentum constraint given in terms of a
potential.

Once the polynomials $v^i_{(k)}$ have been determined, one is left to
analyse the behaviour of the remainder $v^i_R$. The remainder
satisfies the equation
\[
\mathbf{L}_h v^i_R = r^{-2} \left( \Xi^i_R-\sum_{k=8}^m u^i_{(k)R} \right).
\]
Using the methods of theorem 17 in \cite{DaiFri01} one concludes that
the right-hand side of this equation is in
$C^{m+s-2,\alpha}(\mathcal{B}_a)$. Standard results on elliptic
regularity then yield $v^i_R \in C^{m+s,\alpha}(\mathcal{B}_a)$. Since
$m$ is arbitrary, one can conclude, by using a further argument that
$v^i_R \in C^\infty(B_a)$.

\bigskip
Summarising, if $v^i[\mbox{Re}(\lambda^{(2)})/r,\mbox{Im}(\lambda^{(1)})]$ is the solution
to equation (\ref{elliptic:eqn:v}) with with seed given by
$\mathring{\psi}_{ij}[\mbox{Re}(\lambda^{(2)})/r,\mbox{Im}(\lambda^{(1)})]$, then
\[
v^i[\mbox{Re}(\lambda^{(2)})/r,\mbox{Im}(\lambda^{(1)})]=v^i_1[\mbox{Re}(\lambda^{(2)})/r,\mbox{Im}(\lambda^{(1)})] + v_2^i[\mbox{Re}(\lambda^{(2)})/r,\mbox{Im}(\lambda^{(1)})],
\]
with
\[
 v^i_1[\mbox{Re}(\lambda^{(2)})/r,\mbox{Im}(\lambda^{(1)})]\in\mathcal{Q}_{\infty}(\mathcal{B}_a), \quad v^i_1[\mbox{Re}(\lambda^{(2)})/r,\mbox{Im}(\lambda^{(1)})]=O(r^3), \quad
v^i_1[\mbox{Re}(\lambda^{(2)})/r,\mbox{Im}(\lambda^{(1)})]\in C^\infty(\mathcal{B}_a).
\]

\subsubsection{The condition $\psi_{ij}\psi^{ij}\in E^\infty(\mathcal{B}_a)$}
In order to be able to use theorem \ref{thm:DaiFri} with the
$\psi_{ij}$ which has been constructed in the previous paragraphs, one
has to verify that
\[
\psi_{ij}= \mathring{\psi}_{ij}[A,J,Q] +\mathring{\psi}_{ij}[\lambda^{(1)}] + \mathring{\psi}_{ij}[\lambda^{(2)}] + (\mathcal{L}v)_{ij}[A,J,Q] + (\mathcal{L}_hv)_{ij}[\lambda^{(1)}]+ (\mathcal{L}_hv)_{ij}[\lambda^{(2)}],
\]
satisfies $r^8 \psi_{ij} \psi^{ij}\in E^\infty(\mathcal{B}_a)$. A
lengthy calculation using the ideas of section 4.3 in \cite{DaiFri01}
together with lemma \ref{source:momentum} shows that this is the case.


\end{document}